\documentclass[%
 reprint,
superscriptaddress,
 amsmath,amssymb,
 aps,
 prx,
floatfix,
]{revtex4-2}

\usepackage{graphicx}
\usepackage[colorlinks]{hyperref}
\usepackage{physics}
\usepackage{caption}
\usepackage{subcaption}

\newcommand{\Lind}[1]{{\cal D}[#1](\rho)}

\newcommand{\adag}{a^{\dagger}}

\newcommand{\kb}{k_{\textrm B}}
\newcommand{\nbar}{\bar{n}}
\newcommand{\grvdp}{\gamma_{2}^{\textrm{RvdP}}}
\newcommand{\gr}{\gamma_{2}^{\textrm{Rayleigh}}}
\newcommand{\gvdp}{\gamma_{2}^{\textrm{vdP}}}

\begin{document}

\title{Quantum limit-cycles and the Rayleigh and van der Pol oscillators}

\author{Lior Ben Arosh}
\affiliation{Raymond and Beverly Sackler School of Physics and Astronomy, Tel Aviv University, Tel Aviv 69978, Israel}

\author{M.~C.~Cross}
\affiliation{Condensed Matter Physics, California Institute of Technology, Pasadena, California 91125, USA}

\author{Ron Lifshitz}
\email[Corresponding author:\ ]{ronlif@tau.ac.il}
\affiliation{Raymond and Beverly Sackler School of Physics and Astronomy, Tel Aviv University, Tel Aviv 69978, Israel}

\date{\today}


\begin{abstract}
Self-oscillating systems, described in classical dynamics as limit cycles, are emerging as canonical models for driven dissipative nonequilibrium open quantum systems, and as key elements in quantum technology. We consider a family of models that interpolates between the classical textbook examples of the Rayleigh and the van der Pol oscillators, and follow their transition from the classical to the quantum domain, while properly formulating their corresponding quantum descriptions. We derive an exact analytical solution for the steady-state quantum dynamics of the simplest of these models, applicable to any bosonic system---whether mechanical, optical, or otherwise---that is coupled to its environment via single-boson and double-boson emission and absorption. Our solution is a generalization to arbitrary temperature of existing solutions for very-low, or zero, temperature, often misattributed to the quantum van der Pol oscillator. We closely explore the classical to quantum transition of the bifurcation to self-oscillations of this oscillator, while noting changes in the dynamics and identifying features that are uniquely quantum.

\end{abstract}


\maketitle

\section{Introduction}

Self-oscillating systems are ubiquitous---from human-made clocks and transistors, through heart cells and neurons in the living body, to flashing fireflies and circadian rhythms---and are now emerging as canonical models for driven dissipative nonequilibrium open quantum systems, and as key elements in quantum technology. The dynamics of self oscillation are captured mathematically by the notion of a limit-cycle. Here we consider a family of models that interpolates between the Rayleigh~\cite{Rayleigh} and the van der Pol (vdP)~\cite{vdp} oscillators, which are probably the most common textbook examples of limit-cycles in classical nonlinear dynamics. These models consist of a simple harmonic oscillator, driven by a time-independent energy pump in the form of ``negative damping.'' When the pumping rate exceeds that of the normal damping rate, self-oscillations develop, which are then saturated by a nonlinear form of damping. The frequency of the oscillation is set by the physical parameters of the oscillator, while the magnitude of the oscillation is set by the ratio of the linear to the nonlinear damping rates. This provides a convenient knob with which to transition the oscillator from large-amplitude classical behavior to small-amplitude quantum behavior, which is our focus here.

Existing models for quantum limit cycles~\cite{lorch14} consist of a harmonic, or possibly anharmonic, quantum oscillator, with linear as well as nonlinear coupling to the environment, which are expressed in terms of quantum Lindblad operators. These models are currently being used to study quantum entrainment~\cite{walter14},  synchronization~\cite{lee13,*lee14,walter15,*lorch16,davis18} and the phenomenon of ``oscillation collapse'' or ``amplitude death''~\cite{ishibashi17,amitai18} in systems of coupled self-sustained oscillators, as well as the nonequilibrium spectral properties~\cite{scarlatella19}, and the critical response to external drive~\cite{dutta19}, of single oscillators.

Our current focus is more basic. The classical Rayleigh and vdP oscillators are known for exhibiting a Hopf bifurcation, from a state of no motion at all to a state of self-oscillations at a fixed amplitude. We seek to characterize this bifurcation as the system transitions from the classical to the quantum domain. Our goal is to find answers to such questions as: How exactly should one model the Rayleigh and vdP oscillators in quantum mechanics? Can the quantum model analytically be solved, at least in its steady state? Is the quantum bifurcation different from the classical one? What experimentally observable indications are there to distinguish between quantum and classical behavior? What would be the first corrections to classical dynamics as one approaches the quantum domain?

Answers to these questions are relevant to a broad range of physical systems exhibiting quantum behavior, including lasers, or more generally photonic systems with nonlinear loss~\cite{simaan75, *simaan78, dodonov97, leghtas15}, as well as trapped ions~\cite{lee13, Akerman10} and electronic or superconducting circuits~\cite{Siddiqi04}. Particularly interesting is the attempt to observe such quantum behavior in nanotechnology-based human-made mechanical systems~\cite{Roukes07}. 
Indeed, modern nanomechanical resonators show exceptional behavior, as they routinely operate in the GHz range~\cite{huang03,*huang05}. With nano-electromechanical systems (NEMS)~\cite{Craighead00,*Roukes01,*ekinci05} and nano-optomechanical systems (NOMS)~\cite{aspelmeyer14} it is now possible to perform ultrasensitive measurements of physical quantities~\cite{li07, *arash15} such as single spins~\cite{rugar04}, minute charges~\cite{Cleland98}, and tiny masses~\cite{Ilic04, *yang06, *buks06, *Lassagne08, *jensen08}. Relatively weak drive is needed in order for nonlinearity to be evident in the dynamics of nanomechanical systems~\cite{lifshitz08,lifshitz12}, which is experimentally observed~\cite{aldridge05,*kozinsky06,*Kozinsky07,*Karabalin09,*Matheny13} and also exploited for applications~\cite{Karabalin11, *Kenig12, *Villanueva13}. Most importantly, at GHz frequencies, one need only cool to temperatures on the order of tens to even hundreds of mK for the thermal energy to become comparable to the quantum energy-level spacing of the mechanical resonator. This allows now to cool mechanical resonators down to their quantum ground state~\cite{lahaye04,*Rocheleau10,*OConnell10,*Teufel11, *Chan11, *qiu20}, and to start investigating fundamental physical questions on the borderline between the quantum and the classical worlds~\cite{Leggett02,*Blencowe04,*Schwab05,*Meystre13,*Chen13}, as it applies to human-made macroscopic nonlinear mechanical objects. This, in turn, requires a well-based quantum theoretical framework. 

We employ a phase-space approach to study the correspondence between  classical and quantum limit-cycles. Since classical notions like a particle trajectory do not have a straightforward quantum analog, it is reasonable to compare quantum expectation values with classical statistical ensemble averages. We do so by solving the classical equations of motion for many different initial conditions (typically $N = 10^4$) taken from a Gaussian distribution, and keeping track of the different trajectories, thus representing a statistical distribution over phase space. The width of the initial distribution in phase space is taken to be the same as the quantum uncertainties $\Delta x$ and $\Delta p$ of an initial coherent-state wave function. In addition to expectation values, we also compare the full classical distribution with the quantum Wigner function $W(x, p)$. The quantum dynamics are those of an open quantum system, and therefore described by a density matrix and its master equation, which dictates the steady state, and more generally, the dynamics of the quantum system.  

We begin in section~\ref{Sec:Classical} with theoretical background for the classical dynamics of a family of models described by a generalized Rayleigh-van der Pol equation of motion~\eqref{Eq:GRvdP}, which interpolates continuously between the pure Rayleigh oscillator and the pure vdP oscillator. We provide a perturbative steady-state solution for limit cycles that are nearly-circular in phase space, obtained for weak driving just above the Hopf bifurcation to the oscillatory state. Moreover, we note that this solution is exact, and the limit cycles are always circular, for the model that lies exactly halfway between the pure Rayleigh and pure vdP oscillators, which we call \emph{the} Rayleigh-van der Pol (RvdP) oscillator. In section~\ref{sec:quantum_models} we introduce three quantum models, differing in the form of the nonlinear coupling of the oscillator to the environment. We discuss the basic features of these quantum models, and show that, for weak driving, their classical limits correspond to the RvdP oscillator (sec.~\ref{sec:quantum_RvdP}), and to the pure vdP (sec.~\ref{sec:quantum_vdP}), and pure Rayleigh oscillators (sec.~\ref{sec:quantum_Rayleigh}). In sec.~\ref{Sec:correlations} we employ time correlation functions to elucidate some of the differences between these models. In section~\ref{sec:Analytic_sol} we derive an exact analytical solution for the steady-state dynamics of the quantum RvdP oscillator, which is a generalization to arbitrary temperature of existing solutions for very-low, or zero, temperature, often misattributed to the quantum vdP oscillator. In section~\ref{sec:QCT} we consider in some detail the transition from classical to quantum dynamics of the RvdP oscillator, identifying dynamical behavior that is unique to the quantum domain. We conclude with a few summarizing remarks in section~\ref{Sec:Conclusion}.

\section{The Classical Rayleigh and van der Pol Oscillators}
\label{Sec:Classical}

Consider the following classical equation of motion, describing a harmonic oscillator with effective mass $m$ and natural frequency $\omega$,  
\begin{equation}\label{Eq:classical-models-unscaled}
    m\dv[2]{\tilde{x}}{\tilde{t}} + m{\omega}^2\tilde{x} 
    = \left(\tilde{\kappa}_1 - \tilde{\gamma}_1\right)
    \dv{\tilde{x}}{\tilde{t}}  
    - \tilde{\eta}\tilde{x}^2\dv{\tilde{x}}{\tilde{t}}
    - \tilde{\zeta} \left(\dv{\tilde{x}}{\tilde{t}}\right)^3,
\end{equation}
where tildes denote physical parameters that are soon to be rescaled. The oscillator is driven by a velocity-dependent force or ``negative damping'', with coefficient $\tilde{\kappa}_1\geq0$, as described earlier. It also experiences normal linear damping, with coefficient $\tilde{\gamma}_1\geq0$, which is unavoidable in most physical systems, as well as two types of nonlinear damping mechanisms: \emph{vdP damping} with coefficient $\tilde{\eta}\geq0$, which is proportional to the velocity and the squared displacement of the oscillator, and \emph{Rayleigh damping} with coefficient $\tilde{\zeta}\geq0$, which is proportional to the cubed velocity of the oscillator.

To obtain a dimensionless equation of motion we (a) measure mass in units of $m$, effectively setting $m$ in Eq.~\eqref{Eq:classical-models-unscaled} to unity; (b) measure inverse time in units of the oscillator frequency $\omega$ by defining
\begin{equation}\label{Eq:timescale}
    t = \omega\tilde{t},
\end{equation}
which effectively sets $\omega$ to unity; (c) measure length in units of $x_0=\sqrt{\hbar/m\omega}$ by setting
\begin{equation}\label{Eq:lengthscale}
    x = \frac{\tilde{x}}{x_0}
    = \tilde{x}\sqrt{\frac{m\omega}{\hbar}},
\end{equation}
in anticipation of the quantum treatment below; and consequently, (d) measure the pumping and damping rates with respect to the chosen units of mass and time, by defining
\begin{equation}\label{Eq:scale_classical}
    \kappa_1 =  \frac{\tilde{\kappa}_1}{m\omega};\
    \gamma_1 =  \frac{\tilde{\gamma}_1}{m\omega};\
    \gamma_2\eta = \frac{\hbar\tilde{\eta}}{m^2 \omega^2};
    \textrm{\ and\ }
    \gamma_2\zeta =  \frac{\hbar\tilde{\zeta}}{m^2};
\end{equation}
where $\gamma_2>0$ is an overall dimensionless nonlinear damping rate, and $\eta$ and $\zeta$ are numerical factors, indicating the relative contributions of the two nonlinear damping mechanisms. Without loss of generality, one can set the larger of the two to unity, and the smaller to a number between 0 and 1. 

\begin{figure}
    \begin{subfigure}[t]{0.15\textwidth}
    \includegraphics[width=1\textwidth]{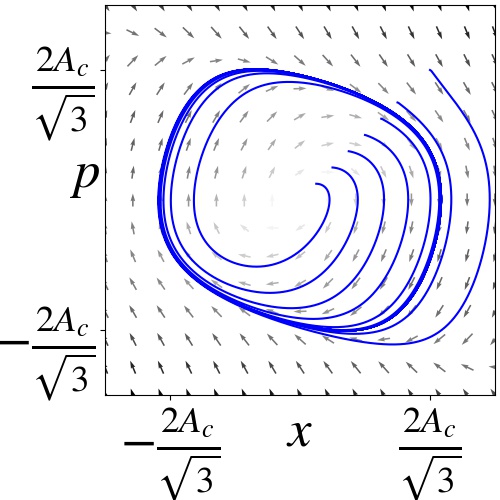}
    \caption{$\eta=0$; $\zeta=1$}
    \label{}
    \end{subfigure}
    \hfill
    \begin{subfigure}[t]{0.15\textwidth}
    \includegraphics[width=1\textwidth]{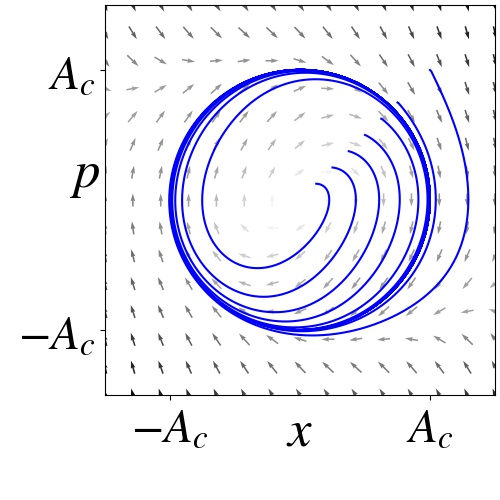}
    \caption{$\eta=\zeta=1$}
    \label{}
    \end{subfigure}
    \hfill
    \begin{subfigure}[t]{0.15\textwidth}
    \includegraphics[width=1\textwidth]{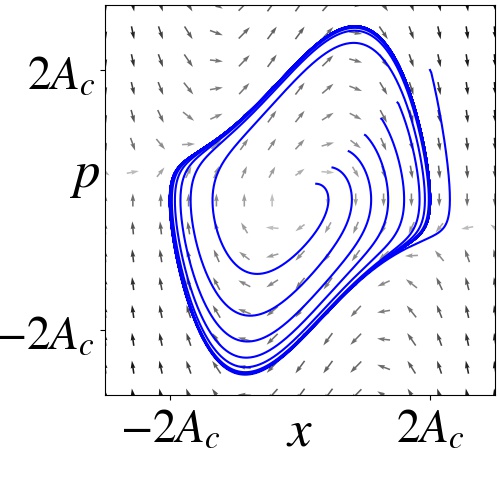}
    \caption{$\eta=1$; $\zeta=0$}
    \label{}
    \end{subfigure}
    \hfill
    \begin{subfigure}[b]{0.15\textwidth}
    \includegraphics[width=1\textwidth]{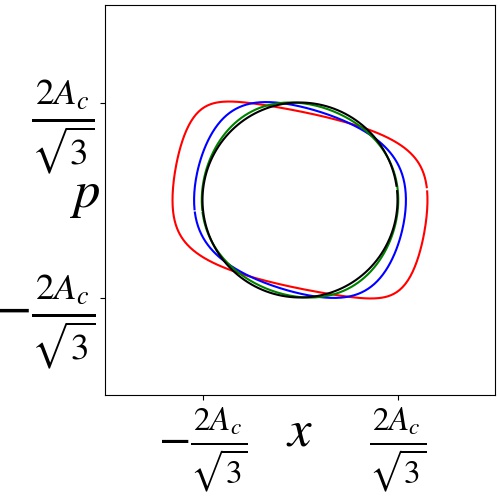}
    \caption{$\eta=0$; $\zeta=1$}
    \label{}
    \end{subfigure}
    \hfill
    \begin{subfigure}[b]{0.15\textwidth}
    \includegraphics[width=1\textwidth]{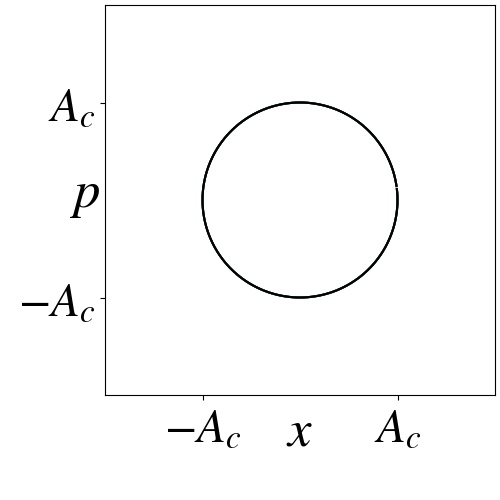}
    \caption{$\eta=\zeta=1$}
    \label{}
    \end{subfigure}
    \hfill
    \begin{subfigure}[b]{0.15\textwidth}
    \includegraphics[width=1\textwidth]{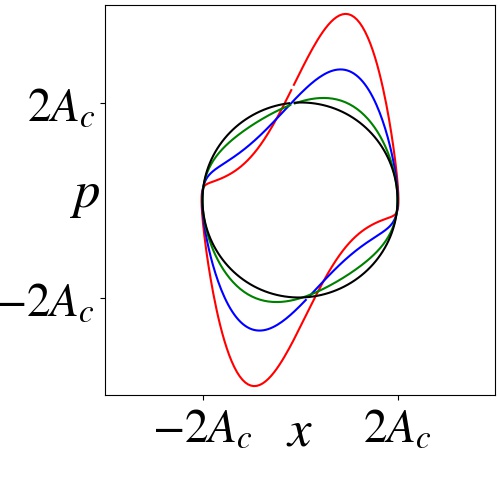}
    \caption{$\eta=1$; $\zeta=0$}
    \label{}
    \end{subfigure}
    \hfill
    \caption{
    From left to right: Limit cycles of the Rayleigh, the Rayleigh-van der Pol, and the van der Pol oscillators, as given by Eq.~\eqref{Eq:GRvdP}.
    Top row: Convergence to the limit cycles from different initial conditions, with $\epsilon=1$.
    Bottom row: Scaled limit cycles for different values of $\epsilon=0.01,0.3,1,2$, demonstrating that the RvdP oscillator remains circularly symmetric for all values of $\epsilon$.}
    \label{fig:classical_limit_cycles}
\end{figure}

Finally, we divide the original equation of motion~\eqref{Eq:classical-models-unscaled} by the characteristic unit of force, $m\omega^2x_0$, yielding a scaled dimensionless equation of the form
 \begin{equation}\label{Eq:GRvdP}
    \Ddot{x} + x -\epsilon \dot{x} + \gamma_2\left(\eta x^2 + \zeta \dot{x}^2\right)\dot x = 0, 
 \end{equation}
where $\epsilon = \kappa_1-\gamma_1$, and dots denote derivatives with respect to the dimensionless time $t$. 
This \emph{generalized Rayleigh-van der Pol equation} is usually studied in one of its following limiting cases:
(1) the Rayleigh oscillator~\cite{Rayleigh} with $\eta=0,\zeta=1$; (2) the van der Pol (vdP) oscillator~\cite{vdp} with $\eta=1,\zeta=0$; and (3) the Rayleigh-van der Pol (RvdP) oscillator with $\eta=\zeta=1$, which is sometimes refered to as the harmonic RvdP oscillator~\cite{Nathan77}.
All these variants are known to generate steady-state limit cycles for positive $\epsilon$, as shown in Fig.~\ref{fig:classical_limit_cycles}. 

In the weak-drive limit of small $\epsilon$, with nearly circular orbits, one can use secular perturbation theory~\cite{lifshitz08, lifshitz12} to obtain an approximate solution for the generalized RvdP equation of motion~\eqref{Eq:GRvdP}, and determine the amplitude of limit-cycle oscillations. The solution is written as a slow modulation of harmonic oscillations at unit frequency, with $\epsilon$-dependent corrections
\begin{equation}\label{Eq:power}
    x(t) = \frac{1}{2}\left[A(T) e^{-it} + c.c.\right] + \epsilon x_1(t) + \order{\epsilon^2},
\end{equation} 
where $T=\epsilon t$ is a slow time scale, characteristic of the rate of relaxation toward the limit cycle, as opposed to the fast time scale $t$ of the oscillations themselves. As usual, $c.c.$ stands for the complex conjugate.

\begin{figure}[b]
    \begin{subfigure}[t]{0.507\linewidth}
    \centering
    \includegraphics[width=1\linewidth ]{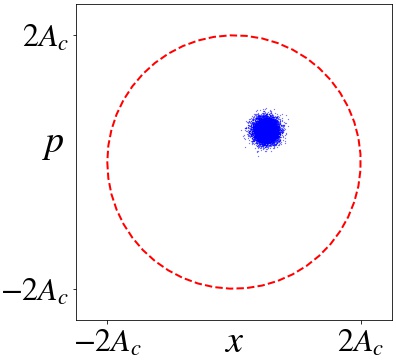}
    \caption{$t=0$}
    \label{}
    \end{subfigure}
    \begin{subfigure}[t]{0.46\linewidth}
    \centering
    \includegraphics[width=1\linewidth ]{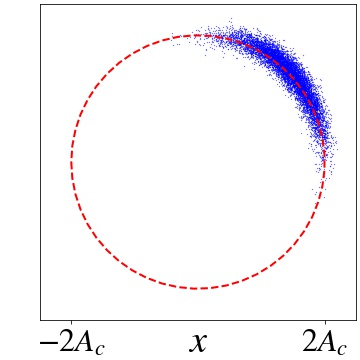}
    \caption{$t=20\pi$}
    \label{}
    \end{subfigure}
    \begin{subfigure}[t]{0.507\linewidth}
    \centering
    \includegraphics[width=1\linewidth ]{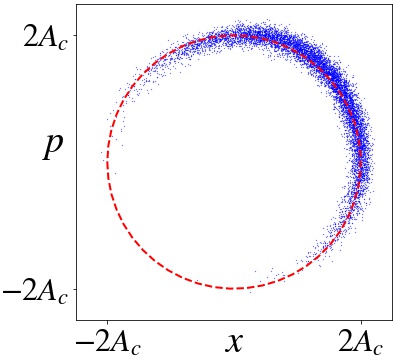}
    \caption{$t=200\pi$}
    \label{}
    \end{subfigure}
    \begin{subfigure}[t]{0.46\linewidth}
    \centering
    \includegraphics[width=1\linewidth ]{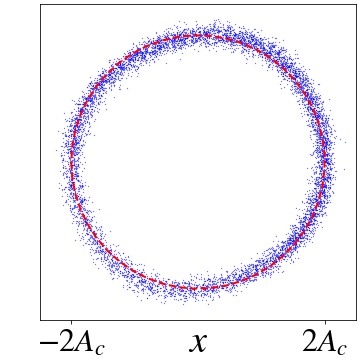}
    \caption{$t=2000\pi$}
    \label{}
    \end{subfigure}
    \caption{Ensemble of $10^4$ classical vdP oscillators, initially sampled from a Gaussian distribution around $(A_c/2,A_c/2)$ with standard deviation $0.1 A_c$. Dynamics are simulated according to Eq.~\eqref{Eq:GRvdP} with $\eta=1$, $\zeta=0$, $\epsilon=0.1$, and delta-correlated Gaussian white noise, corresponding to a dimensionless temperature $T=0.01{A_c}^2$, or to thermal energy on the order of $0.01$ of the oscillator energy. The dashed red circles have radius $2A_c$. }
    \label{Fig:classic_ensemble_w_t}
\end{figure}

The slow time variation of the complex amplitude $A(T)$ also provides the freedom to eliminate secular terms, and to ensure that the perturbative correction $x_1(t)$, as well as all higher-order corrections do not diverge. Substituting the solution~\eqref{Eq:power} into the equation of motion~\eqref{Eq:GRvdP} indeed generates such a secular term~\cite[Section 11.4]{lifshitz12}, which when required to vanish leads to a first-order differential equation for determining the slowly varying amplitude,
\begin{equation}\label{Eq:amp}
    \dv{A}{T} = \frac{1}{2}\left(1 - \frac{\eta+3\zeta}{4 A_c^2}|A|^2\right)A.
\end{equation}
The parameter $A_c=\sqrt{\epsilon/\gamma_2}$ sets the overall scale of the oscillations, but each variant has its own particular saturated oscillation-amplitude, depending on the relative contributions of the Rayleigh and van der Pol damping mechanisms. Steady-state oscillations are obtained when Eq.~\eqref{Eq:amp} is set to zero, and the amplitude satisfies
\begin{equation}\label{Eq:Steady-state-Amp}
    |A| = \frac{2A_c}{\sqrt{\eta+3\zeta}}
    =
    \begin{cases}
        2A_c & \textrm{vdP},\\
        2A_c/\sqrt3 & \textrm{Rayleigh},\\
        A_c & \textrm{RvdP}.
    \end{cases}
\end{equation}

Note that in the small-amplitude slow limit, without a particular model at hand, it is difficult to discern the nonlinear terms from one another, as they merely combine into a single effective coefficient $\eta_{\textrm{eff}} = \eta+3\zeta$. However, in the large-amplitude strong-drive limit, with $\epsilon\gg 1$, as can be seen in Fig.~\ref{fig:classical_limit_cycles}, the limit cycles look qualitatively very different. In particular, the RvdP oscillator, with $\eta=\zeta$, is unique in that it is invariant under phase-space rotations, producing circularly-symmetric limit cycles, or harmonic oscillations~\cite{Nathan77}, for arbitrary drive strength $\epsilon$. In fact, one can easily verify that the zeroth-order term of the expanded solution~\eqref{Eq:power}, gives the exact steady-state solution, $x(t)=A_c\cos t$, for the RvdP oscillator, with all higher-order corrections cancelling out. As we shall see below, the RvdP oscillator is also the simplest to treat quantum mechanically.

Finally, as expected for an autonomous or time-independent equation of motion~\eqref{Eq:GRvdP}, the complex amplitude equation~\eqref{Eq:amp} is independent of phase, which drops out of both sides. This implies that with purely deterministic dynamics the oscillator will maintain any initial arbitrary phase, but in the presence of thermal, or any other source of noise, the phase of the oscillator will diffuse over time. This is demonstrated numerically in Fig. \ref{Fig:classic_ensemble_w_t} for the vdP oscillator with weak thermal noise, where an initial Gaussian-distributed ensemble of independent oscillators quickly relaxes to the expected amplitude $2A_c$, and eventually spreads over the whole limit cycle.


\section{Zero-Temperature Quantum Limit Cycles}\label{sec:quantum_models}
\subsection{The Quantum Rayleigh-van der Pol Oscillator}\label{sec:quantum_RvdP}

The simplest quantum model of a limit cycle---which is often mistaken for ``the quantum vdP oscillator''---employs standard Lindblad formalism to describe the interaction of the oscillator with its environment, whereby the energy pump, or negative damping, is implemented in terms of single-phonon absorption, and the nonlinear damping is described as two-phonon emission (``phonon'' should be replaced with ``photon'', ``polaron'', ``magnon'', or any other bosonic excitation, depending on the particular physical realization of the oscillator). The physical realization we have in mind follows the framework that was introduced by Dykman and Krivoglaz in the 1970's, whereby the nonlinear damping~\cite{dykman75} appears as a result of nonlinear interaction of the oscillator with a continuum of bath oscillators, while energy injection~\cite{dykman1978} is introduced in the form of an off-resonance pump, detuned a frequency $\Delta_1$ away from the oscillator frequency $\omega$. Within this realization, and as expected in most other alternative realizations, the coupling of the oscillator to the bath inevitably will induce normal linear damping with single-phonon emission, in addition to the two-phonon processes above. 

Consequently, the master equation for the density matrix $\rho$ of the oscillator---considered at $T=0$, for the time being---contains three Lindblad operators of the form
\begin{equation}\label{Eq:LindOp}
 \Lind{C} = C\rho C^{\dagger} 
 - \frac12\left(C^{\dagger}C\rho+\rho C^{\dagger}C\right),
\end{equation}
and is given by
\begin{equation}\label{Eq:master_unscaled}
    \dot\rho
    = \frac{1}{i\hbar}\comm{H_0}{\rho} 
    + \tilde{\kappa}_1\Lind{\adag}
    + \tilde{\gamma}_1\Lind{a}
    + \tilde{\gamma}_2\Lind{a^2},
\end{equation}
where $H_0=\hbar\omega (\adag a+1/2)$ is the Hamiltonian of the harmonic oscillator, and $a$ is its annihilation operator. 

This master equation~\eqref{Eq:master_unscaled} differs conceptually from those that are commonly used in the literature~\cite{lee13, *lee14, walter14, walter15, *lorch16, davis18, ishibashi17, amitai18, scarlatella19}. Common models assume that as in the classical regime the effect of the pump, or negative linear damping, combines with the normal linear damping to give one physical process, with coefficient $(\tilde{\kappa}_1-\tilde{\gamma}_1)=m\omega_0\epsilon$. Thus they either omit the first Lindblad operator below the threshold of self oscillations, when $\tilde{\kappa}_1<\tilde{\gamma}_1$, or omit the second Lindblad operator above threshold, for $\tilde{\kappa}_1>\tilde{\gamma}_1$.  Consequently, as will become evident below, even though they obtain limit cycles in the steady state, they miss important physical effects in the quantum regime, related to the fact that at zero temperature there are \emph{three} rather than only two sources of spontaneous quantum processes that affect the quantum oscillator and its phase stability. 

In order to facilitate the direct comparison between classical and quantum dynamics of limit cycles, we use the same scaling here for the quantum master equation~\eqref{Eq:master_unscaled} as we did earlier for the classical equation of motion~\eqref{Eq:classical-models-unscaled}. This, again, amounts to using the effective mass $m$ of the oscillator as the unit of mass, and its inverse frequency $1/\omega$ as the unit of time, thereby effectively setting both $m$ and $\omega$ to unity. The choice of $x_0=\sqrt{\hbar/m\omega}$ as the unit of length, and correspondingly $p_0=\sqrt{m\hbar\omega}$ as the unit of momentum, amounts to using $\hbar$ as the unit of action with which phase-space area is measured, thereby effectively setting $\hbar$ to unity \footnote{An alternative choice of scaling would have the units of length and momentum as $x_{\rm zp} = \sqrt{\hbar/2m\omega} = x_0/\sqrt{2}$ and $p_{\rm zp} = \sqrt{m\hbar\omega/2} = p_0/\sqrt{2}$, respectively, associated with the zero-point motion of a simple harmonic oscillator, but this effectively sets $\hbar=2$.}. With this choice of scaling, energy is measured in units of $\hbar\omega$, the Hamiltonian becomes $H =  \left(p^2 + x^2\right)/2 = \adag a +1/2,$
where the creation and annihilation operators are defined as
\begin{equation}\label{Eq:a-adag-def}
    a = \frac{1}{\sqrt{2}} \left(x + ip\right); \qquad \adag = \frac{1}{\sqrt{2}} \left(x - ip\right);
\end{equation}
and the commutator $\comm{x}{p}=i$. 

\begin{figure}
    \begin{subfigure}[t]{0.506\linewidth}
    \centering
    \includegraphics[width=1\linewidth ]{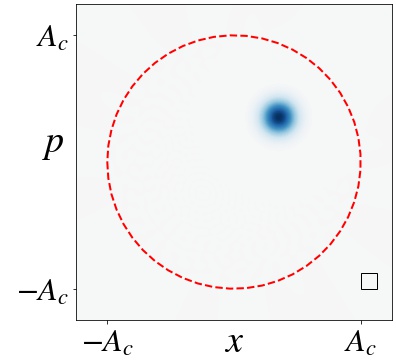}
    \caption{$t=0$}
    \label{}
    \end{subfigure}
    \begin{subfigure}[t]{0.46\linewidth}
    \centering
    \includegraphics[width=1\linewidth ]{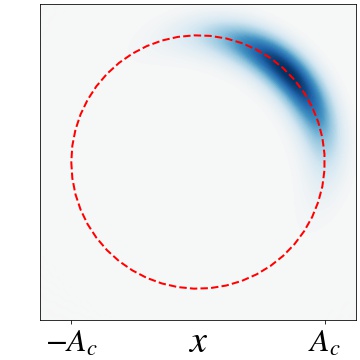}
    \caption{$t=20\pi$}
    \label{}
    \end{subfigure}
    \begin{subfigure}[t]{0.506\linewidth}
    \centering
    \includegraphics[width=1\linewidth ]{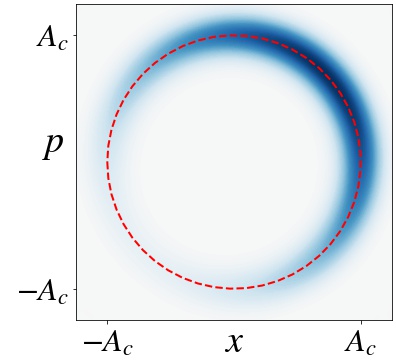}
    \caption{$t=200\pi$}
    \label{}
    \end{subfigure}
    \begin{subfigure}[t]{0.46\linewidth}
    \centering
    \includegraphics[width=1\linewidth ]{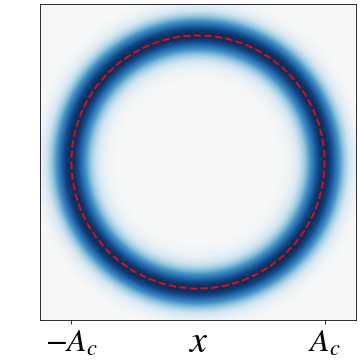}
    \caption{$t=2000\pi$}
    \label{}
    \end{subfigure}
    \caption{Time evolution the Wigner function of a quantum limit cycle, calculated numerically using the RvdP master equation~\eqref{Eq:scaledmaster}, at $T=0$ with $\kappa_1 =0.1$, $\gamma_1 =0$, $\gamma_2=1/640$, and therefore $A_c =8$, starting at $t=0$ with an initial coherent state with $\alpha=(1+i) A_c/4$. The Wigner function approaches the limit cycle, and then loses its initial phase over time. A square of area $\hbar$ is shown in panel (a). The dashed red circles have radius $A_c$.}
    \label{fig:quantum_dynamics}
\end{figure}

The resulting dimensionless zero-temperature master equation
\begin{equation}\label{Eq:scaledmaster}
    \dot \rho = -i[\adag a,\rho] 
    + \kappa_1\Lind{\adag} 
    + \gamma_1\Lind{a}
    + \gamma_2\Lind{a^2}
\end{equation}
can be used to study the dynamics of the density matrix itself, or any dynamical quantity that can be derived from it. For example, Fig.~\ref{fig:quantum_dynamics} shows the characteristic behavior of the time evolution of the Wigner function
\begin{equation}\label{Eq:wig_def}
  W(x,p) = \frac{1}{\pi \hbar}\int_{-\infty}^{\infty} \mel{x+y}{\rho}{x-y} e^{-2ipy/\hbar}dy,  
\end{equation}
calculated numerically\footnote{All the numerical analysis of the quantum dynamics in this work was performed using the QuTip Python package~\cite{qutip}.}, for an oscillator initiated as a coherent state with $\alpha=0.25(1+i) A_c$. As in the classical case, shown in Fig.~\ref{Fig:classic_ensemble_w_t}, one can see how the quantum oscillator first approaches the fixed-amplitude orbit of the limit cycle and only later loses its phase. Note that the amplitude of the quantum limit cycle is $A_c$ rather than $2A_c$, which according to Eq.~\eqref{Eq:Steady-state-Amp} seems to indicate that this limit cycle may in fact be the quantum version of the RvdP oscillator, and not that of the vdP oscillator.

One may use the master equation~\eqref{Eq:scaledmaster} to obtain the equation of motion for any expectation value, $\expval{O}=\Tr{\rho O}$. In the Schr\"odinger picture, where operators are time-independent, one has
\begin{equation}\label{dot-expval}
    \dv{t}\expval{O}=\Tr{\dot\rho O}.
\end{equation}
Thus, for the annihilation operator $a$---using the fact that the trace of a product of operators is invariant under their cyclic permutations---the scaled master equation~\eqref{Eq:scaledmaster} gives the zero-temperature equation of motion
\begin{equation}\label{Eq:dot-expval-a}
    \dv{\expval{a}}{t} = -i\expval{a} + \frac{\kappa_1-\gamma_1}{2}\expval{a} -\gamma_2\expval{\adag aa}. 
\end{equation}

We see that the nonlinear term is proportional to $(\adag a)a$, or $(x^2+p^2)a$, again, as one would expect for the RvdP oscillator rather than the vdP oscillator. To see this more clearly, we take the semiclassical limit where $\expval{\adag a}\gg1$, and therefore $\expval{\adag a} \simeq \expval{a \adag}$. The semiclassical amplitude equation for $\alpha=\expval{a}$ is then readily derived from Eq.~\eqref{Eq:dot-expval-a} by replacing $\expval{\adag aa}$ with $|\alpha|^2\alpha$ to give
\begin{equation}\label{Eq:dot-alpha}
\dv{\alpha}{t} = -i \alpha +\frac{\epsilon}{2}\left(1-\frac{2|\alpha |^2}{A_c^2}\right)\alpha,
\end{equation} 
where $A_c^2 = \epsilon/\gamma_2$, as defined earlier. In order to use an equivalent ansatz to the classical one in Eq.~\eqref{Eq:power} we note that, according to the definition of the creation and annihilation operators in Eq.~\eqref{Eq:a-adag-def}, $\alpha$ is a factor of $\sqrt{2}$ smaller than the complex amplitude of the oscillator. We therefore take
\begin{equation}
    \alpha(t) = \frac{A(T)}{\sqrt{2}}e^{-it},
\end{equation}
with $T=\epsilon t$ as before, and find that the slow amplitude equation is given by
\begin{equation}\label{Eq:qslow_scaled_A}
    \dv{A}{T} = \frac{1}{2}\left(1-\frac{|A|^2}{A_c^2}\right)A,
\end{equation}
which corresponds to the classical amplitude equation~\eqref{Eq:Steady-state-Amp} as long as one takes $\eta_\textrm{eff}=4$, or $\eta=\zeta=1$, as expected for the RvdP oscillator, and in agreement with the amplitude of the limit cycle observed numerically in Fig.~\ref{fig:quantum_dynamics}.

Finally, using the definition of $a$ in Eq.~\eqref{Eq:a-adag-def}, we can take the real and imaginary parts of Eq.~\eqref{Eq:dot-alpha} to obtain the equations of motion for the expectation values of the position and momentum operators \cite[Section 7.4]{bowen15},
\begin{subequations}\label{Eq:hamilton-eq}
\begin{eqnarray}\label{Eq:hamilton-eq1}
    \expval{\dot{x}} &= &\frac{\epsilon}{2}\left(1 - \frac{\expval{x}^2+\expval{p}^2}{A_c^2}\right) \expval{x} +\expval{p},\\\label{Eq:hamilton-eq2}
    \expval{\dot{p}} &= &\frac{\epsilon}{2}\left(1 - \frac{\expval{x}^2+\expval{p}^2}{A_c^2}\right) \expval{p}
    -\expval{x}.
\end{eqnarray}
\end{subequations}
Differentiating Eq.~\eqref{Eq:hamilton-eq1} with respect to time, and substituting Eq.~\eqref{Eq:hamilton-eq2} for $\expval{\dot{p}}$, yields a second-order equation of motion for $\expval{x}$ of the form
\begin{equation}
    \expval{\Ddot{x}} + \expval{x} 
    = \epsilon\left(1 - \frac{\expval{x}^2+\expval{\dot{x}}^2}{A_c^2}\right) \expval{\dot{x}} + \order{\epsilon^2}.
\end{equation}
Neglecting corrections of order $\epsilon^2$, this explicitly agrees with the classical equation of motion~\eqref{Eq:GRvdP} for the Rayleigh-van der Pol oscillator, with $\eta=\zeta=1$. 

We wish to emphasize the circular symmetry of the steady-state Wigner function in Fig.~\ref{fig:quantum_dynamics}(d). In order for the steady-state Wigner function to lack such symmetry, the steady-state density matrix must contain off-diagonal elements that do not decay to zero. This can be seen by noting that the Wigner function~\eqref{Eq:wig_def} is a linear function of the density matrix, which can be expressed as a sum of its elements, and that the Wigner function of a diagonal element---given by~\cite[Section 4.4.4]{gardiner2004} \begin{equation}\label{Eq:Wigner-Fockstate}
    W_{n,n}(\alpha,\alpha^*) = \frac{(-1)^n}{\pi\hbar}L_n(4|\alpha|^2)e^{-2|\alpha|^2},
\end{equation}
where $L_n(x)$ is the $n^{th}$ Laguerre polynomial---is rotationally invariant. 

\begin{figure}
\begin{subfigure}[b]{0.15\textwidth}
    \includegraphics[width=1\textwidth]{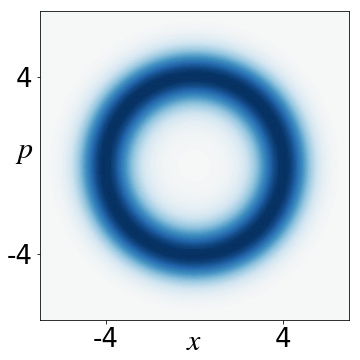}
    \caption{$\epsilon=0.01$}
    \label{}
    \end{subfigure}
    \hfill
    \begin{subfigure}[b]{0.15\textwidth}
    \includegraphics[width=1\textwidth]{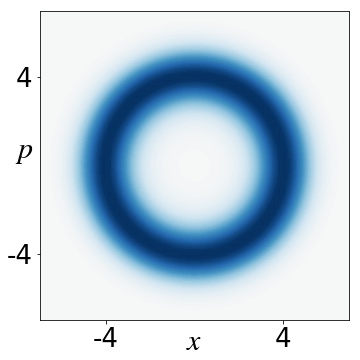}
    \caption{$\epsilon=0.3$}
    \label{}
    \end{subfigure}
    \hfill
    \begin{subfigure}[b]{0.15\textwidth}
    \includegraphics[width=1\textwidth]{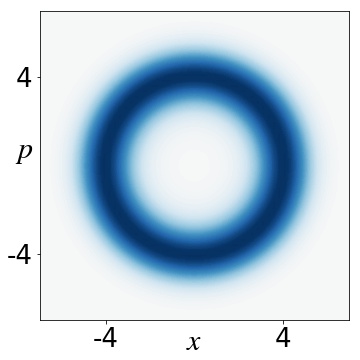}
    \caption{$\epsilon=1$}
    \label{}
    \end{subfigure}
    
    \hfill
    \begin{subfigure}[t]{0.15\textwidth}
    \includegraphics[width=1\textwidth]{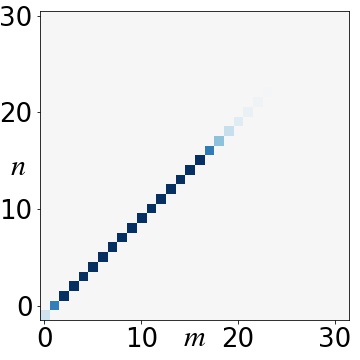}
    \caption{$\epsilon=0.01$}
    \label{}
    \end{subfigure}
    \hfill
    \begin{subfigure}[t]{0.15\textwidth}
    \includegraphics[width=1\textwidth]{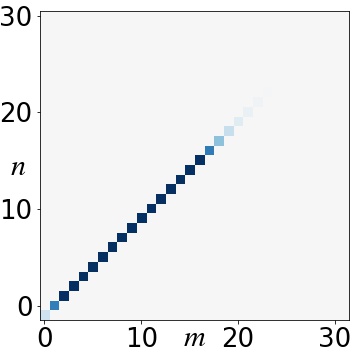}
    \caption{$\epsilon=0.3$}
    \label{}
    \end{subfigure}
    \hfill
    \begin{subfigure}[t]{0.15\textwidth}
    \includegraphics[width=1\textwidth]{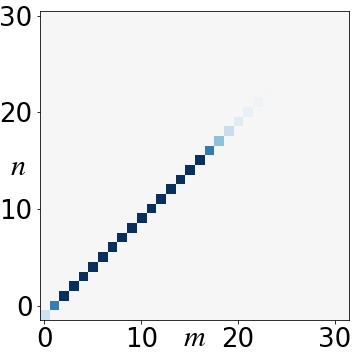}
    \caption{$\epsilon=1$}
    \label{}
    \end{subfigure}
    \hfill
    \caption{Steady state Wigner functions (top row) and absolute values of the density-matrix elements $\rho_{n,m}$ (bottom row) at $T=0$, obtained numerically for the quantum RvdP master equation~\eqref{Eq:scaledmaster}, with $\gamma_1=0$, $\gamma_2=\epsilon/16$, and therefore $A_c=4$ for all different values of $\epsilon=\kappa_1$. All off-diagonal matrix elements decay to zero in the steady state, yielding the same circular limit cycle, independent of $\epsilon$ for constant $A_c$. Compare with Figs.~\ref{fig:vdp_dm_elements} and \ref{fig:Rayleigh_dm_elements} below for the quantum vdP and Rayleigh oscillators.}
    \label{fig:Rvdp_dm_elements}
\end{figure}

As previous authors~\cite{simaan78,scarlatella19} have noted, the master equation~\eqref{Eq:scaledmaster} for the RvdP oscillator does not couple density-matrix elements that are not on the same diagonal. To see this, it is helpful to relabel the matrix elements $\rho_{n,n+m}= \mel{n}{\rho}{n+m}$ according to their degree $m$ of off-diagonality using a transformation, similar to the one used by Simaan and Loudon~\cite{simaan78},
\begin{equation}\label{Eq:transformed}
     \varrho_{n,m}(t) = e^{imt}\sqrt{\frac{(n+m)!}{n!}}\rho_{n,n+m}(t),
     \quad m+n \geq0.
\end{equation}
The rate equations for the transformed matrix elements are then
\begin{eqnarray}
\label{Eq:me-trans_element1}
    \dot \varrho_{n,m}   \nonumber
    &=& \kappa_1 \left\{ (n+m)\varrho_{n-1,m}-\frac{1}{2}\left(2n+m+2\right)\varrho_{n,m} \right\}\\ \nonumber
    &+& \gamma_1 \left\{ (n+1)\varrho_{n+1,m}-\frac{1}{2}\left(2n+m\right)\varrho_{n,m} \right\}\\ 
    &+& \gamma_2 \left\{ (n+1)(n+2)\varrho_{n+2,m}\phantom{\frac{1}{2}} \right. \\ 
    &-& \left.\frac{1}{2}\left(n\left(n-1\right)+(n+m)\left(n+m-1\right)\right)\varrho_{n,m} \right\}, \nonumber
\end{eqnarray}
where evidently, matrix elements are coupled only if they have the same degree $m$ of off-diagonally. Thus, each diagonal can be considered as a separate ``block'' of the density matrix, evolving independently of all the others, allowing the off-diagonal elements to decay to zero, as one expects, independent of the principal diagonal elements, which are the only ones to survive in the steady state. This is confirmed numerically in Fig.~\ref{fig:Rvdp_dm_elements}.

\begin{figure*}
    \begin{subfigure}[b]{0.2\linewidth}
    \includegraphics[width=1\linewidth]{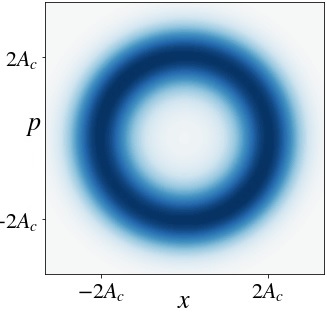}
    \caption{$\epsilon=0.01$}
    \label{}
    \end{subfigure}
    \begin{subfigure}[b]{0.2\linewidth}
    \includegraphics[width=1\linewidth]{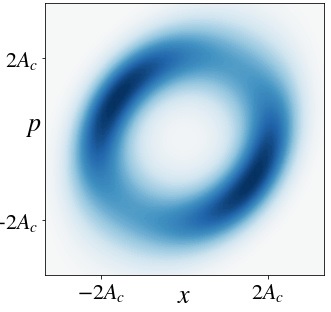}
    \caption{$\epsilon=0.3$}
    \label{}
    \end{subfigure}
    \begin{subfigure}[b]{0.2\linewidth}
    \includegraphics[width=1\linewidth]{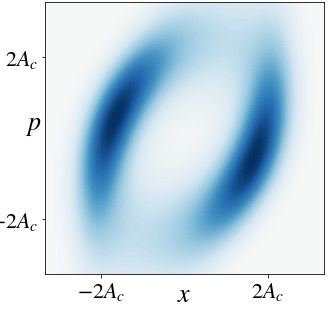}
    \caption{$\epsilon=1$}
    \label{}
    \end{subfigure}
    \begin{subfigure}[b]{0.2\linewidth}
    \includegraphics[width=1\linewidth]{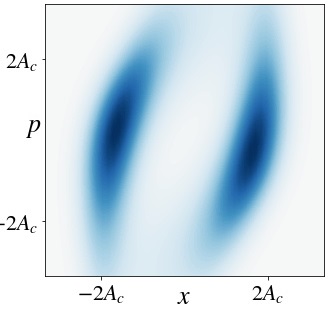}
    \caption{$\epsilon=2$}
    \label{}
    \end{subfigure}
    \begin{subfigure}[b]{0.2\linewidth}
    \includegraphics[width=1\linewidth]{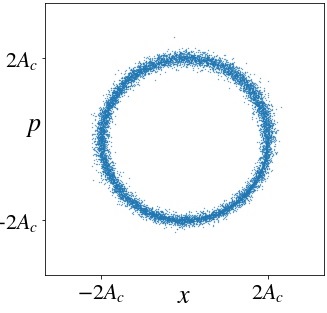}
    \caption{$\epsilon=0.01$}
    \label{}
    \end{subfigure}
    \begin{subfigure}[b]{0.2\linewidth}
    \includegraphics[width=1\linewidth]{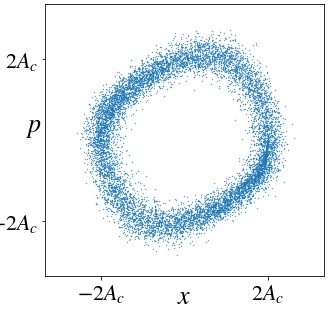}
    \caption{$\epsilon=0.3$}
    \label{}
    \end{subfigure}
    \begin{subfigure}[b]{0.2\linewidth}
    \includegraphics[width=1\linewidth]{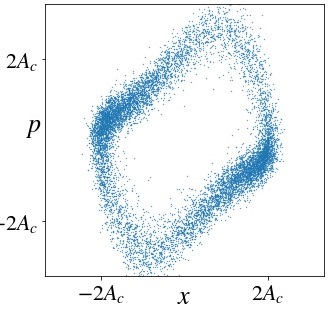}
    \caption{$\epsilon=1$}
    \label{}
    \end{subfigure}
    \begin{subfigure}[b]{0.2\linewidth}
    \includegraphics[width=1\linewidth]{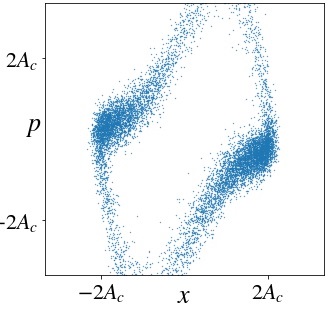}
    \caption{$\epsilon=2$}
    \label{}
    \end{subfigure}
    \caption{
    Steady-state Wigner functions of the quantum vdP master equation~\eqref{Eq:new_master} at $T=0$ (top row), and phase-space distributions of the classical vdP equation~\eqref{Eq:GRvdP} with $\eta=1$ and $\zeta=0$, for an ensemble of $10^4$ oscillators at $T=0.1$ (bottom row), both with $A_c=\sqrt{10}$ and $\gamma_1=0$. Figures (a) and (e) show the nearly circular limit cycles, obtained for small values of $\epsilon$, while in (b) and (f) one can begin to see small differences between the quantum and the classical models, that are expected to differ from each other on the order of $\epsilon^2$. For larger values of $\epsilon$, clear deviations appear between the quantum and the classical limit cycles, yet they are both non-circular, lingering for large fractions of the period where the Wigner functions and the classical distributions are peaked.}
    \label{fig:new_lind1}
\end{figure*}

\subsection{The Quantum van der Pol Oscillator}
\label{sec:quantum_vdP}

One can obtain a master equation whose classical limit gives the vdP oscillator, at least to first order in $\epsilon$, and is capable of producing quantum limit cycles that are non-circular in phase space. This is done by changing the Lindblad operator for the nonlinear damping term in Eq.~\eqref{Eq:scaledmaster} from $\gamma_2\mathcal{D}[a^2]$ to $\gamma_2\mathcal{D}[xa/\sqrt{2}]$, breaking the rotational symmetry in phase space. The zero-temperature master equation then becomes
\begin{equation}\label{Eq:new_master}
    \dot \rho = -i[\adag a,\rho] 
    + \kappa_1\Lind{\adag}
    + \gamma_1\Lind{a}
    + \frac{\gamma_2}{2}\Lind{xa},
\end{equation}
where we recall that $x=(a+\adag)/\sqrt{2}$. Consequently, the nonlinear term in Eq.~\eqref{Eq:dot-expval-a} for the dynamics of $\expval{a}$ becomes $-\gamma_2\expval{x^2 a}/4$, which in the semiclassical limit, where $\expval{x^2a}\simeq\expval{x}^2\alpha$, yields
\begin{equation}\label{Eq:amp_new_lind}
    \frac{d\alpha}{dt} = -i\alpha + \frac{\epsilon}{2}\left(1 -\frac{\expval{x}^2}{2 A_c^2} \right)\alpha,
\end{equation}
in place of Eq.~\eqref{Eq:dot-alpha}. Finally, by taking the real and imaginary parts of Eq.~\eqref{Eq:amp_new_lind},
\begin{subequations}\label{Eq:alt_hamilton-eq}
\begin{eqnarray}\label{Eq:alt_hamilton-eq1}
    \expval{\dot{x}} &= &\frac{\epsilon}{2}\left(1 - \frac{\expval{x}^2}{2A_c^2}\right) \expval{x} +\expval{p},\\\label{Eq:alt_hamilton-eq2}
    \expval{\dot{p}} &= &\frac{\epsilon}{2}\left(1 - \frac{\expval{x}^2}{2A_c^2}\right) \expval{p}
    -\expval{x},
\end{eqnarray}
\end{subequations}
and as in Eqs.~\eqref{Eq:hamilton-eq}, differentiating Eq.~\eqref{Eq:alt_hamilton-eq1} with respect to time, and substituting Eq.~\eqref{Eq:alt_hamilton-eq2} for $\expval{\dot{p}}$, we obtain a second-order equation of motion for $\expval{x}$ of the form
\begin{equation}
    \expval{\Ddot{x}} + \expval{x} 
    = \epsilon\left(1 - \frac{\expval{x}^2}{A_c^2}\right) \expval{\dot{x}} + \order{\epsilon^2},
\end{equation}
which to within corrections of $\order{\epsilon^2}$, is indeed the classical equation of motion~\eqref{Eq:GRvdP}, for the van der Pol oscillator, with $\eta=1$ and $\zeta=0$.

Figure~\ref{fig:new_lind1} shows the steady-state Wigner functions that are obtained numerically from the vdP master equation~\eqref{Eq:new_master} for different values of $\epsilon$ at $T=0$. A comparison with the phase-space distributions of $10^4$ classical van der Pol oscillators at $T=0.1$, confirms that for small values of $\epsilon$ the quantum and classical models agree very well. For large values of $\epsilon$, the quantum master equation clearly deviates from the classical vdP behavior, as expected, yet it retains the non-circular limit cycles associated with the relaxation-oscillation behavior of large-amplitude vdP oscillators. 

\begin{figure}
\begin{subfigure}[b]{0.15\textwidth}
    \includegraphics[width=1\textwidth]{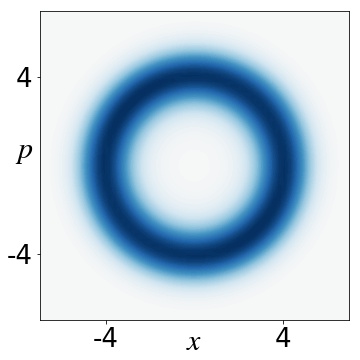}
    \caption{$\epsilon=0.01$}
    \label{}
    \end{subfigure}
    \hfill
    \begin{subfigure}[b]{0.15\textwidth}
    \includegraphics[width=1\textwidth]{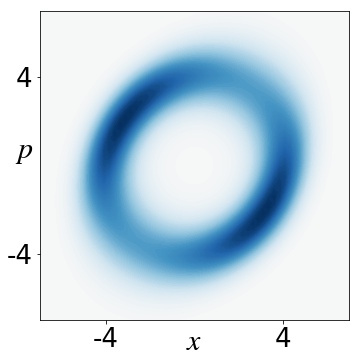}
    \caption{$\epsilon=0.3$}
    \label{}
    \end{subfigure}
    \hfill
    \begin{subfigure}[b]{0.15\textwidth}
    \includegraphics[width=1\textwidth]{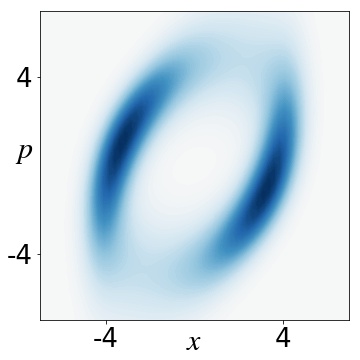}
    \caption{$\epsilon=1$}
    \label{}
    \end{subfigure}
    \hfill
    \begin{subfigure}[t]{0.15\textwidth}
    \includegraphics[width=1\textwidth]{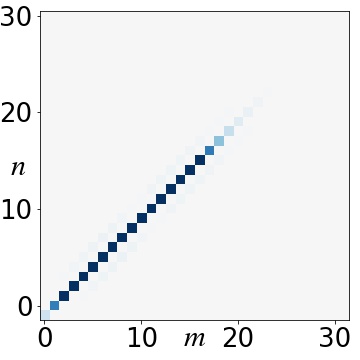}
    \caption{$\epsilon=0.01$}
    \label{}
    \end{subfigure}
    \hfill
    \begin{subfigure}[t]{0.15\textwidth}
    \includegraphics[width=1\textwidth]{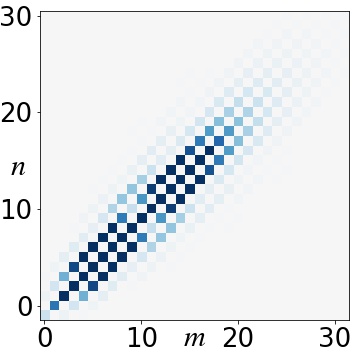}
    \caption{$\epsilon=0.3$}
    \label{}
    \end{subfigure}
    \hfill
    \begin{subfigure}[t]{0.15\textwidth}
    \includegraphics[width=1\textwidth]{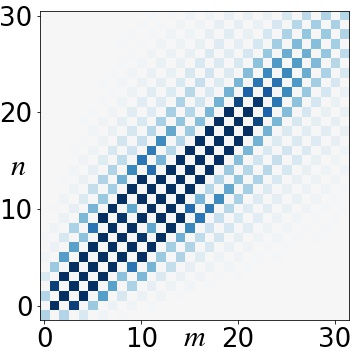}
    \caption{$\epsilon=1$}
    \label{}
    \end{subfigure}
    \hfill
    \caption{Steady state Wigner functions (top row) and absolute values of the density-matrix elements $\rho_{n,m}$ (bottom row) at $T=0$, obtained numerically for the quantum vdP master equation~\eqref{Eq:new_master}, with $A_c=2$ and $\gamma_1=0$. All the odd diagonals are free to decay to zero, while the even diagonals, which are coupled to the principal diagonal, are not. Compare with Fig.~\ref{fig:Rvdp_dm_elements} above for the quantum RvdP oscillator.
    }
    \label{fig:vdp_dm_elements}
\end{figure}

\begin{figure}
\begin{subfigure}[b]{0.15\textwidth}
    \includegraphics[width=1\textwidth]{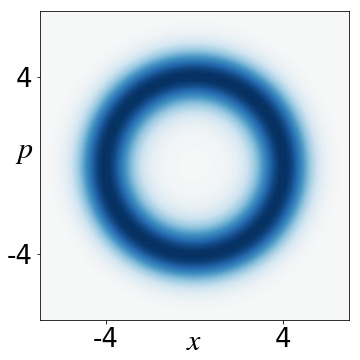}
    \caption{$\epsilon=0.01$}
    \label{}
    \end{subfigure}
    \hfill
    \begin{subfigure}[b]{0.15\textwidth}
    \includegraphics[width=1\textwidth]{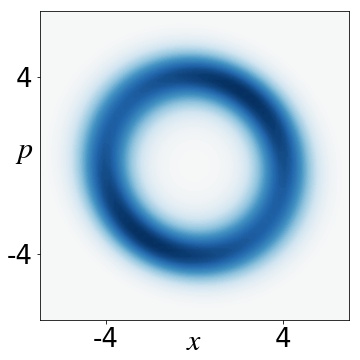}
    \caption{$\epsilon=0.3$}
    \label{}
    \end{subfigure}
    \hfill
    \begin{subfigure}[b]{0.15\textwidth}
    \includegraphics[width=1\textwidth]{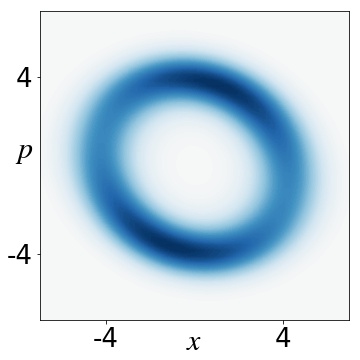}
    \caption{$\epsilon=1$}
    \label{}
    \end{subfigure}
    
    \hfill
    \begin{subfigure}[t]{0.15\textwidth}
    \includegraphics[width=1\textwidth]{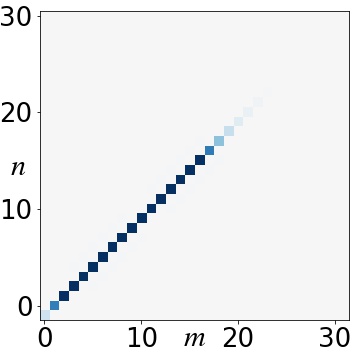}
    \caption{$\epsilon=0.01$}
    \label{}
    \end{subfigure}
    \hfill
    \begin{subfigure}[t]{0.15\textwidth}
    \includegraphics[width=1\textwidth]{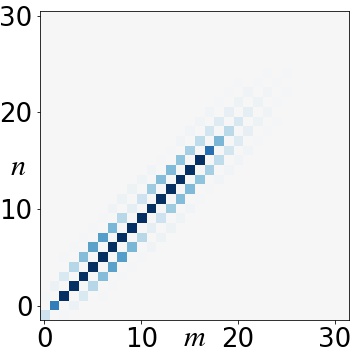}
    \caption{$\epsilon=0.3$}
    \label{}
    \end{subfigure}
    \hfill
    \begin{subfigure}[t]{0.15\textwidth}
    \includegraphics[width=1\textwidth]{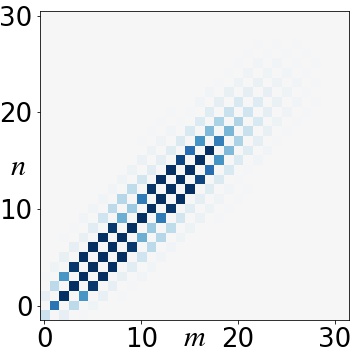}
    \caption{$\epsilon=1$}
    \label{}
    \end{subfigure}
    \hfill
    \caption{As in Fig.~\ref{fig:vdp_dm_elements}, but for the quantum Rayleigh master equation~\eqref{eq:Rayleigh_me}, with $A_c=2\sqrt{3}$ and $\gamma_1=0$.  Compare with Fig.~\ref{fig:Rvdp_dm_elements} above for the quantum RvdP oscillator.}
    \label{fig:Rayleigh_dm_elements}
\end{figure}

The rate equations for the transformed density-matrix elements~\eqref{Eq:transformed}, obtained from the vdP master equation~\eqref{Eq:new_master}, take the form
\begin{eqnarray}\label{Eq:vdP_rate}\nonumber
    \dot \varrho_{n,m}
    &=& \kappa_1 \left\{ (n+m)\varrho_{n-1,m}-\frac{1}{2}\left(2n+m+2\right)\varrho_{n,m} \right\}\\ \nonumber
    &+& \gamma_1 \left\{ (n+1)\varrho_{n+1,m}-\frac{1}{2}\left(2n+m\right)\varrho_{n,m} \right\}\\ \nonumber
    &+& \frac{\gamma_2}{8} \bigg\{\left(2nm-2(n+m)^2+(2n+m)\right)\varrho_{n,m} \\ \nonumber
    &&\phantom{\frac12}+2(n+2)(n+1)\varrho_{n+2,m}\\ \nonumber
    &&\phantom{\frac12}+e^{-2it}\left[\left(n-m\right)\varrho_{n,m+2} -n\varrho_{n-2,m+2}\right]\\\nonumber
    &&\phantom{\frac12}+e^{2it}(2m+n)(n+1)(n+2)\varrho_{n+2,m-2}\\
    &&\left. \phantom{\frac12}-e^{2it}(n+m-1)(n+m)^2\varrho_{n,m-2} \right\}.
\end{eqnarray}
One can see that matrix elements on the $m^{th}$ diagonal are now coupled  to elements from the $m\pm 2$ diagonals, thus coupling the even diagonals to each other, and the odd diagonals to each other. Given the fact that the principle $m=0$ diagonal cannot decay to zero, the rate equations~\eqref{Eq:vdP_rate} feed the even diagonals that are coupled to it, generically hindering their decay in the steady state.

\begin{figure*}
    \begin{subfigure}[b]{0.2\linewidth}
    \includegraphics[width=1\linewidth]{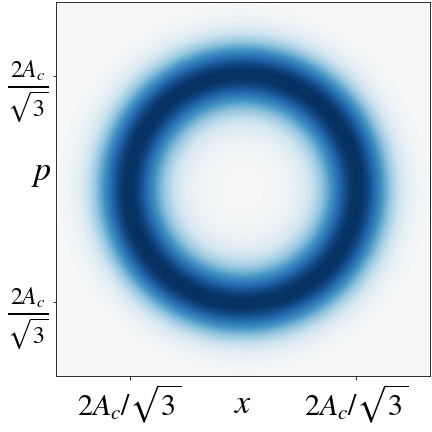}
    \caption{$\epsilon=0.01$}
    \label{}
    \end{subfigure}
    \begin{subfigure}[b]{0.2\linewidth}
    \includegraphics[width=1\linewidth]{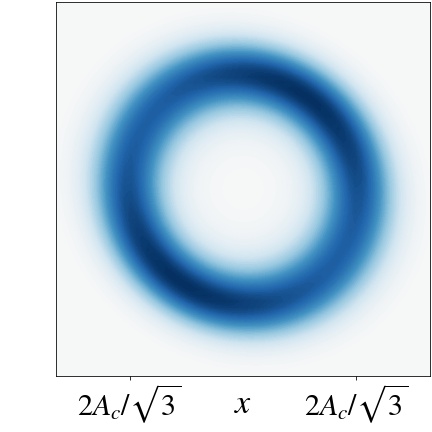}
    \caption{$\epsilon=0.3$}
    \label{}
    \end{subfigure}
    \begin{subfigure}[b]{0.2\linewidth}
    \includegraphics[width=1\linewidth]{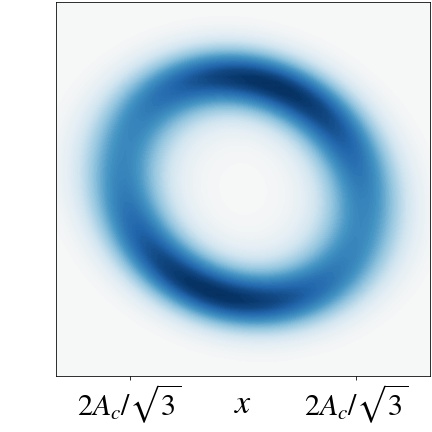}
    \caption{$\epsilon=1$}
    \label{}
    \end{subfigure}
    \begin{subfigure}[b]{0.2\linewidth}
    \includegraphics[width=1\linewidth]{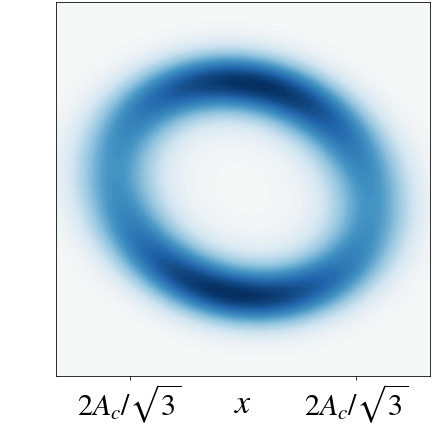}
    \caption{$\epsilon=2$}
    \label{}
    \end{subfigure}
    \begin{subfigure}[t]{0.2\linewidth}
    \includegraphics[width=1\linewidth]{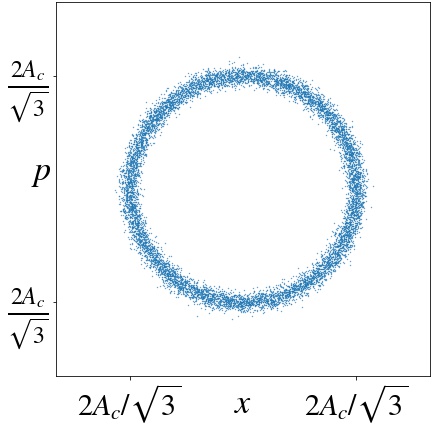}
    \caption{$\epsilon=0.01$}
    \label{}
    \end{subfigure}
    \begin{subfigure}[t]{0.2\linewidth}
    \includegraphics[width=1\linewidth]{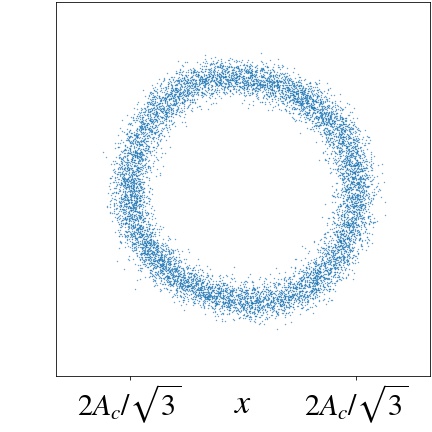}
    \caption{$\epsilon=0.3$}
    \label{}
    \end{subfigure}
    \begin{subfigure}[t]{0.2\linewidth}
    \includegraphics[width=1\linewidth]{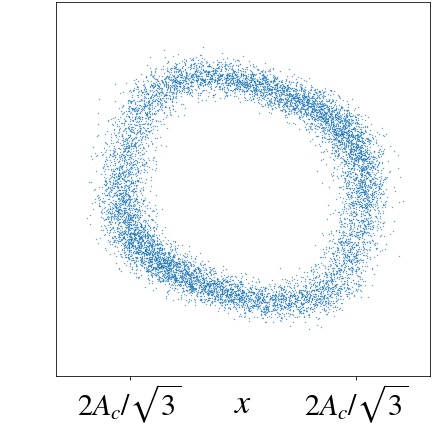}
    \caption{$\epsilon=1$}
    \label{}
    \end{subfigure}
    \begin{subfigure}[t]{0.2\linewidth}
    \includegraphics[width=1\linewidth]{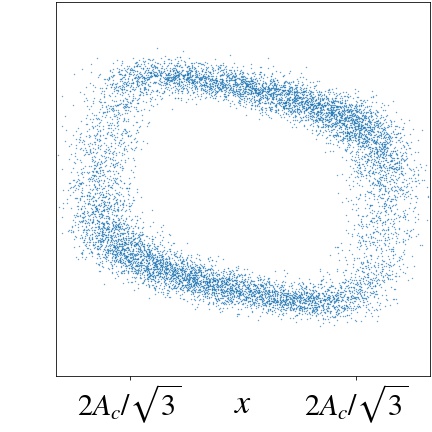}
    \caption{$\epsilon=2$}
    \label{}
    \end{subfigure}
    \caption{
    Steady-state Wigner functions of the quantum Rayleigh master equation~\eqref{eq:Rayleigh_me} at $T=0$ (top row), and phase-space distributions of the classical Rayleigh equation~\eqref{Eq:GRvdP} with $\eta=0$ and $\zeta=1$, for an ensemble of $10^4$ oscillators at $T=0.1$ (bottom row), both with $A_c=4$ and $\gamma_1=0$. Compare with Fig.~\ref{fig:new_lind1} for the vdP oscillator.
    }
    \label{fig:new_Rayleigh}
\end{figure*}

This is demonstrated numerically in Fig.~\ref{fig:vdp_dm_elements}, where we plot the Wigner functions and the absolute values of the density-matrix elements for different values of $\epsilon$, while keeping the ratio between $\epsilon$ and $\gamma_2$, and therefore $A_c$, constant. For small values of $\epsilon$ the coupling between the off-diagonals is relatively weak, making the density matrix nearly diagonal and the Wigner function nearly circular. Increasing $\gamma_2$ increases the coupling between the even diagonals, which become non-zero. Note that the odd diagonals, which are not coupled to the principle diagonal do vanish in the steady state. Compare with the corresponding Fig.~\ref{fig:Rvdp_dm_elements} for the RvdP master equation~\eqref{Eq:scaledmaster}, where the limit cycles remain circular and the density matrix remains diagonal, even for large $\gamma_2$.

\subsection{The Quantum Rayleigh Oscillator}
\label{sec:quantum_Rayleigh}

For completeness, let us present a quantum model whose classical limit at $T=0$ yields the classical Rayleigh oscillator of Eq.~\eqref{Eq:GRvdP}, with $\eta=0$ and $\zeta=1$, to within corrections of $\order{\epsilon^2}$. To do so, we change the Lindblad operator for the nonlinear damping term in Eq.~\eqref{Eq:scaledmaster} from $\gamma_2 \mathcal{D}[a^2]$ to $\gamma_2 \left(\mathcal{D}[xa/\sqrt{2}] + \mathcal{D}[pa]\right)$. The Rayleigh master equation is then given by
\begin{eqnarray}\label{eq:Rayleigh_me}\nonumber
    \dot \rho = -i[\adag a,\rho] 
    &+& \kappa_1 \mathcal{D}[\adag]\rho 
    + \gamma_1 \mathcal{D}[a]\rho\\
    &+& \frac{\gamma_2}{2} \mathcal{D}[xa]\rho + \gamma_2 \mathcal{D}[pa]\rho
\end{eqnarray}

After making this change to the nonlinear term in the master equation, the nonlinear term in Eq.~\eqref{Eq:dot-expval-a} for the dynamics of $\expval{a}$ becomes $-\gamma_2\left(\expval{x^2 a} + 2\expval{p^2 a}\right)/4$, which in the classical limit yields an amplitude equation of the form
\begin{equation}\label{Eq:Rayleigh-amp}
    \frac{d\alpha}{dt} = -i\alpha + \frac{\epsilon}{2}\left(1 -\frac{\expval{x}^2 + 2\expval{p}^2}{2 A_c^2} \right)\alpha,
\end{equation}
in place of Eq.~\eqref{Eq:dot-alpha}. Finally, by taking the real and imaginary parts of \eqref{Eq:Rayleigh-amp},
\begin{subequations}\label{Eq:Rayliegh_hamilton-eq}
\begin{eqnarray}\label{Eq:Rayliegh_hamilton-eq1}
    \expval{\dot{x}} &= &\frac{\epsilon}{2}\left(1 - \frac{\expval{x}^2 + 2\expval{p}^2}{2A_c^2}\right) \expval{x} +\expval{p},\\\label{Eq:Rayliegh_hamilton-eq2}
    \expval{\dot{p}} &= &\frac{\epsilon}{2}\left(1 - \frac{\expval{x}^2 + 2\expval{p}^2}{2A_c^2}\right) \expval{p}
    -\expval{x},
\end{eqnarray}
\end{subequations}
and as in Eqs.~\eqref{Eq:hamilton-eq}, differentiating Eq.~\eqref{Eq:Rayliegh_hamilton-eq1} with respect to time, and substituting Eq.~\eqref{Eq:Rayliegh_hamilton-eq2}, we obtain a second-order equation of motion for $\expval{x}$ of the form
\begin{equation}
    \expval{\Ddot{x}} + \expval{x} 
    = \epsilon\left(1 - \frac{\expval{\dot{x}}^2}{A_c^2}\right) \expval{\dot{x}} + \order{\epsilon^2},
\end{equation}
which up to corrections of $\order{\epsilon^2}$, is the classical Rayleigh equation, given by Eq.~\eqref{Eq:GRvdP} with $\eta=0$ and $\zeta=1$. Wigner functions for the quantum Rayleigh oscillator are plotted in Fig.~\ref{fig:Rayleigh_dm_elements} alongside their density matrices, and comparisons between quantum and classical limit cycles are shown in Fig.~\ref{fig:new_Rayleigh}.


\begin{figure}
\begin{subfigure}[b]{0.48\linewidth}
\includegraphics[width=1\linewidth]{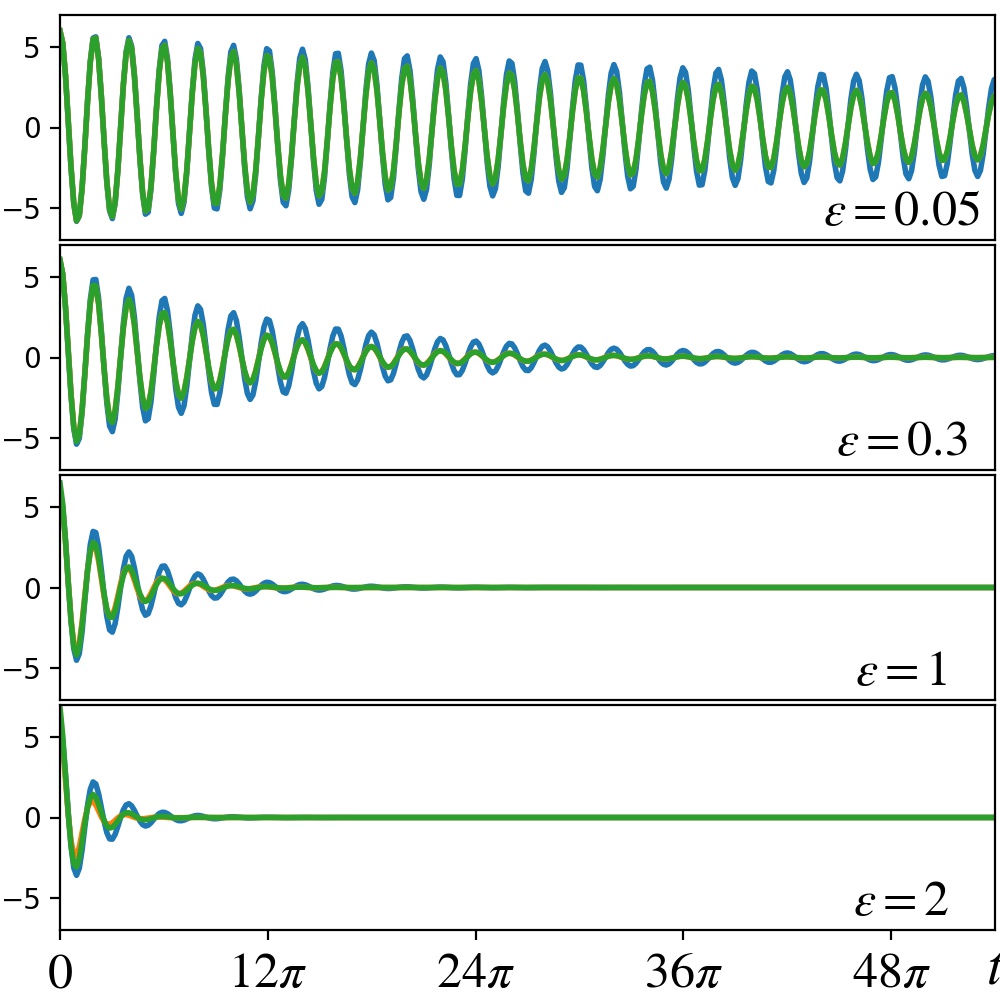}
\caption{$\expval{x(t)x(0)}$}
\label{}
\end{subfigure}
\hfill
\begin{subfigure}[b]{0.48\linewidth}
\includegraphics[width=1\linewidth]{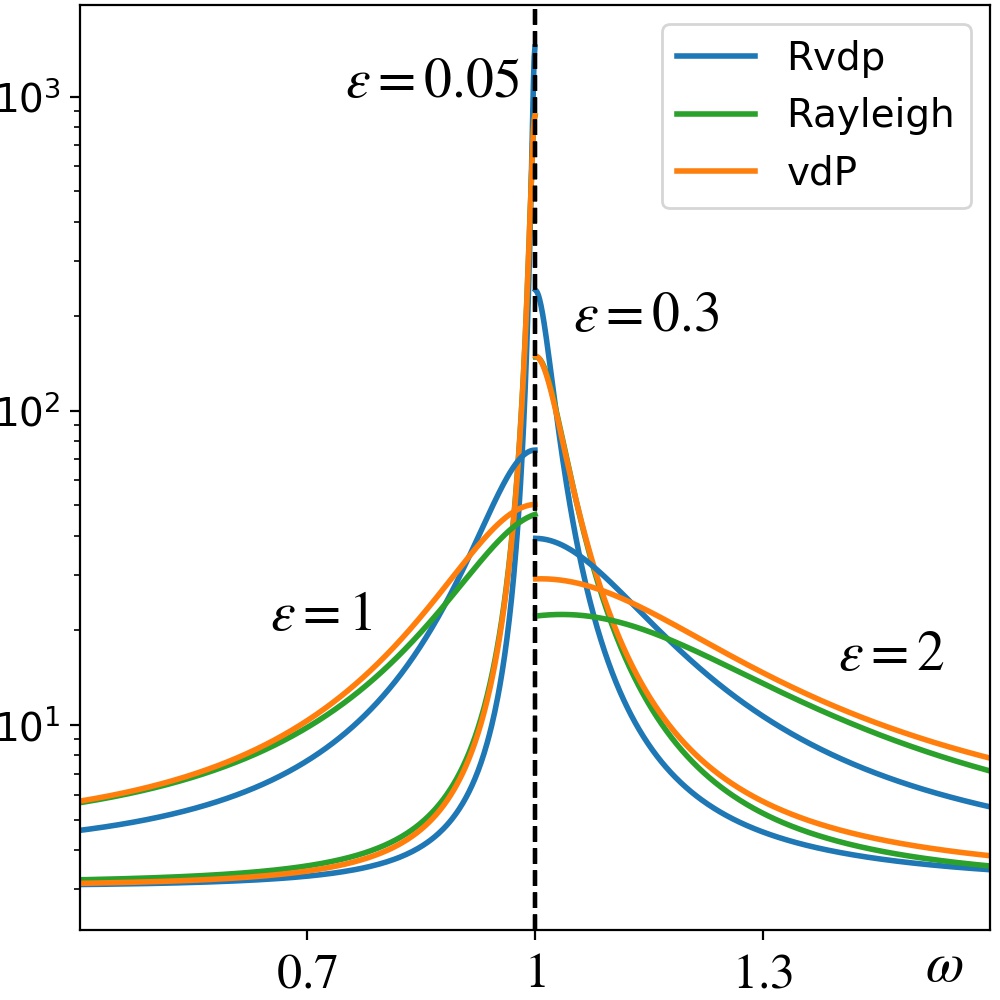}
\caption{$S_{xx}(\omega)$}
\label{}
\end{subfigure}
\caption{(a) Displacement correlation functions $\expval{x(t)x(0)}$, and (b) left or right halves of the corresponding nearly symmetric spectral distributions $S_{xx}(\omega)$ for the three quantum models, with $\epsilon=0.05, 0.3, 1, 2$, and $\gamma_1=0$. Simulations are initiated from the steady state at $t=0$. Parameters are chosen to yield limit cycles with amplitude $A=\sqrt{10}$, for all three models, by setting $\grvdp=\epsilon/10$, $\gvdp=4\epsilon/10$ and $\gr=4\epsilon/30$. 
}
\label{fig:spectra_and_corrs}
\end{figure}

\begin{figure}
    \begin{subfigure}[t]{0.48\linewidth}
    \includegraphics[width=1\linewidth]{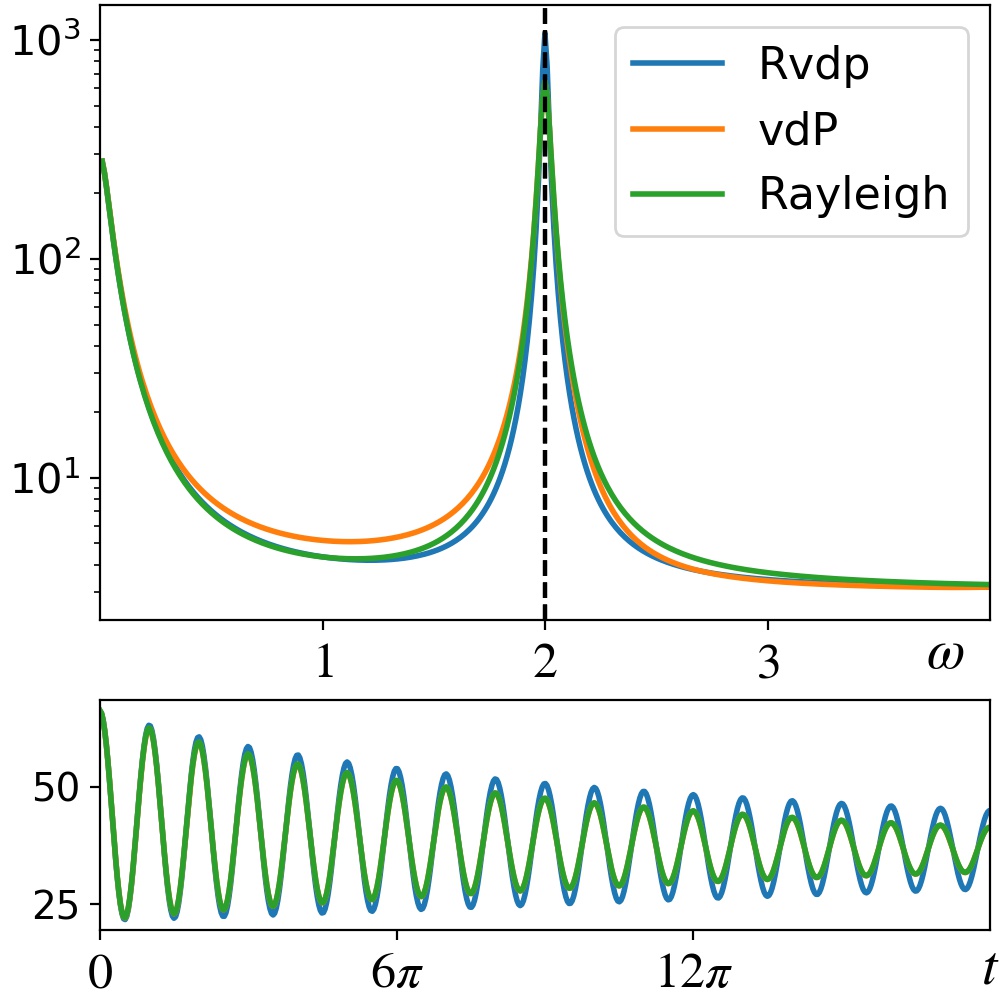}
    \caption{$\epsilon=0.05$}
    \label{}
    \end{subfigure}
    \hfill
    \begin{subfigure}[t]{0.48\linewidth}
    \includegraphics[width=1\linewidth]{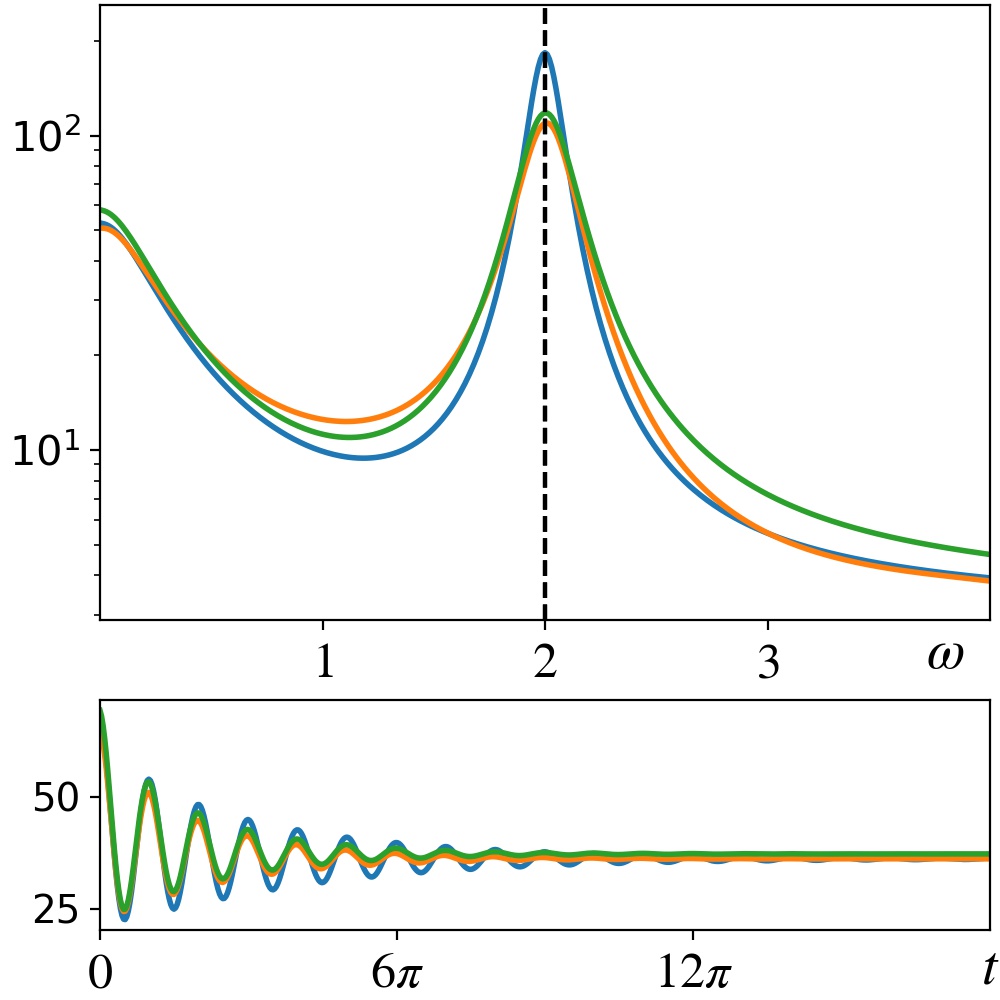}
    \caption{$\epsilon=0.3$}
    \label{}
    \end{subfigure}
    \hfill
    \begin{subfigure}[t]{0.48\linewidth}
    \includegraphics[width=1\linewidth]{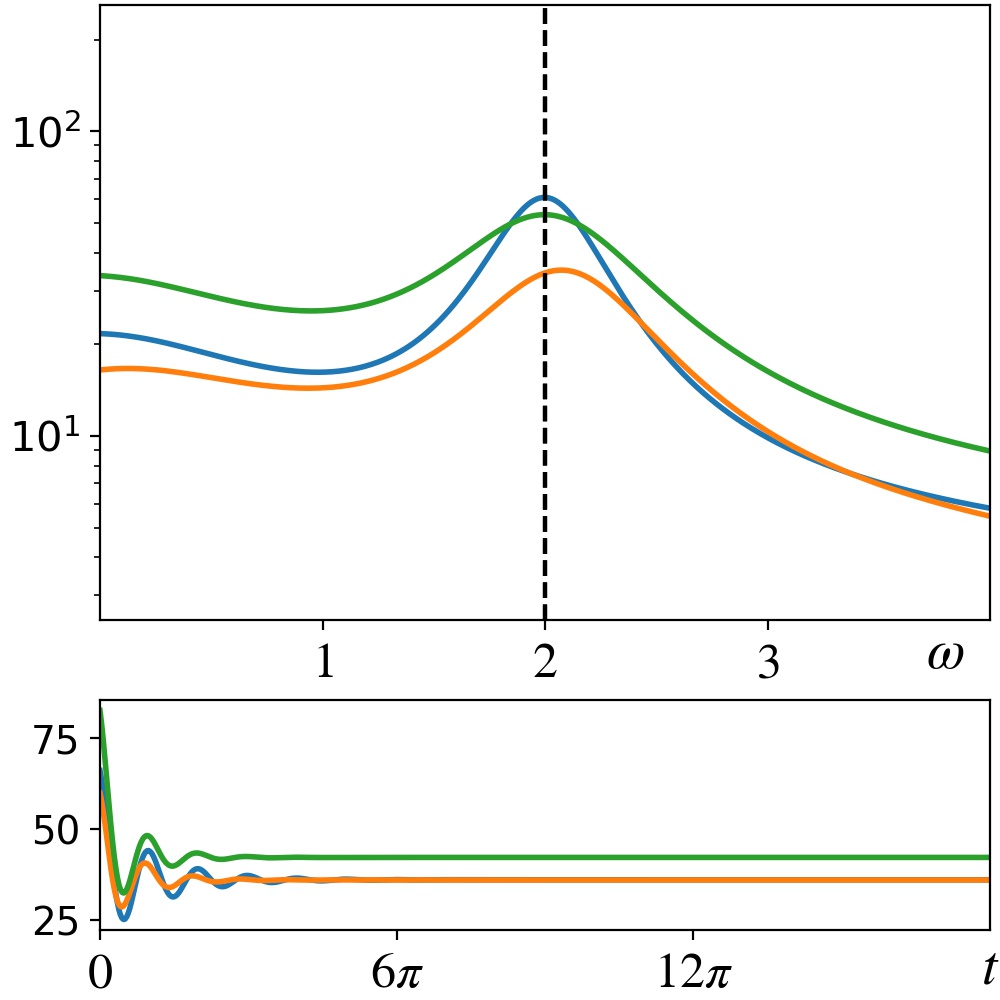}
    \caption{$\epsilon=1$}
    \label{}
    \end{subfigure}
    \hfill
    \begin{subfigure}[t]{0.48\linewidth}
    \includegraphics[width=1\linewidth]{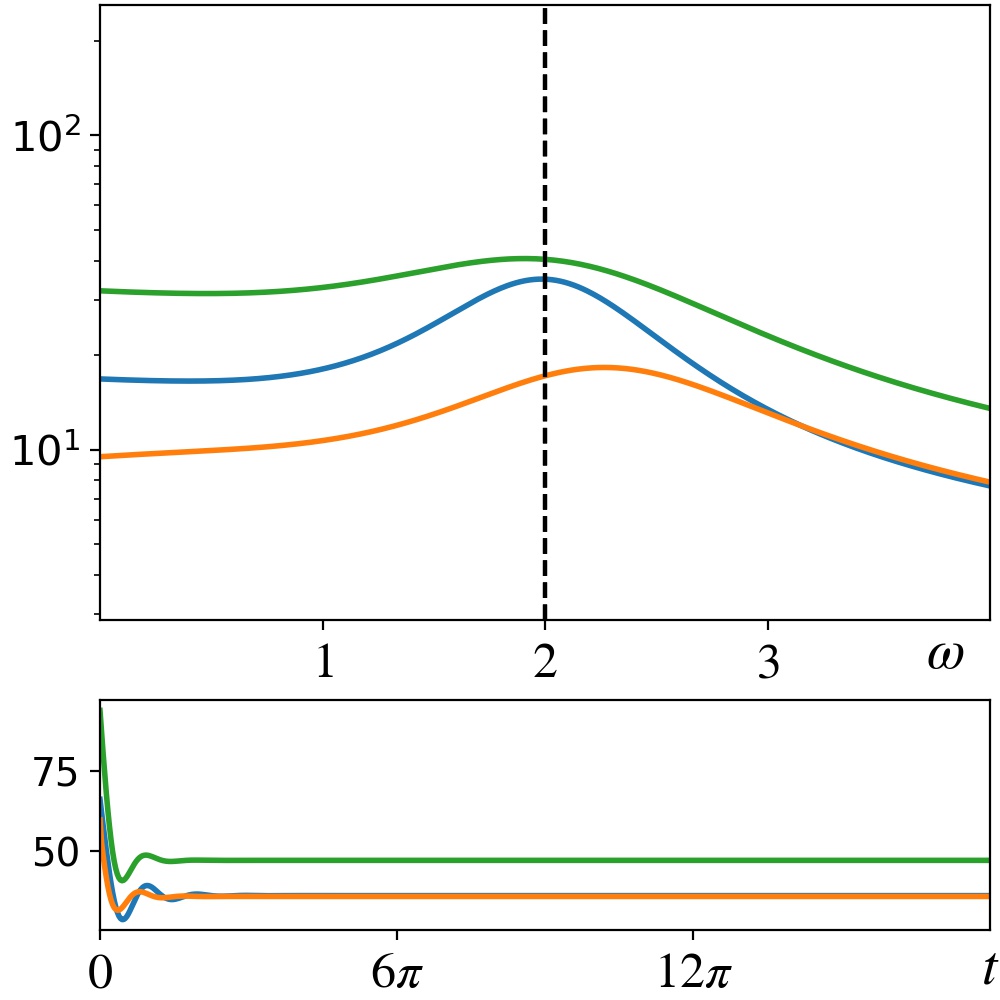}
    \caption{$\epsilon=2$}
    \label{}
    \end{subfigure}
    \caption{Squared-displacement correlation functions $\expval{x^2(t)x^2(0)}$ (bottom panels), and their corresponding spectral distributions $S_{x^2x^2}(\omega)$ (top panels), calculated for the same parameters as in Fig.~\ref{fig:spectra_and_corrs} for the three quantum models. }
    \label{fig:xx_corr}
\end{figure}

\begin{figure}
    \begin{subfigure}[t]{0.48\linewidth}
    \includegraphics[width=1\linewidth]{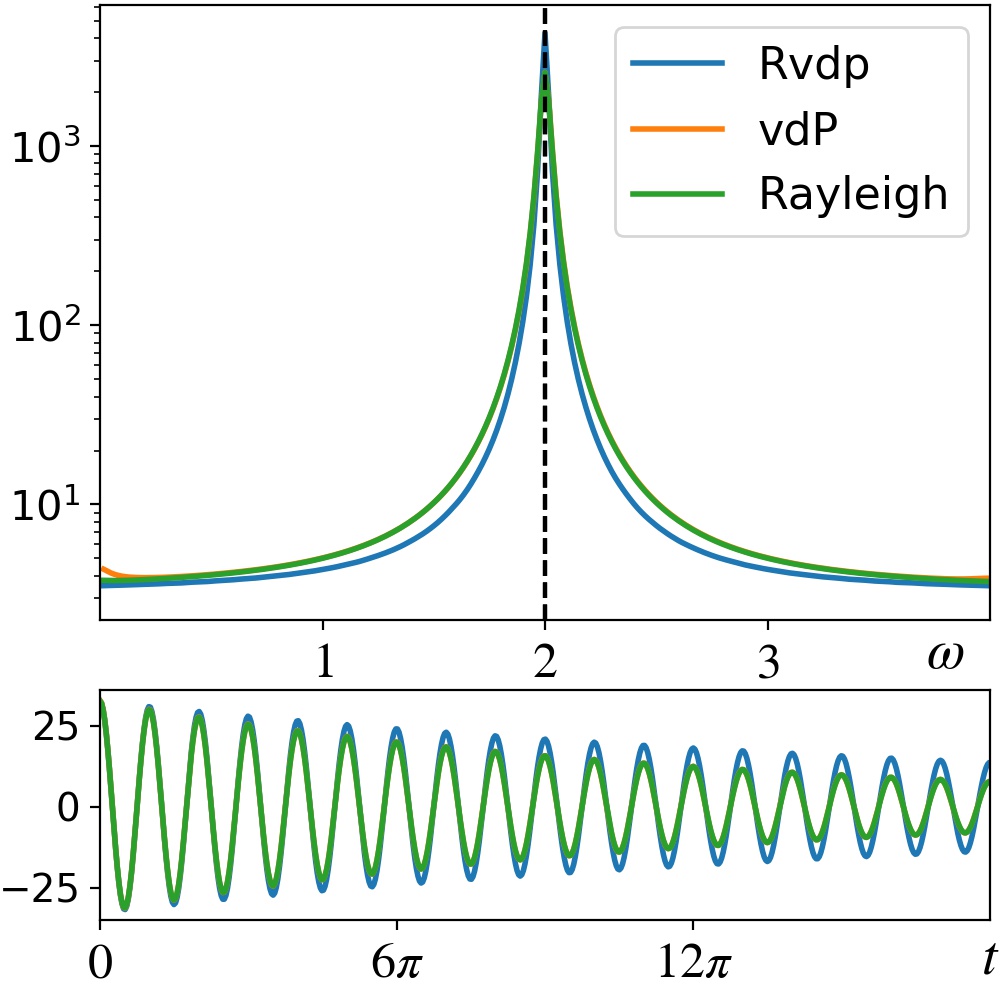}
    \caption{$\epsilon=0.05$}
    \label{}
    \end{subfigure}
    \hfill
    \begin{subfigure}[t]{0.48\linewidth}
    \includegraphics[width=1\linewidth]{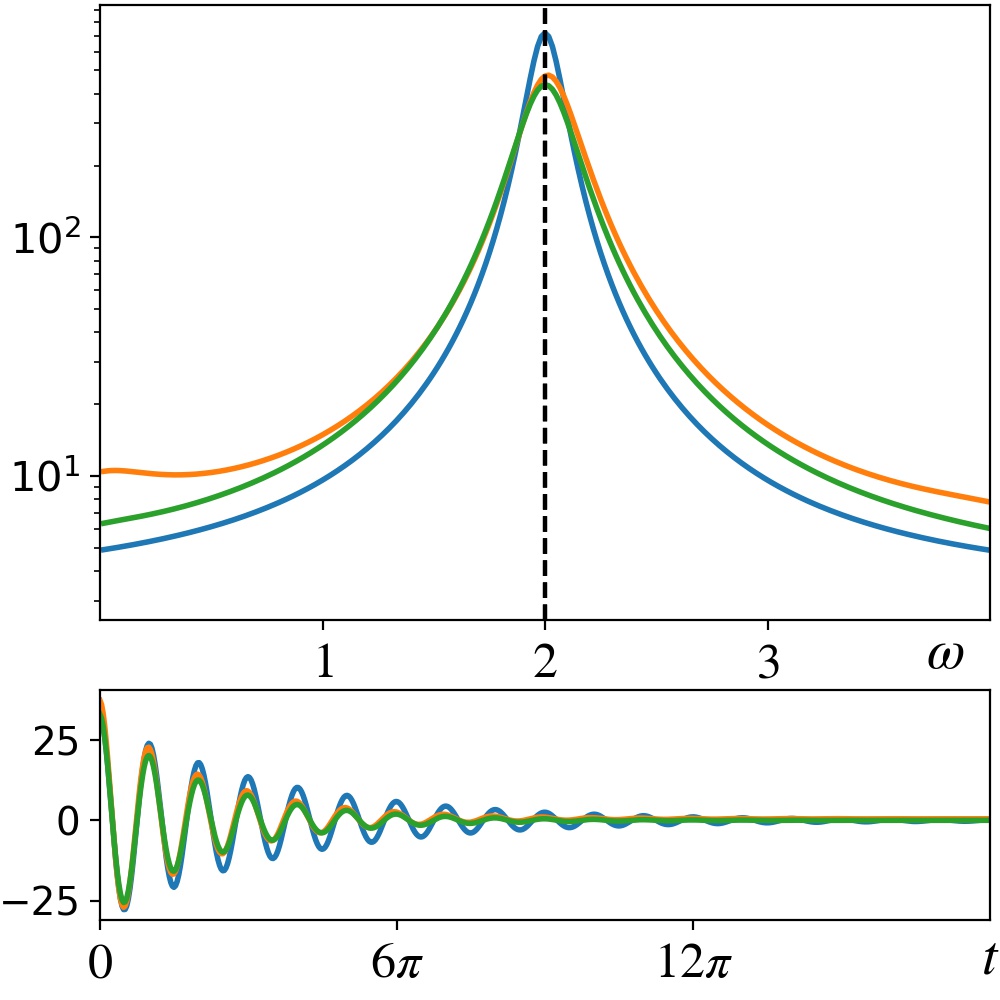}
    \caption{$\epsilon=0.3$}
    \label{}
    \end{subfigure}
    \hfill
    \begin{subfigure}[t]{0.48\linewidth}
    \includegraphics[width=1\linewidth]{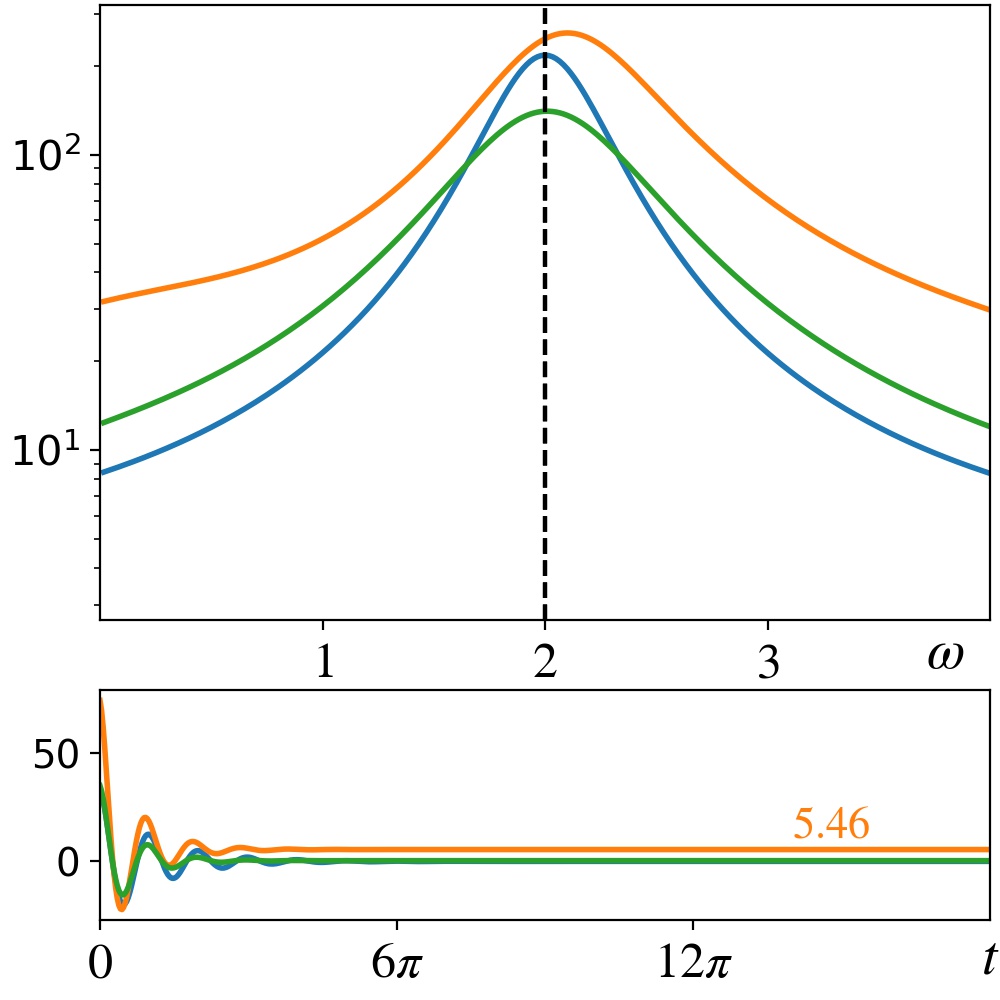}
    \caption{$\epsilon=1$}
    \label{}
    \end{subfigure}
    \hfill
    \begin{subfigure}[t]{0.48\linewidth}
    \includegraphics[width=1\linewidth]{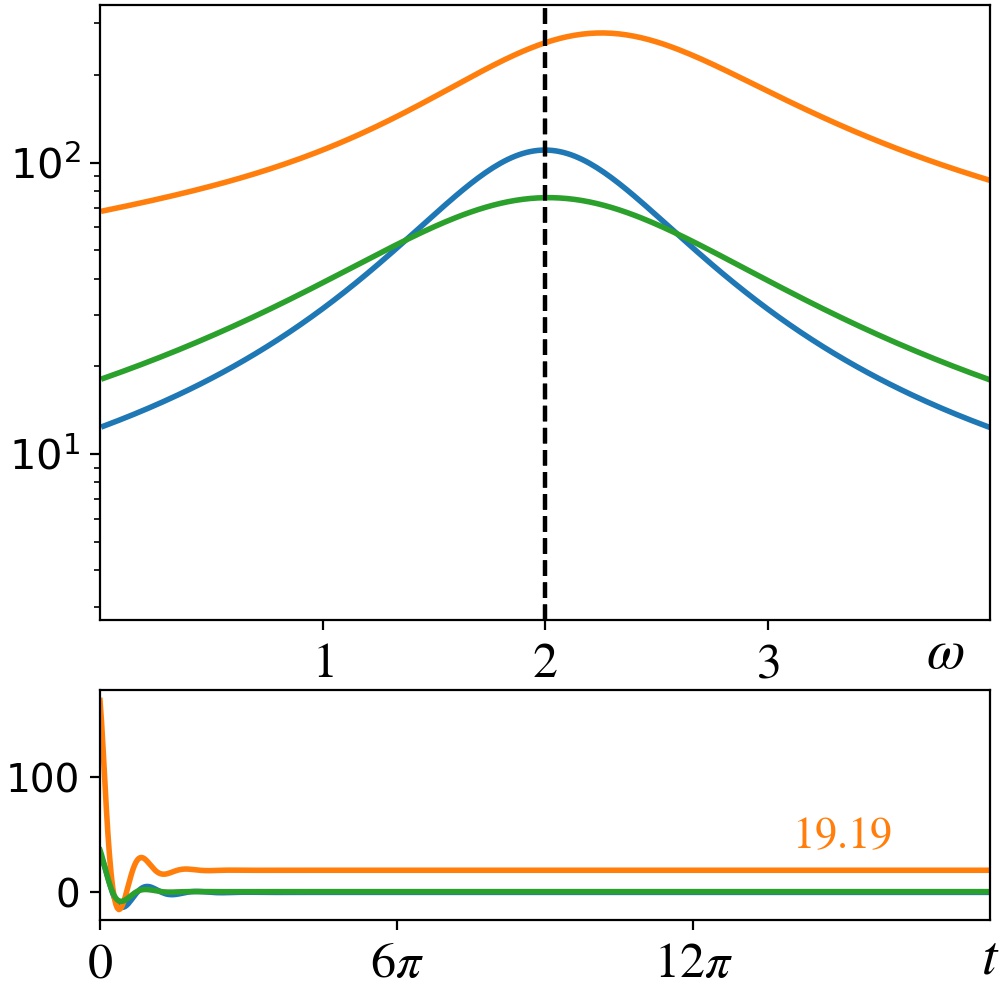}
    \caption{$\epsilon=2$}
    \label{}
    \end{subfigure}
    \caption{Bottom panels: Two-phonon correlation function $\expval{{\adag}^2(t)a^2(0)}$. Top panels: The corresponding spectral distributions $S_{{a}^2{a}^2}(\omega)$. Calculated for the same parameters as in Fig.~\ref{fig:spectra_and_corrs} for the three quantum models.}
    \label{fig:two_phonons}
\end{figure}

\begin{figure}
    \begin{subfigure}[t]{0.48\linewidth}
    \includegraphics[width=1\linewidth]{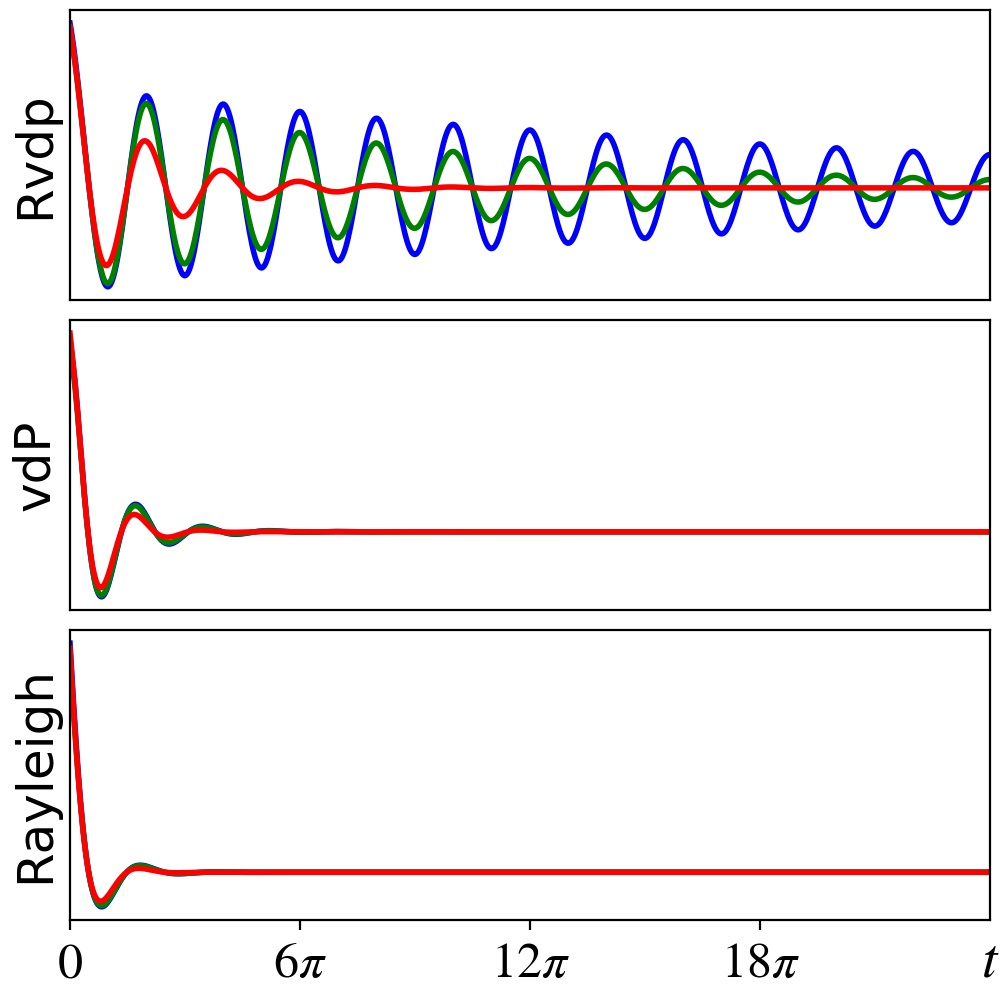}
    \caption{$\expval{x(t)x(0)}$}
    \label{}
    \end{subfigure}
    \hfill
    \begin{subfigure}[t]{0.48\linewidth}
    \includegraphics[width=1\linewidth]{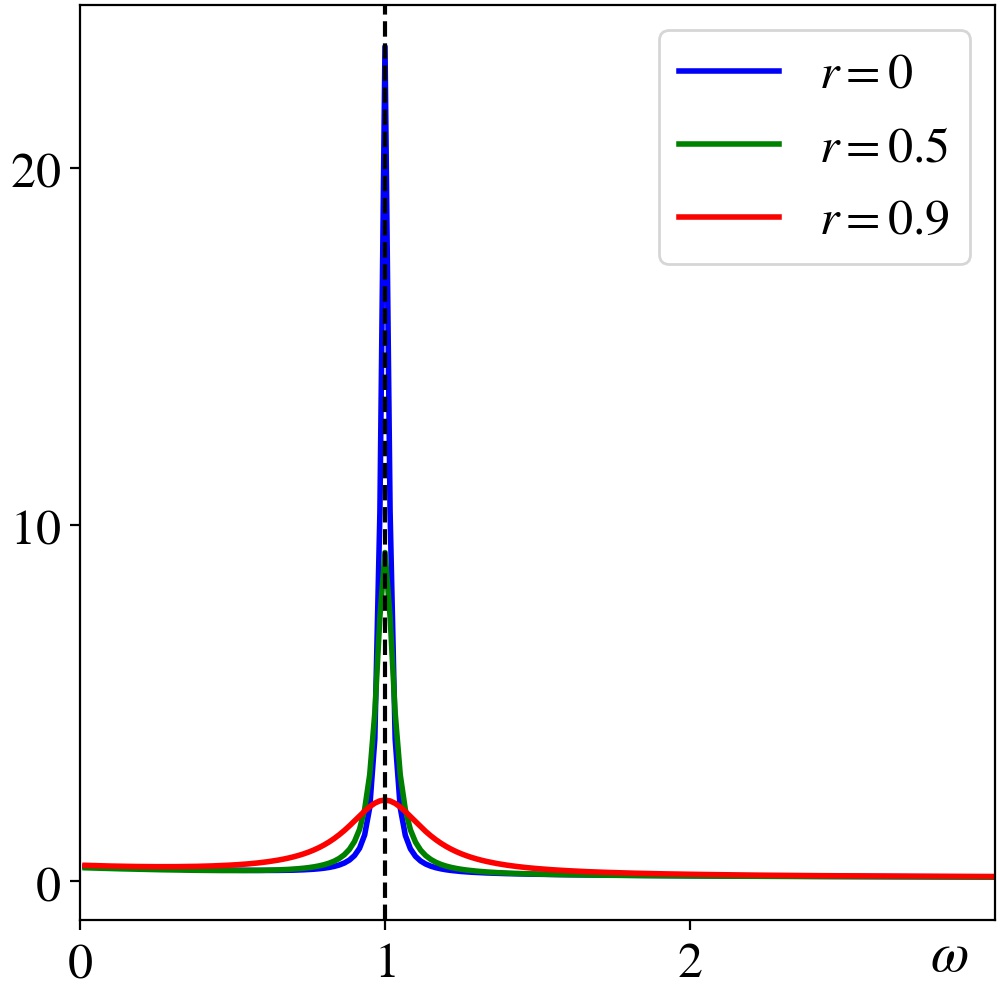}
    \caption{$S_{xx}(\omega)$, RvdP}
    \label{}
    \end{subfigure}
    \hfill
    \begin{subfigure}[t]{0.48\linewidth}
    \includegraphics[width=1\linewidth]{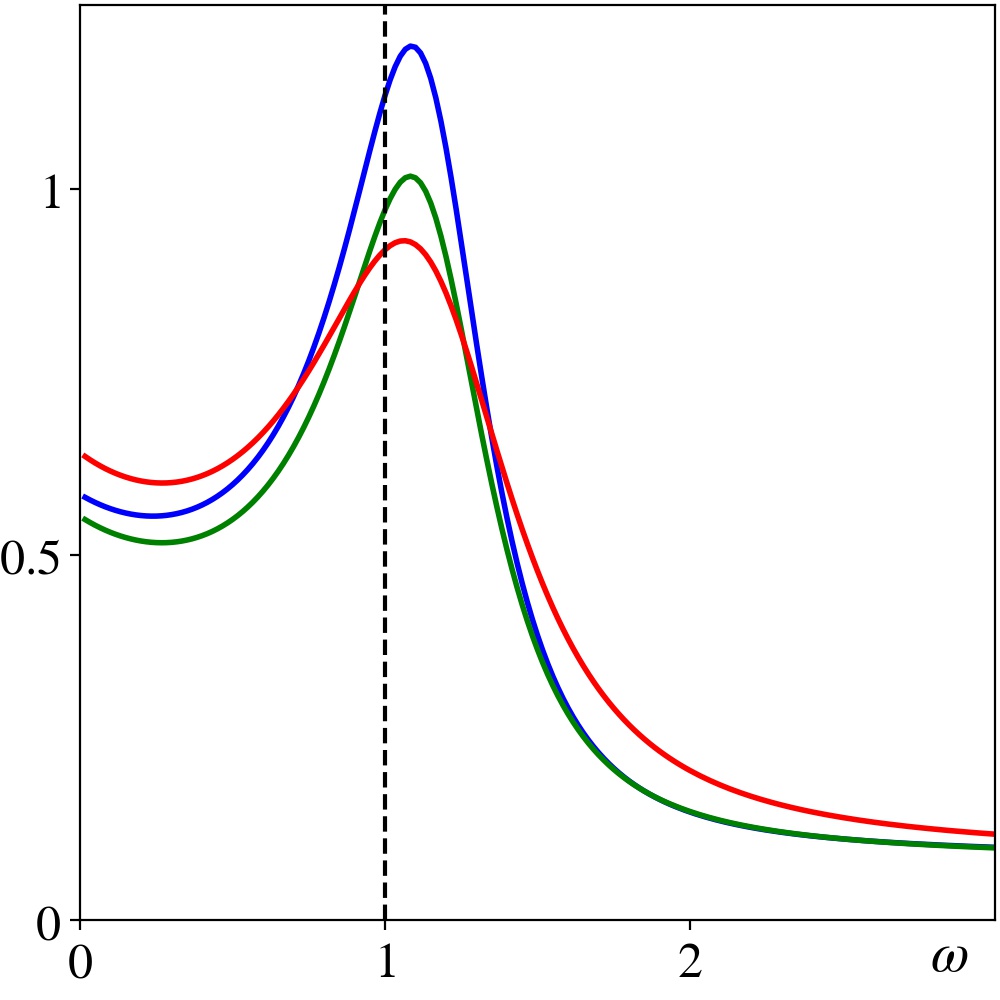}
    \caption{$S_{xx}(\omega)$, vdP}
    \label{}
    \end{subfigure}
    \hfill
    \begin{subfigure}[t]{0.48\linewidth}
    \includegraphics[width=1\linewidth]{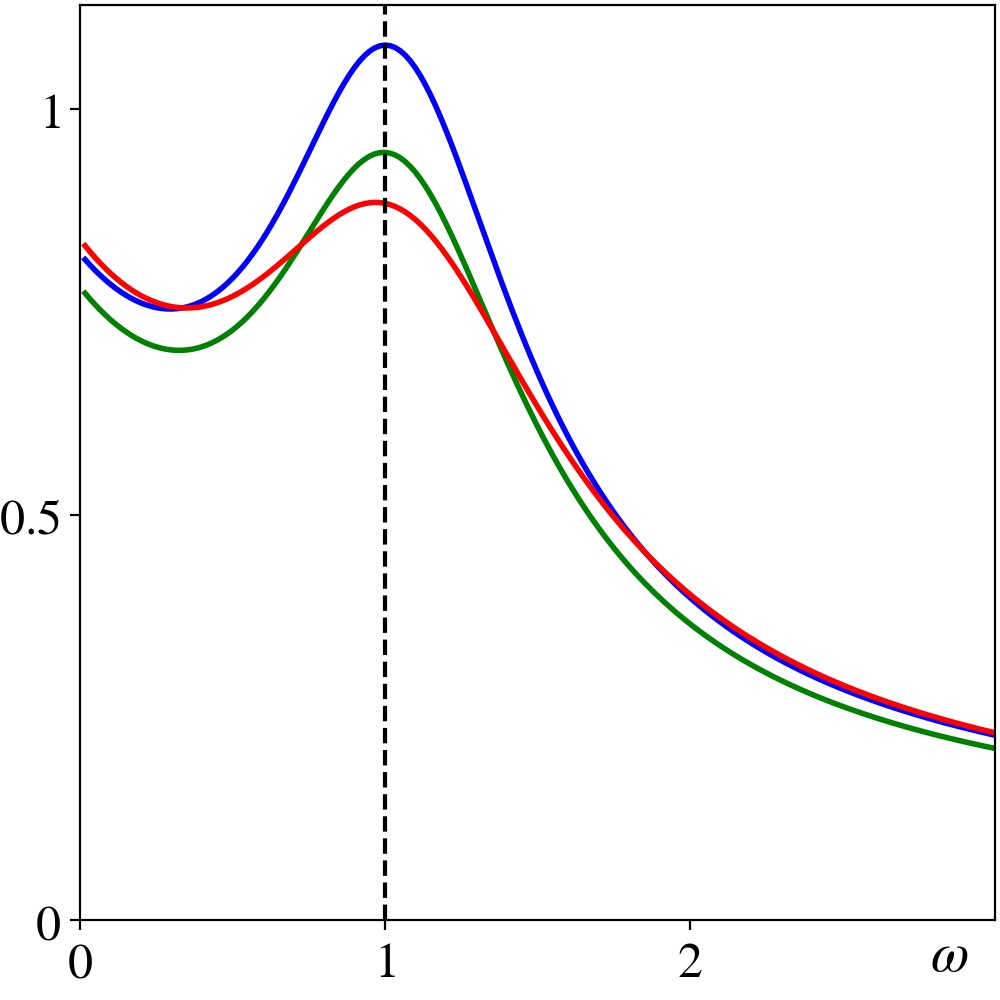}
    \caption{$S_{xx}(\omega)$, Rayleigh}
    \label{}
    \end{subfigure}
    \caption{Displacement correlation functions  $\expval{x(t)x(0)}$ and the corresponding spectral distributions $S_{xx}(\omega)$ for the three models, calculated with $\epsilon=0.01$, for different ratios $r=\gamma_1/\kappa_1$. Parameters are chosen to yield limit cycles with amplitude $A=0.1$, by setting $\grvdp=1$, $\gvdp=4$ and $\gr=4/3$. The decay of correlations in the Rayleigh and vdP models is much faster, governed by the large nonlinear damping rate $\gamma_2$ and only slightly affected by changes in $r$. The decay rate for the RvdP model is unaffected by the nonlinear damping rate and is therefore much smaller, and increasing as $r$ approaches 1, as given by Eq.~\eqref{Eq:RvdPCorrDecay}.}
    \label{fig:quantum_corr}
\end{figure}

\subsection{Correlations and spectral distributions}
\label{Sec:correlations}

It is convenient to consider time correlation functions of various operators, along with their Fourier spectral distributions~\cite{dykman75}, in order to characterize the different quantum limit-cycle models. It is not our intention to provide a thorough analysis of these quantities here, but only to demonstrate that the models do differ in their dynamics. To compare the models side by side we use parameters that generate limit cycles with equal amplitudes $A$, maintaining the same $\kappa_1$ and $\gamma_1$, and varying $\gamma_2$ accordingly, thus setting $\gvdp=4\grvdp$ and $\gr=\gvdp/3$. In all the examples shown here we initiate the dynamics with the steady state density matrices, thus following the decay of correlations, while the oscillators are already in their steady state. Recall that we are still operating at $T=0$, thus the decay of correlations, which results from noise-induced phase diffusion, is caused by quantum rather than thermal fluctuations.

Fig.~\ref{fig:spectra_and_corrs} shows the displacement correlation function $\expval{x(t)x(0)}$, along with its spectral distribution, for limit cycles of moderate amplitude $A=\sqrt{10}$ and different driving strengths $\epsilon=\kappa_1$, with $\gamma_1=0$. We see that for very small $\epsilon$, where the steady-state limit cycles are all circular, the relaxation dynamics are also very similar, with the correlations for the RvdP oscillator decaying only slightly slower than for the other two oscillators. Correspondingly, the RvdP spectral peak at $\omega=1$ is slightly sharper. Recall that the RvdP oscillator is the only one that performs exact simple harmonic motion at frequency $\omega=1$, for any value of $\epsilon$. As $\epsilon$ increases, as shown in  Figs.~\ref{fig:new_lind1} and \ref{fig:new_Rayleigh}, the vdP and Rayleigh limit cycles deviate from perfect circles, and the differences between the three spectral peaks become more evident.

Because the steady-state density matrices of the vdP and Rayleigh oscillators contain non-zero elements in their even off-diagonals, as shown in figures \ref{fig:vdp_dm_elements} and \ref{fig:Rayleigh_dm_elements}, it is interesting to examine the squared-displacement correlation function $\expval{x^2(t)x^2(0)}$, shown in Fig.~\ref{fig:xx_corr}, whose calculation involves simultaneous creation or annihilation of pairs of phonons. Again, at very small $\epsilon$, the behavior is quite similar in all three models, showing two spectral peaks, at $\omega=0$ and $\omega=2$, as expected from the squaring of $x(t)$. As $\epsilon$ increases, deviations between the models quickly become noticeable. In particular, note the rather large shift to higher frequencies of the spectral peak of the vdP oscillator, as the limit cycle becomes less and less circular. Also note the different asymptotic values of the correlation functions, which tend to ${\expval{x^2(0)}}^2$, which is greatest for the Rayleigh oscillator owing to its larger r.m.s.\ displacement (see the classical limit-cycle shapes in Fig.~\ref{fig:classical_limit_cycles}). 

Compare the squared-displacement correlation function with the two-phonon correlation $\expval{{\adag}^2(t)a^2(0)}$, shown in Fig.~\ref{fig:two_phonons}. This is only one of the terms appearing in the calculation of the squared-displacement correlation function, annihilating two phonons at time equal 0, and recreating them at a later time $t$. It directly probes the $m=2$ off-diagonal of the density matrix~\cite{scarlatella19}, which are non-zero for the vdP and Rayleigh models. Indeed, for these models the two phonon correlations decay to a non-zero asymptotic value $\expval{{\adag}^2(0)}\expval{a^2(0)}$, which is greater for the vdP oscillator.

As a final example, we wish to demonstrate that both $\kappa_1$ and $\gamma_1$, and not only their difference $\epsilon$ affect the dynamics of the oscillators as independent sources of quantum noise, especially as one approaches the quantum regime. For this purpose we return to the displacement correlation function, and consider limit cycles of a smaller amplitude $A=0.1$, weakly driven with $\epsilon=0.01$, at $\grvdp=1\gg\epsilon$. The correlation functions and their spectral distributions are shown in Fig.~\ref{fig:quantum_corr}, for three different values of the ratio $r=\gamma_1/\kappa_1$ between the linear damping and the pumping rates. One sees very clearly that the RvdP oscillator exhibits a much slower decay of its displacement correlation function, as well as a higher sensitivity to the value of $r$, than the other two oscillators. To see why this is so, consider the rate equations~\eqref{Eq:vdP_rate} for the density matrix elements of the vdP oscillator, and notice that all the off-diagonal ($m\neq0$) terms have some negative coupling due to the nonlinear damping $\gamma_2$, which causes these terms quickly to decay for the large values of $\gamma_2$ in the quantum limit. The same holds for the Rayleigh oscillator. On the other hand, inspection of the rate equations~\eqref{Eq:me-trans_element1} for the the RvdP density matrix, reveals that it has a single off-diagonal element $\rho_{0,1}$ that does not have a negative coupling term proportional to $\gamma_2$. Assuming that the large nonlinear damping rate quickly depletes all other off-diagonal matrix elements, one remains with this last element, whose decay rate,
\begin{equation}\label{Eq:RvdPCorrDecay}
    \frac{\dot{\rho}_{0,1}}{\rho_{0,1}} \simeq -\frac{3\kappa_1+\gamma_1}{2} = 
    -\frac{\epsilon}{2} \frac{3+r}{1-r},
\end{equation}
which is governed by the much smaller rates $\gamma_1$ and $\kappa_1$, is indifferent to the nonlinear damping rate, and increases as $r$ approaches 1. Also note that the contributions of noise in the energy pump and noise in the linear damping mechanism to the decay rate are additive. In particular, the decay rate at the bifurcation, where $\epsilon=0$, tends to $2\kappa_1=2\gamma_1$. Thus, the oscillator experiences critical slowing down as it crosses the bifurcation only if $\kappa_1$ and $\gamma_1$ are both zero, which may be difficult to arrange experimentally.

\section{Analytical solution for the steady-state Density Matrix of the Rayleigh-van der Pol oscillator}
\label{sec:Analytic_sol}

An analytical solution for the steady state of the $T=0$ quantum RvdP oscillator can be found in previous work \cite{dykman1978, bandilla76, hildred80, dodonov97}, along with approximate solutions for $T\geq0$ in the limit of $\kb T \ll \hbar\omega$ \cite{dykman1978,dodonov97}. Here we provide a general analytical solution for arbitrary temperature. In doing so, we consider a slightly more general physical system than the one described by our master equation~\eqref{Eq:scaledmaster} above, by adding to the model a process of two phonon absorption at rate $\kappa_2$. This additional process, while only recently demonstrated in a micromechanical system~\cite{dong18}, might be quite relevant for other physical systems, such as optical ones, where two-photon absorption might be as likely as two-photon emission. 

The revised temperature-dependent master equation is then written as
\begin{eqnarray}\label{Eq:master_with_k2}\nonumber
    \dot {\rho} &=&\frac{1}{i\hbar}[H_0,\rho] \\ \nonumber
    &+& \kappa_1 \left\{\left(1+\nbar(\Delta_1)\right) D[\adag]\rho+ \nbar(\Delta_1) D[a]\rho\right\}\\ \nonumber
    &+& \gamma_1 \left\{ (1+\nbar(\omega))D[a]\rho + \nbar(\omega)D[\adag] \rho \right\}\\ \nonumber
    &+& \gamma_2 \left\{ (1+\nbar(2\omega))D[a^2]+ \nbar(2\omega)D[{\adag}^2] \rho \right\}\\
    &+& \kappa_2 \left\{ (1+\nbar(\Delta_2))D[{\adag}^2]+ \nbar(\Delta_2)D[a^2] \rho \right\},
\end{eqnarray}
where the last line is responsible for two phonon absorption, and $\nbar(\omega)=(e^{\hbar\omega/\kb T}-1)^{-1}$ is the Bose-Einstein distribution through which the temperature $T$ is introduced. 

We start by defining four temperature-dependent effective rates, that reduce back to the original rates in the limit of $T\to 0$, 
\begin{subequations}\label{Eq:total_rates}
\begin{eqnarray}
    \tilde{\Gamma}_1 &\equiv &\left(1+\nbar(\omega)\right)\gamma_1
    + \nbar(\Delta_1) \kappa_1,\\
    \tilde{K}_1 &\equiv &\left(1+\nbar(\Delta_1)\right)\kappa_1+\nbar(\omega)\gamma_1,\\
    \tilde{\Gamma}_2 &\equiv &\left(1+\nbar(2\omega)\right)\gamma_2+\nbar(\Delta_2)\kappa_2,\\
    \tilde{K}_2 &\equiv &\left(1+\nbar(\Delta_2)\right)\kappa_2 + \nbar(2\omega)\gamma_2.
 \end{eqnarray}
\end{subequations}
Using these, we rewrite the revised RvdP master equation~\eqref{Eq:master_with_k2} more compactly as,
\begin{eqnarray}\label{Eq:master_with_k2_2}
  \dot\rho &= &\frac{1}{i\hbar}[H_0,\rho]\\ \nonumber
  &+ &{\tilde{\Gamma}_1}\Lind{a}
  + {\tilde{K}_1}\Lind{\adag}
  +{\tilde{\Gamma}_2}\Lind{a^2}
  + {\tilde{K}_2}\Lind{{\adag}^2}.
\end{eqnarray}
As discussed earlier, the off-diagonal elements of the RvdP density matrix decay to zero in the steady state, as they are decoupled from the principal diagonal. The remaining rate equations for the diagonal elements $P_n\equiv \varrho_{n,0}= \rho_{nn}= \expval{\rho}{n}$ are given by
\begin{eqnarray}\label{Eq:pn}\nonumber
    \frac{1}{\tilde{\Gamma}_2}\dot P_n 
    &=& K_1\left[nP_{n-1}-\left(n+1\right)P_n\right]\\ \nonumber
    &+& \Gamma_1\left[\left(n+1\right)P_{n+1}-nP_n\right]\\ \nonumber
    &+& K_2\left[n\left(n-1\right)P_{n-2}-\left(n+2\right)\left(n+1\right)P_n\right]\\
    &+&\left[\left(n+2\right)\left(n+1\right)P_{n+2}-n\left(n-1\right)P_n\right],
\end{eqnarray}
where we have rescaled all the rates by the nonlinear damping coefficient $\tilde{\Gamma}_2$,
\begin{equation}\label{Eq:scaled-rates}
    \Gamma_1=\frac{\tilde{\Gamma}_1}{\tilde{\Gamma}_2},\quad
    K_1=\frac{\tilde{K}_1}{\tilde{\Gamma}_2},\quad
    K_2=\frac{\tilde{K}_2}{\tilde{\Gamma}_2}.
\end{equation} 

In the steady state, with $\dot P_n =0$, the set of equations~\eqref{Eq:pn} provide recurrence relations for the Fock-state probabilities $P_n$, giving the steady-state value of each level in terms of the four levels preceding it. Dykman~\cite{dykman1978} and others~\cite{bandilla76,hildred80,dodonov97} solve these recurrence relations for special limiting cases, by using the method of generating functions, which yields a second-order differential equation for the generating function. We use the same method here, but before doing so we note that when summing consecutive rate equations, one obtains a telescopic sum in which many terms cancel out. Thus by summing the first $n+1$ equations~\eqref{Eq:pn}, from 0 to $n$, and dividing by an overall factor of $(n+1)$, we obtain a simpler equation to solve,
\begin{eqnarray}\label{Eq:tel_sum}\nonumber
    \frac{1}{n+1}\frac{1}{\tilde{\Gamma}_2}\sum_{m=0}^n{\dot P_m} 
    &=& \left[\left(n+2\right)P_{n+2}+nP_{n+1}\right]\\ \nonumber
    &-& K_2\left[\left(n+2\right)P_{n}+nP_{n-1}\right] \\
    &+& \Gamma_{1} P_{n+1} - K_1 P_{n} = 0,
\end{eqnarray}
where the maximum power of $n$ is 1 rather than 2, reducing the corresponding differential equation from second to first order.

We solve these simplified recurrence relations using the generating function
\begin{equation}\label{Eq:A-expansion}
    A(x) = \sum_{n=0}^\infty{P_n x^n},\quad \textrm{with}\quad 
    A'(x) = \sum_{n=0}^\infty{(n+1)P_{n+1} x^n}.
\end{equation}
By multiplying Eq.~\eqref{Eq:tel_sum} by $x^{n+1}$, and summing from $n=0$ to $\infty$, we replace the infinite set of recurrence relations with a single differential equation with respect to the auxiliary variable $x$,
\begin{eqnarray}\label{Eq:GF}
    \left[\left(1-K_2 x^2\right)\left(1+x\right)\right]A'(x)&-&\\ \nonumber
    \left[1+K_2\left(2x+x^2\right)+K_1 x-\Gamma_1\right]A(x)
    &=&\left(\Gamma_1-1\right)P_0 + P_1. 
\end{eqnarray}
This nonhomogeneous first-order differential equation can be solved in a standard manner, using an integrating factor. It should be noted, though, that the apparent constant term on the right-hand side of the equation depends linearly on the solution itself, with $P_0=A(0)$ and $P_1=A'(0)$ \footnote{In fact, any arbitrary function $A(x)$ satisfies the equation at $x=0$.}. Therefore, the solution of the associated homogeneous equation as well as any particular solution of the full nonhomogeneous equation are both determined only to within a multiplicative factor. As a consequence, the space of solutions is a 2-dimensional vector space, and we still require two constraints, or boundary conditions, to pin down the physically relevant solution. We shall use the fact that the coefficients $P_n$ in the expansion~\eqref{Eq:A-expansion} of $A(x)$ are probabilities. As such, their values are constrained to be between 0 and 1; they are normalized such that their sum
\begin{equation}\label{Eq:normalization}
  \sum_{n=0}^{\infty}P_n=A(1)=1;
\end{equation} 
and their alternating sum lies a distance not greater than unity away from the origin,
\begin{equation}\label{Eq:alt_sum}
    \left|\sum_{n=0}^{\infty} (-1)^{n}P_n\right| 
    =\left|A(-1)\right| \leq 1.
\end{equation}

Before solving Eq.~\eqref{Eq:GF} we perform the substitution
\begin{equation}\label{Eq:substitution}
    A(x) = \frac{f\left(z(x)\right)}{1+ax},
\end{equation}
with 
\begin{equation}\label{Eq:z-def}
    z(x)=\frac{2a}{1+a}\frac{1+x}{1+ax},
\end{equation}
where for convenience we set $a=\sqrt{K_2}$. After some algebra, we obtain a differential equation for $f(z)$ of the form
\begin{eqnarray}\label{Eq:f-eq}\nonumber
    z(1-z)f'(z) &+& \left(\frac{\Gamma_1+K_1}{1-a^2} -1 - \frac{a\Gamma_1+K_1}{2a(1-a)}z\right)f(z)\\
    &=& \frac{\left(\Gamma_1-1\right)P_0 + P_1}{1+a},
\end{eqnarray}
which after defining
\begin{equation}\label{Eq:b-c}
     b = \frac{a\Gamma_1+K_1}{2a\left(1-a\right)},\ \ 
     \textrm{and\ \ } 
     c = \frac{\Gamma_1+K_1}{1-a^2},\ 
\end{equation}
becomes
\begin{equation}\label{Eq:f-eq-final}
    z(1-z)f'(z) + (c-1-bz)f(z) = C_1,
\end{equation}
and we remember that $C_1$ is a constant to be determined through the boundary conditions. The general solution to this equation is given by
\begin{equation}\label{Eq:sol-f(z)}
    f(z) = \frac{C_1}{c-1}\ {_2F_1}(1,b;c;z) + C_2 z^{1-c} (1-z)^{c-1-b},
\end{equation}
where $C_2$ is a constant of integration multiplying the solution of the associated homogeneous equation, and
\begin{equation}\label{Eq:hypergeometric-a}
    _2F_1(\alpha,\beta;\gamma;z) = \sum_{n=0}^{\infty}{\frac{(\alpha)_n(\beta)_n}{(\gamma)_n}\frac{z^n}{n!}},
\end{equation}
is the hypergeometric function, where $(x)_n$ is the so-called Pochhammer symbol, denoting the rising factorial,
\begin{equation}
  (x)_n\equiv x\left(x+1\right)\left(x+2\right)\ldots\left(x+n-1\right).
\end{equation}
Because $(1)_n=n!$, the expansion~\eqref{Eq:hypergeometric-a} in our case reduces to
\begin{equation}\label{Eq:hypergeometric-1}
    _2F_1(1,b;c;z) = \sum_{n=0}^{\infty}\frac{(b)_n}{(c)_n}z^n,
\end{equation}
and the solution~\eqref{Eq:sol-f(z)} can equivalently be expressed as
\begin{equation}\label{Eq:sol-f(z)-beta}
    f(z) = z^{1-c} (1-z)^{c-1-b} \left[C_1 B_z(c-1,b-c+1) + C_2\right],
\end{equation}
where
\begin{equation}\label{Eq:inc-beta}
    B_z(\alpha,\beta) = \int_0^z t^{\alpha-1}(1-t)^{\beta-1} dt
\end{equation}
is the incomplete beta function.

Although the power series~\eqref{Eq:hypergeometric-1} diverges for $|z(x)|\geq1$, we need only to evaluate its derivatives at $x=0$, and the condition $|z(0)|<1$ is fulfilled as long as the non-negative parameter $a<1$. Recall that $a^2=K_2$ is the ratio between nonlinear absorption and emission, thus the physical interpretation of $a<1$ is that there is no steady-state solution when the nonlinear gain is stronger than the nonlinear damping, which is indeed the case.

In terms of the original variable, the solution~\eqref{Eq:sol-f(z)} for the generating function becomes
\begin{equation}\label{Eq:sol-A(x)}
\begin{split}
        A(x) = &\frac{D_1}{1+ax}\ {_2F_1}(1,b;c;z(x))\\ 
        + &D_2 (1+ax)^{b-1}(1-ax)^{c-b-1}(1+x)^{1-c}, 
\end{split}
\end{equation}
where the new constants $D_1$ and $D_2$ still need to be determined. 
Clearly, for $c>1$, the solution of the associated homogeneous equation has a singularity at $x=-1$, in contradiction to the condition of Eq.~\eqref{Eq:alt_sum} that the alternating sum be bounded, requiring us to set $D_2=0$. The normalization condition~\eqref{Eq:normalization} then yields the final form of the generating function 
\begin{equation}\label{Eq:gf_sol}
    A(x)= \frac{C}{1+ax}\,{_2F_1}\left(1,b;c;z(x)\right),
\end{equation}
where the normalization constant
\begin{equation}
    C=\frac{1+a}{{_2F_1}\left(1,b;c;\frac{4a}{(1+a)^2}\right)}.
\end{equation}

\begin{figure}
    \begin{subfigure}[t]{0.32\linewidth}
    \includegraphics[width=1\linewidth]{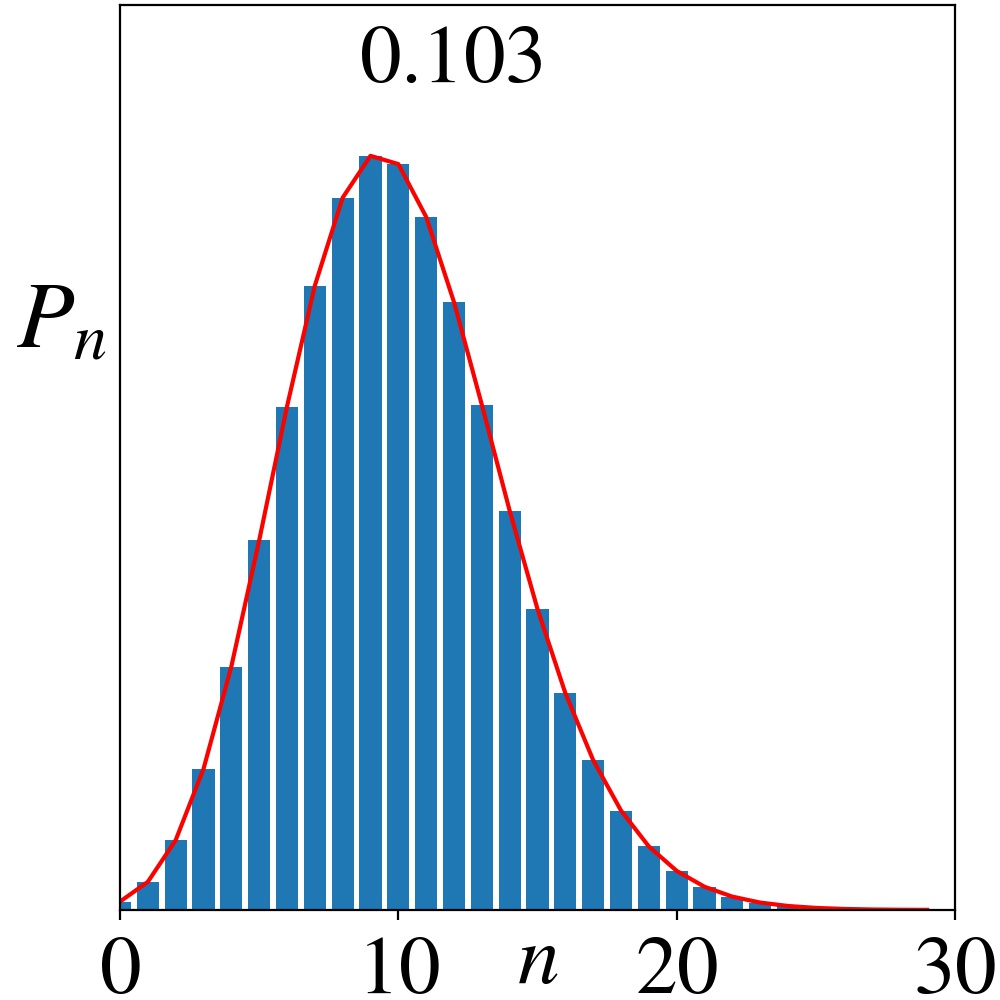}
    \caption{$T=0$}
    \label{}
    \end{subfigure}
    \hfill
    \begin{subfigure}[t]{0.32\linewidth}
    \includegraphics[width=1\linewidth]{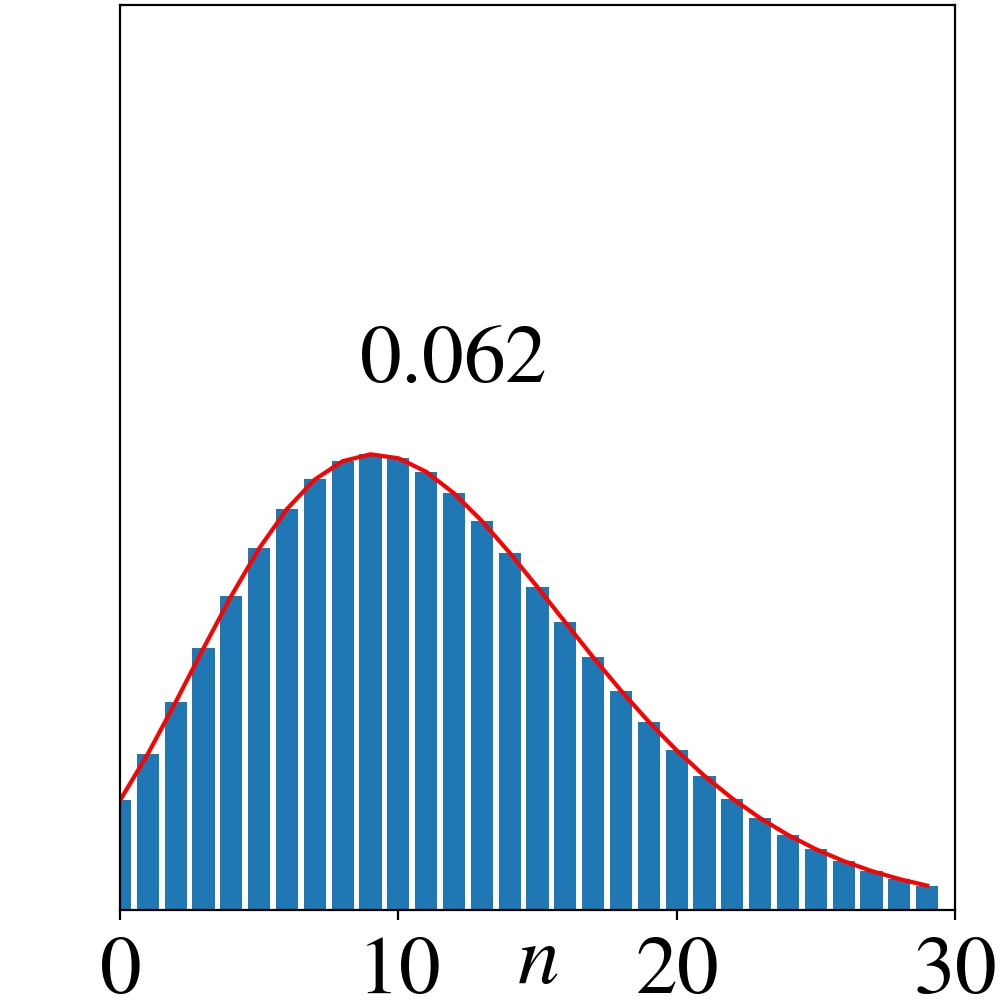}
    \caption{$T=2$}
    \label{}
    \end{subfigure}
    \hfill
    \begin{subfigure}[t]{0.32\linewidth}
    \includegraphics[width=1\linewidth]{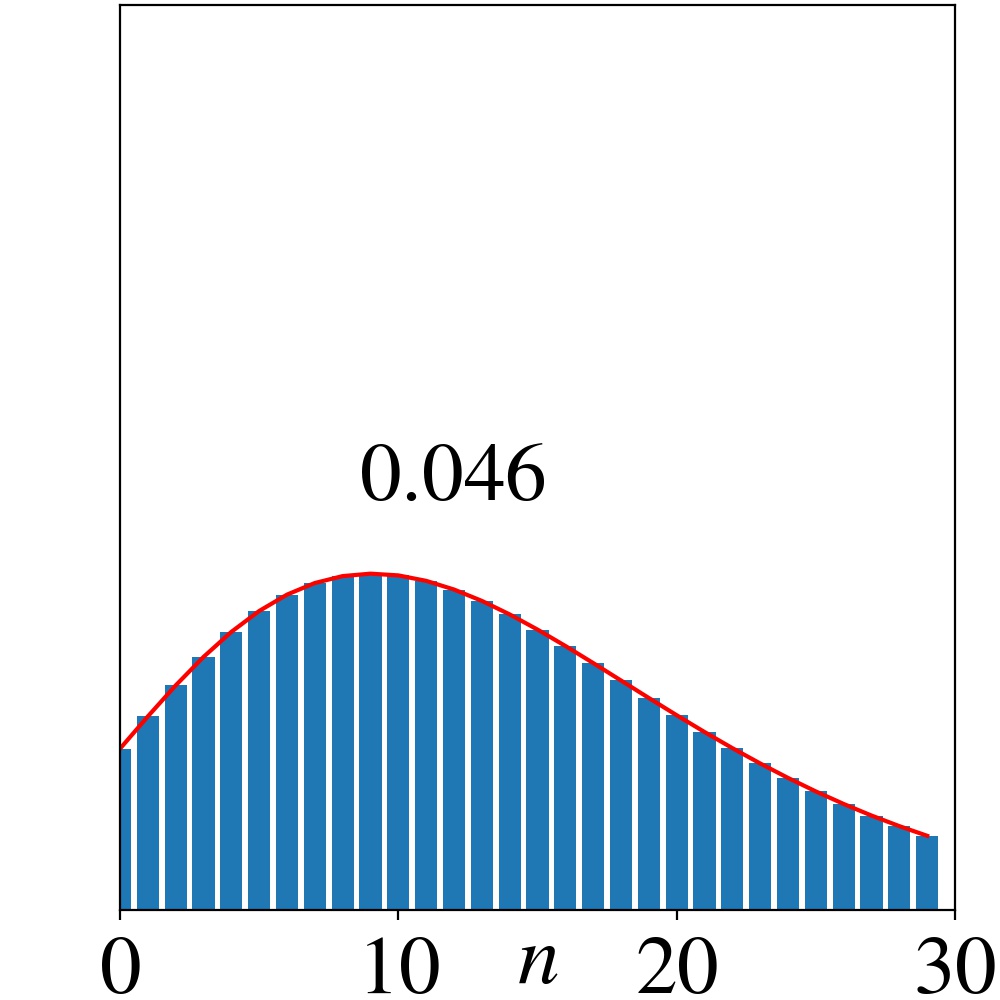}
    \caption{$T=4$}
    \label{}
    \end{subfigure}
    \hfill
    \begin{subfigure}[b]{0.32\linewidth}
    \includegraphics[width=1\linewidth]{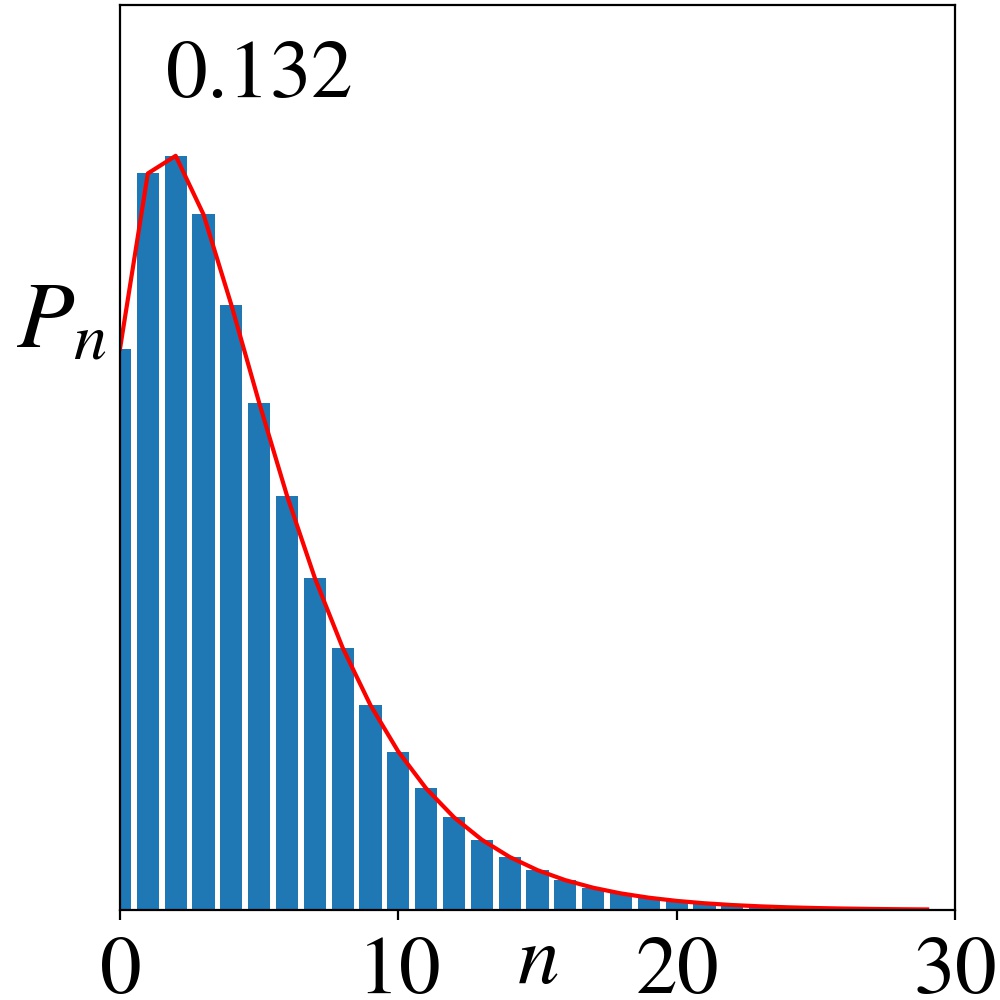}
    \caption{$T=0$}
    \label{}
    \end{subfigure}
    \hfill
    \begin{subfigure}[b]{0.32\linewidth}
    \includegraphics[width=1\linewidth]{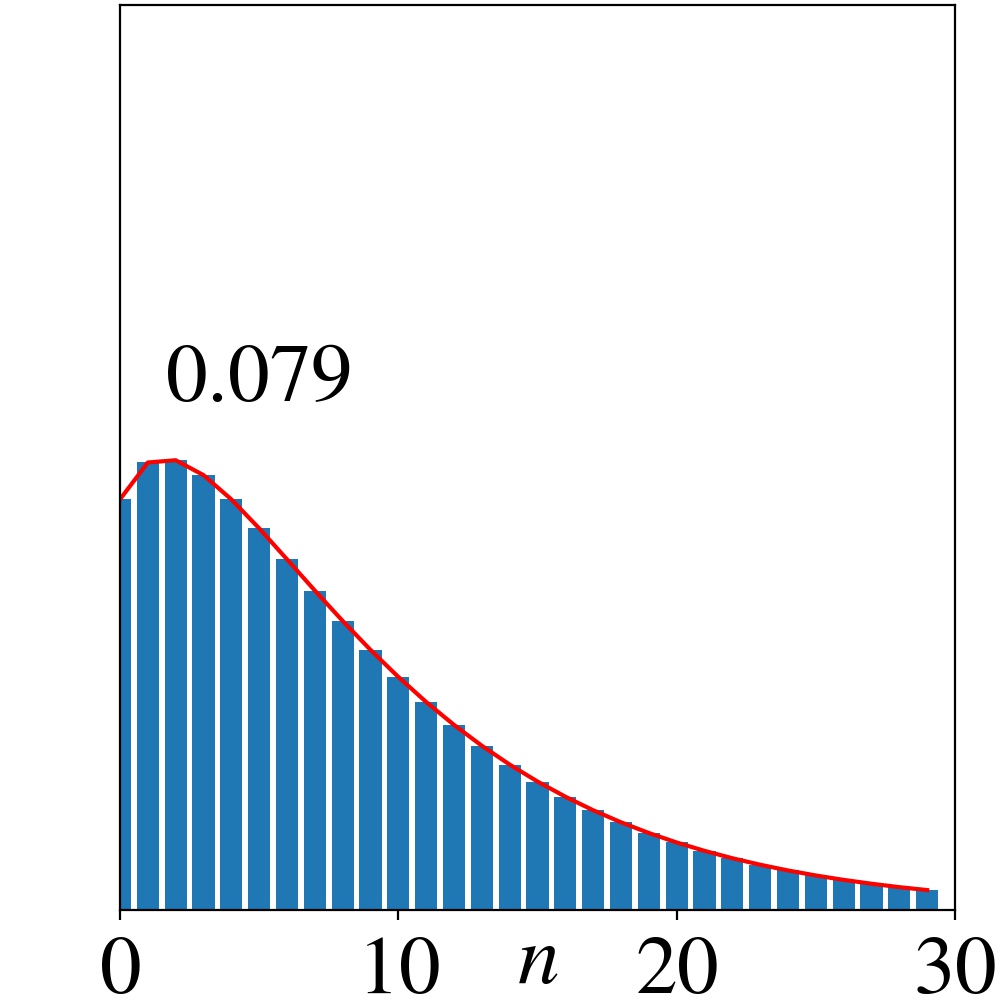}
    \caption{$T=2$}
    \label{}
    \end{subfigure}
    \hfill
    \begin{subfigure}[b]{0.32\linewidth}
    \includegraphics[width=1\linewidth]{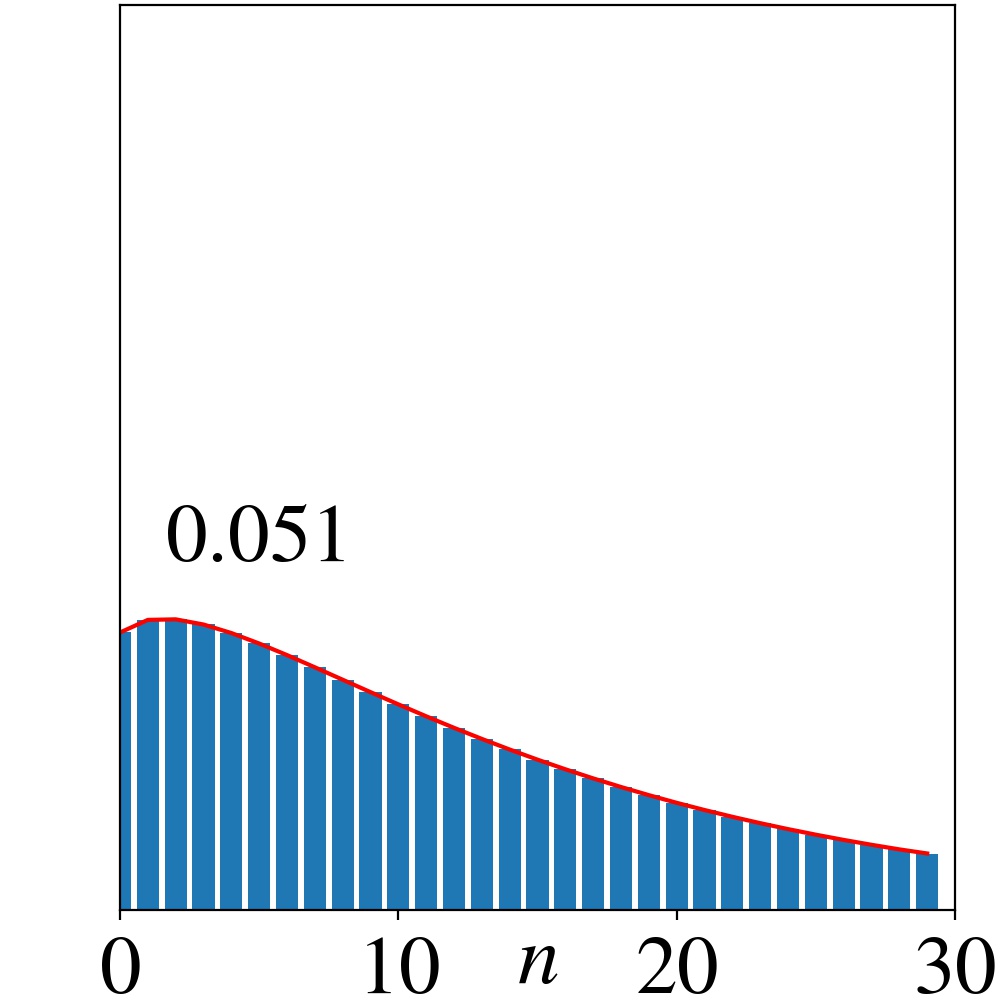}
    \caption{$T=4$}
    \label{}
    \end{subfigure}
    \hfill
    \caption{Analytical (red lines) and numerical (blue bars) solutions for the Fock-state probability distributions of the steady-state RvdP oscillator. Top row : $\kappa_1=20, \kappa_2=0, \gamma_1=1$, and $\gamma_2=1$. Bottom row: $\kappa_1=3, \kappa_2=0.5, \gamma_1=1$, and $\gamma_2=1$. Note that temperature is measured in units of $\hbar\omega/\kb$. The peak values of the distributions are indicated inside the panels.}
    \label{fig:numerics_vs_analytics1}
\end{figure}

With the generating function at hand, we can calculate the probabilities
\begin{equation}\label{Eq:pn_sol}
  P_n = \frac{1}{n!}\frac{\partial^n A}{\partial x^n}\bigg|_{x=0}
    =C\sum_{k=0}^n \frac{\left(-a\right)^{n-k}}{k!} {_2F_1}^{(k)}\left(1,b;c;z(0)\right), 
\end{equation}
where $f^{(k)}(x)$ denotes the $k^{th}$ derivative of $f(x)$. Finally, the derivatives of ${_2F_1}$ can be evaluated using the relation \cite[see their equation (5.2.2)]{abramowitz1948} 
\begin{equation}
    {_2F_1}^{(k)}(\alpha,\beta;\gamma;z)=\frac{(\alpha)_k(\beta)_k}{(\gamma)_k}{_2F_1}(\alpha+k,\beta+k;\gamma+k;z),
\end{equation}
to give 
\begin{equation}\label{Eq:final_sol_pn}
\begin{split}
    P_n = C \left(-a\right)^{n} \sum_{k=0}^n \bigg[ &\binom{n}{k} \frac{(b)_k}{(c)_k}
    \left(\frac{2(a-1)}{1+a}\right)^k \\
    &{_2F_1}\left(1+k,b+k;c+k;
    \frac{2a}{1+a}\right)\bigg]. 
\end{split}
\end{equation}
In Fig.~\ref{fig:numerics_vs_analytics1} we plot the analytical solution given by Eq.~\eqref{Eq:final_sol_pn}, alongside numerical calculations of the steady-state solutions of the temperature-dependent RvdP master equation \eqref{Eq:master_with_k2} for different parameter values, showing perfect agreement.

As noted above, previous authors~\cite{dodonov97,bandilla76,dykman1978,hildred80} used Eq.~\eqref{Eq:pn} directly, without the telescopic sum~\eqref{Eq:tel_sum}. Instead of our first-order nonhomogeneous equation~\eqref{Eq:GF}, they obtained a homogeneous second-order differential equation of the form
\begin{equation}\label{Eq:2_order_GF}
\begin{split}
    \left[\left(1-K_2 x^2\right)\left(1+x\right)\right]&A''(x)\\
    +\left[\Gamma_1-K_1x- 4K_2x(1+x) \right]&A'(x)\\
    -\left[2K_2(1+x) +K_1 \right]&A(x) = 0.
\end{split}
\end{equation}
Differentiation of the first-order equation \eqref{Eq:GF} yields this second-order equation \eqref{Eq:2_order_GF}, thus solutions to the first-order equation solve the second-order equation as well. Previous work examined the special case of $K_2=0$, obtained at $T=0$ with $\kappa_2=0$, where the second-order equation reduces to the so-called Kummer equation
\begin{equation}\label{Eq:Kummer}
    \left(1 + x\right)A''(x)
    +\left(\Gamma_1 - K_1x\right)A'(x)
    - K_1 A(x) = 0,
\end{equation}
whose solution 
\begin{equation}\label{Eq:confluent}
    A(x)=\frac{_1F_1(1;K_1+\Gamma_1;K_1(1+x))}{_1F_1(1;K_1+\Gamma_1;2K_1)}, 
\end{equation}
involves the confluent hypergeometric function\footnote{Note that in most relevant literature the confluent hypergeometric function is denoted as $\Phi(\alpha;\gamma;z)$, however the current notation of $_1F_1(\alpha;\gamma;z)$ helps to clarify the relation to the full solution in Eq.~\eqref{Eq:gf_sol}.} 
\begin{equation}\label{Eq:con_hypergeometric}
    _1F_1(\alpha;\gamma;z) = \sum_{n=0}^{\infty}{\frac{(\alpha)_n}{(\gamma)_n}\frac{z^n}{n!}},
\end{equation}
and where the probabilities are then given by \cite[see their equation (13.4.9)]{abramowitz1948} 
\begin{equation}
    P_n= \frac{{K_1}^n}{(K_1+\Gamma_1)_n}\frac{_1F_1(1+n;K_1+\Gamma_1+n;K_1)}{_1F_1(1;K_1+\Gamma_1;2K_1)}.
\end{equation}
One can obtain the solution~\eqref{Eq:confluent}, for the special case of zero temperature and no two-phonon absorption, from our general solution~\eqref{Eq:gf_sol} by taking the limit of $a\to 0$, where
\begin{equation}
\begin{split}
    \lim_{a\to 0}(b) =  \infty&,
    \quad \lim_{a\to 0}(z(x))=0,\\
    \lim_{a\to 0}(b z(x)) =  K_1(1+x)&,\quad
    \textrm{and} \lim_{a\to 0}(c)  =  \Gamma_1+K_1.\\
\end{split}
\end{equation}
In this limit, the hypergeometric function ${_2F_1(a,b;c;z)}$ reduces to the confluent hypergeometric function $_1F_1(a;c;z)$,
\begin{equation}
    \lim_{b\to\infty}{_2F_1(a,b;c;{z}/{b})}= {_1F_1}(a;c;z),
\end{equation}
such that
\begin{equation}
    \lim_{a\to 0} C\frac{{_2F_1}\left(1,b;c;z(x)\right)}{1+ax} = \frac{_1F_1\left(1;K_1+\Gamma_1;K_1(1+x)\right)}{_1F_1\left(1;K_1+\Gamma_1;2K_1\right)},
\end{equation}
as expected.

\section{Classical to Quantum Transition of the Rayleigh-van der Pol oscillator}
\label{sec:QCT}

We saw in section~\ref{sec:quantum_RvdP} that when taken to its classical limit, with $A_c\gg1$, the quantum RvdP limit-cycle forms self-oscillations at an amplitude given by $A_c$. However, as $A_c$ is reduced toward unity, in terms of the quantum unit of length $x_0$, the oscillator approaches zero-point motion, and one expects quantum effects to take place. In this section we examine how the limit cycle behaves as the oscillator transitions into this quantum regime.

\begin{figure}
    \begin{subfigure}[b]{0.32\linewidth}
    \includegraphics[width=1\linewidth]{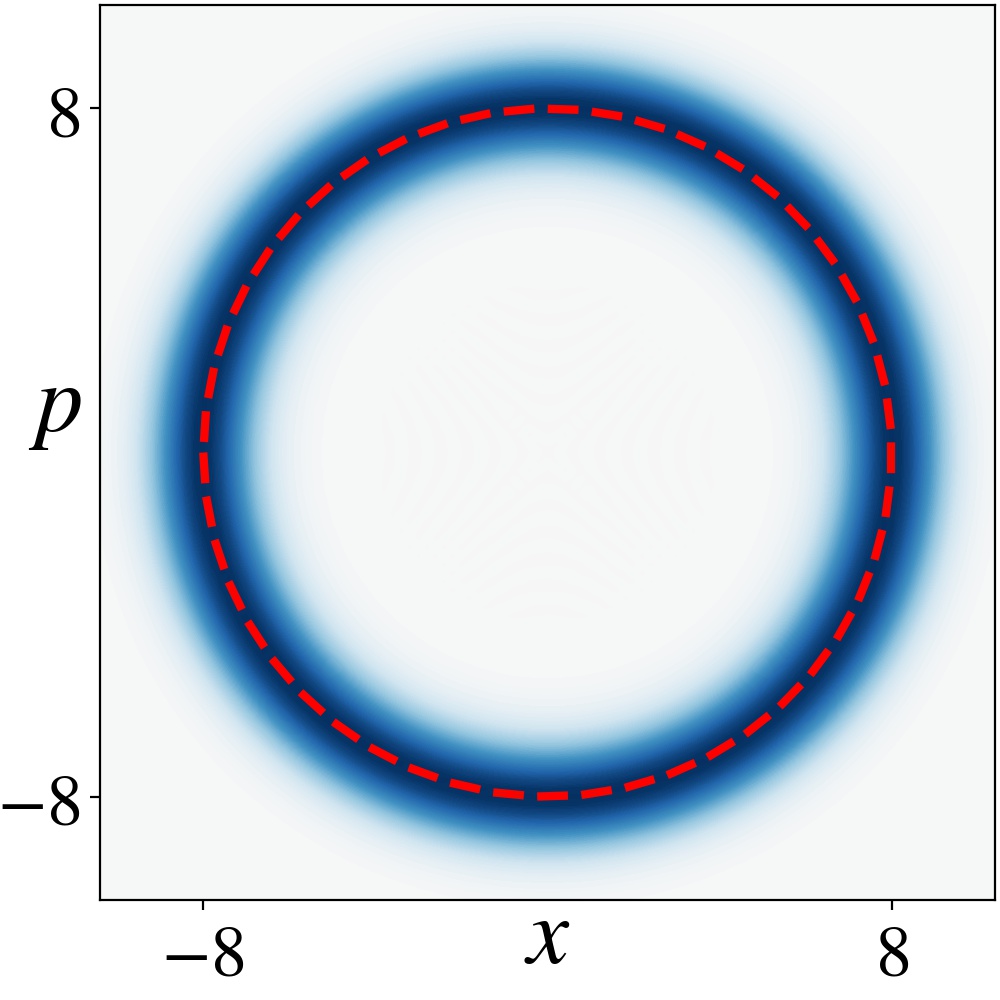}
    \caption{$A_c=8$}
    \label{}
    \end{subfigure}
    \hfill
    \begin{subfigure}[b]{0.32\linewidth}
    \includegraphics[width=1\linewidth]{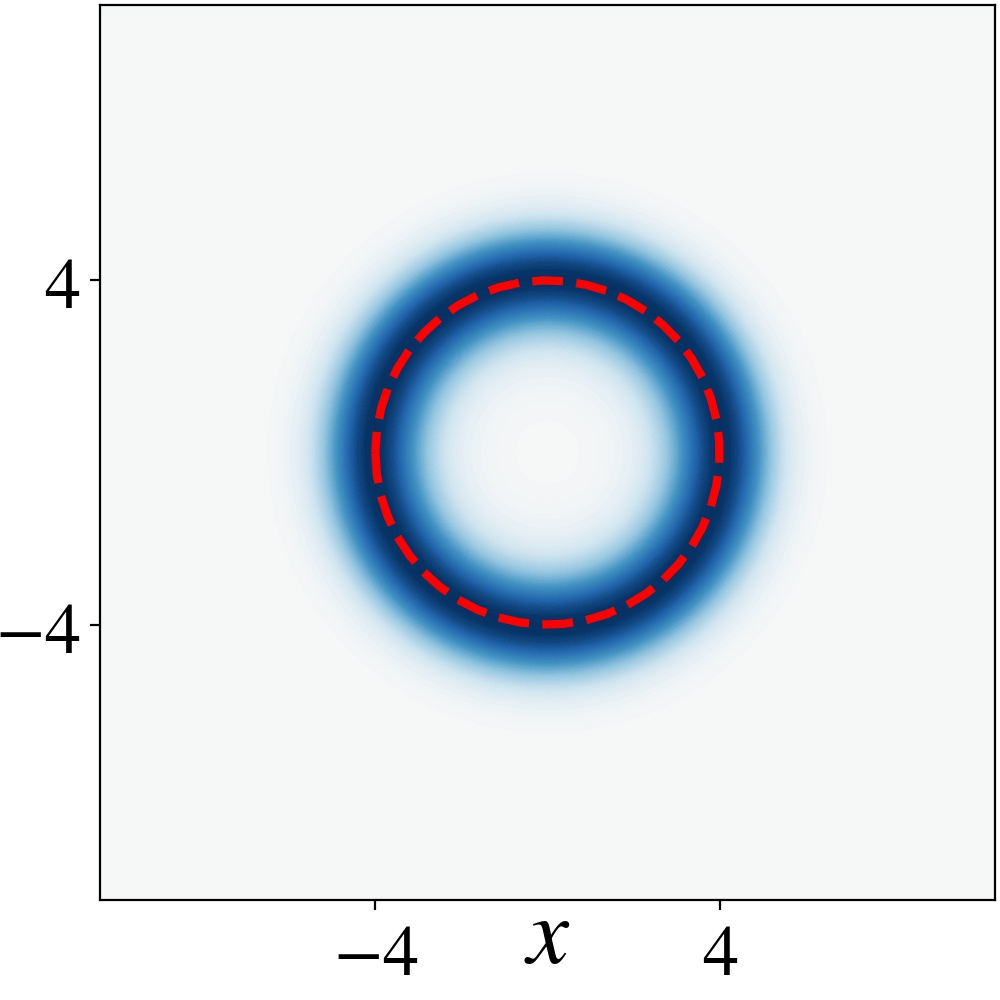}
    \caption{$A_c=4$}
    \label{}
    \end{subfigure}
    \hfill
    \begin{subfigure}[b]{0.32\linewidth}
    \includegraphics[width=1\linewidth]{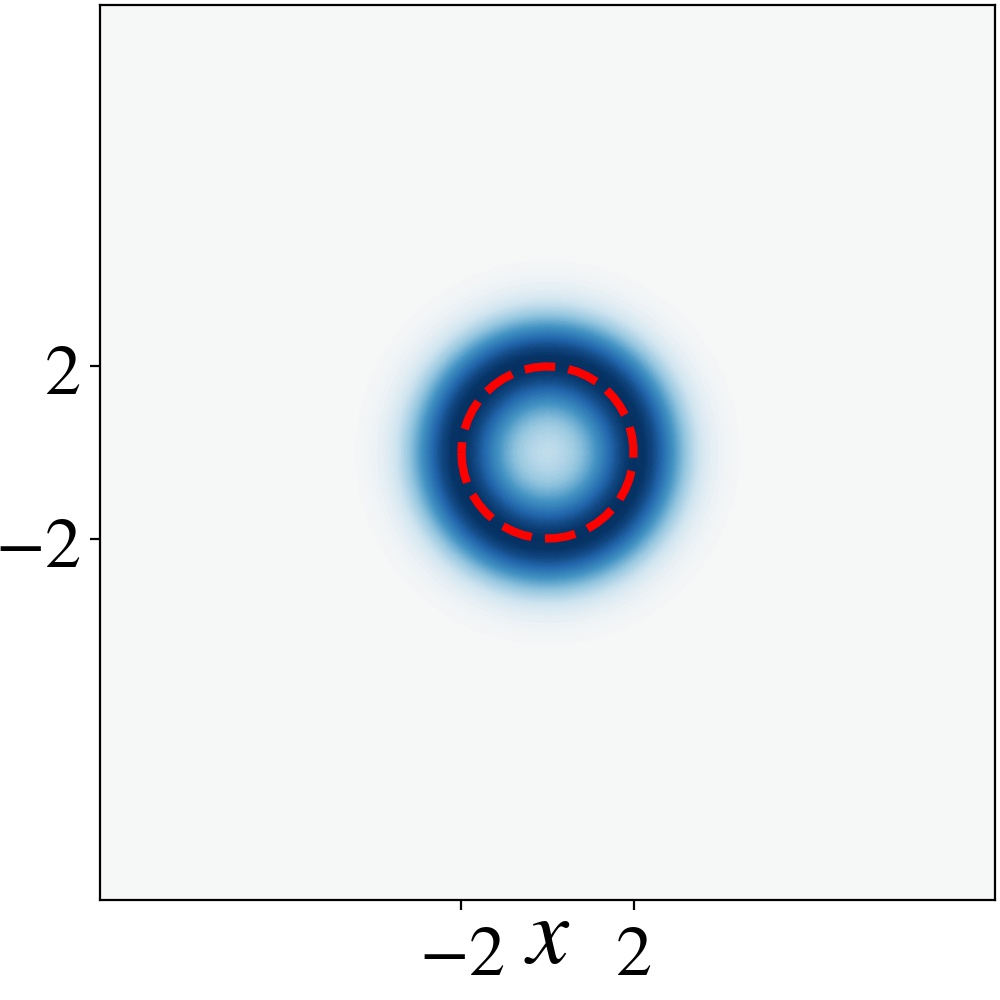}
    \caption{$A_c=2$}
    \label{}
    \end{subfigure}
    \hfill
    \begin{subfigure}[b]{0.32\linewidth}
    \includegraphics[width=1\linewidth]{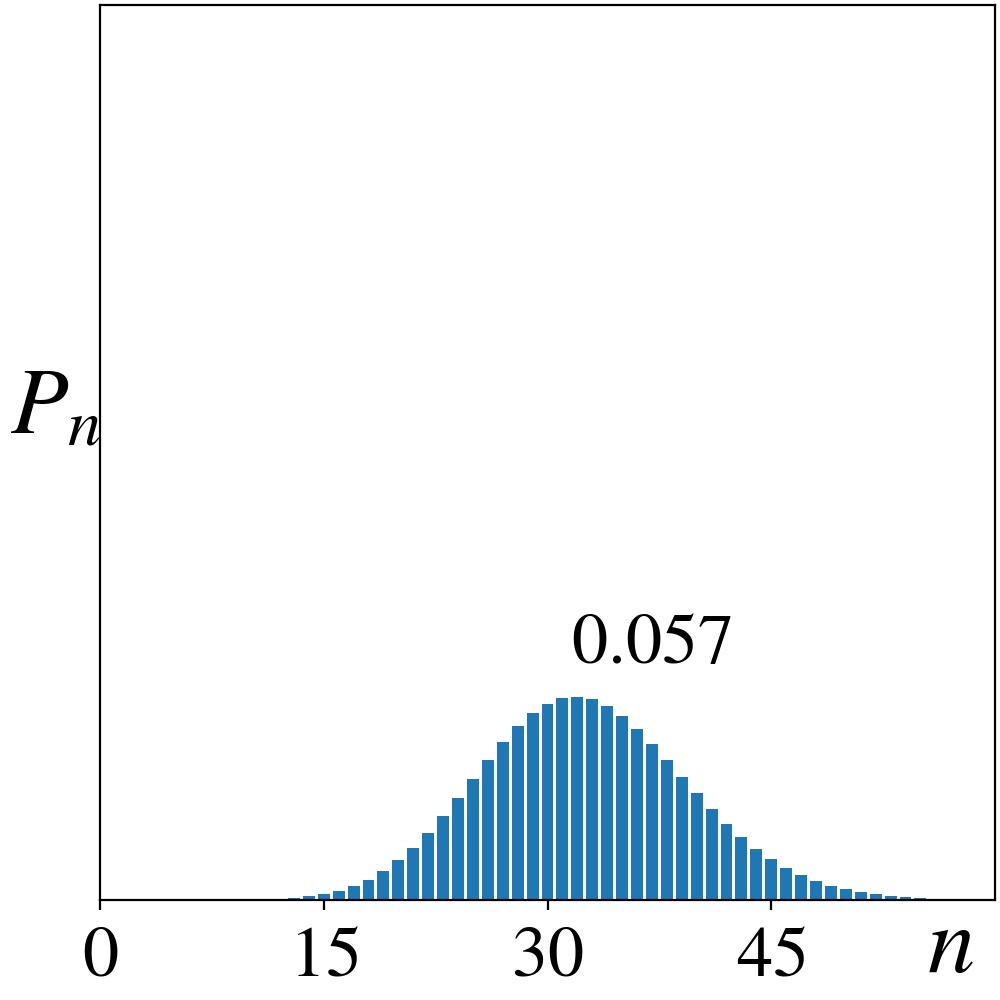}
    \caption{$A_c=8$}
    \label{}
    \end{subfigure}
    \hfill
    \begin{subfigure}[b]{0.32\linewidth}
    \includegraphics[width=1\linewidth]{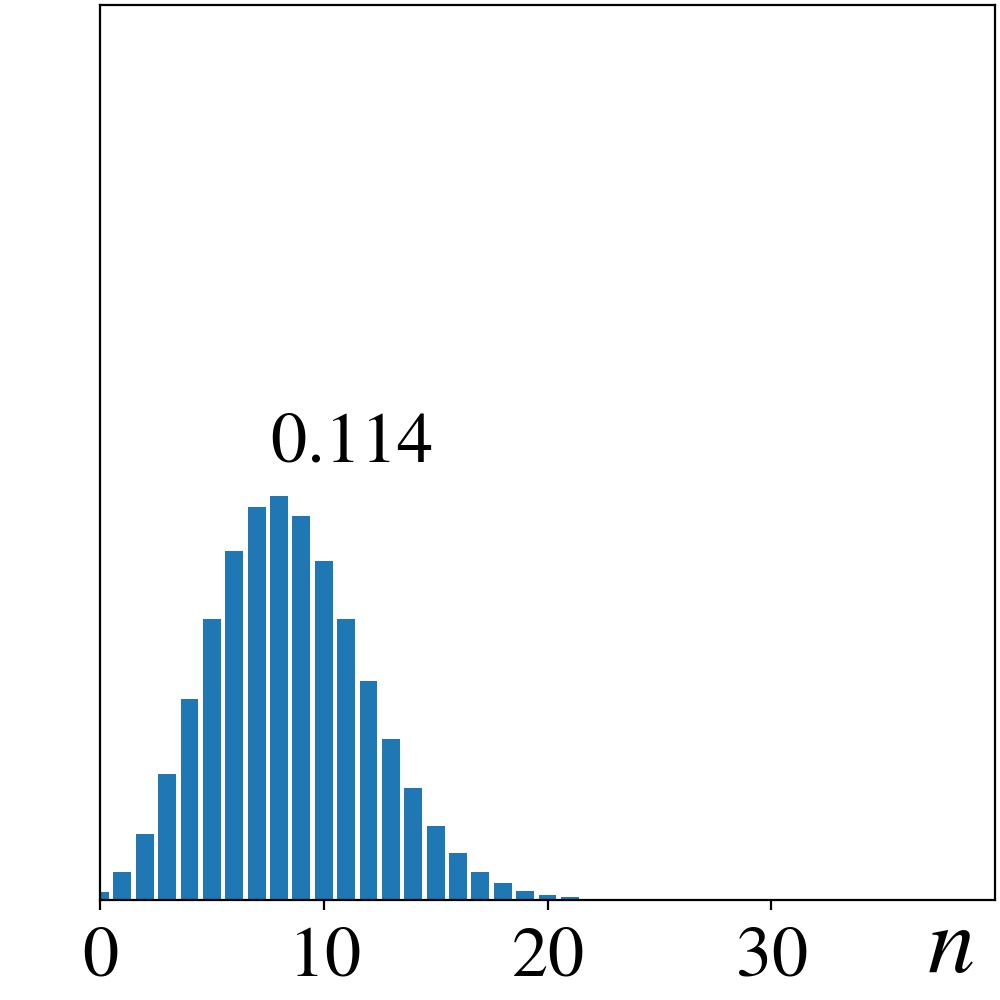}
    \caption{$A_c=4$}
    \label{}
    \end{subfigure}
    \hfill
    \begin{subfigure}[b]{0.32\linewidth}
    \includegraphics[width=1\linewidth]{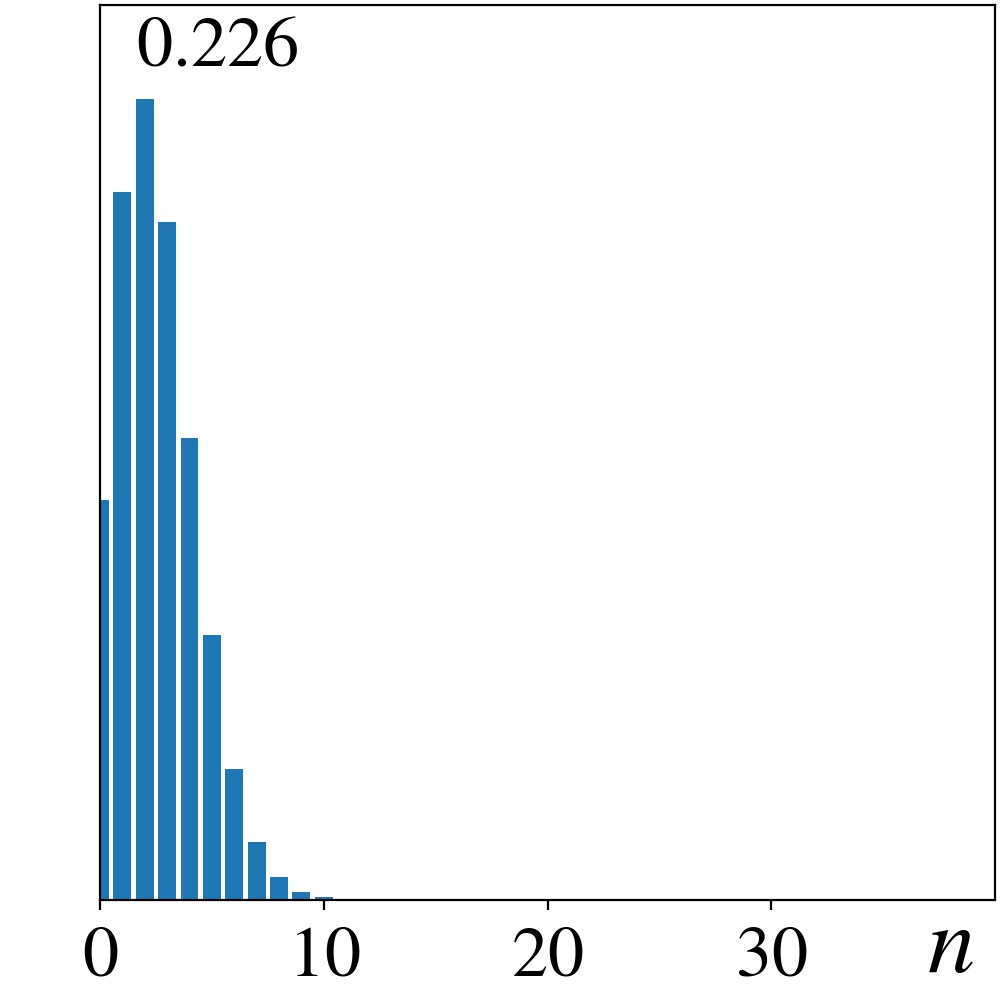}
    \caption{$A_c=2$}
    \label{}
    \end{subfigure}
    \caption{Steady-state Wigner functions and Fock-state distributions at $T=0$, with $\kappa_1=0.1$ and $\gamma_1=0$ for different values of $\gamma_2$ in the classical regime. Peak values of the distributions are written inside the panels. Dashed red circles in the top panels all have radius $A_c$, indicating that this is indeed the amplitude of the limit cycle as long as one remains within the classical limit. 
    }
    \label{fig:wig_fock1}
\end{figure}

\begin{figure}
\hfill
    \begin{subfigure}[b]{0.32\linewidth}
    \includegraphics[width=1\linewidth]{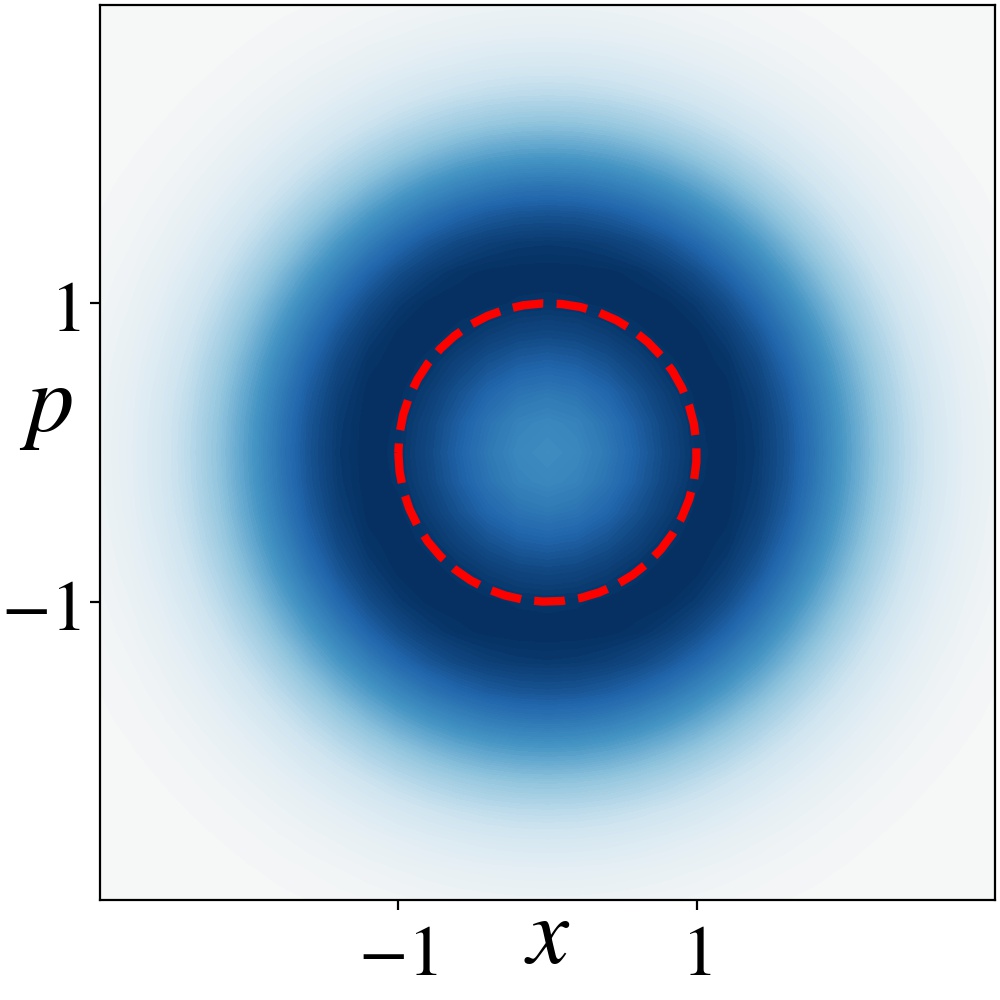}
    \caption{$A_c=1$}
    \label{}
    \end{subfigure}
    \hfill
    \begin{subfigure}[b]{0.32\linewidth}
    \includegraphics[width=1\linewidth]{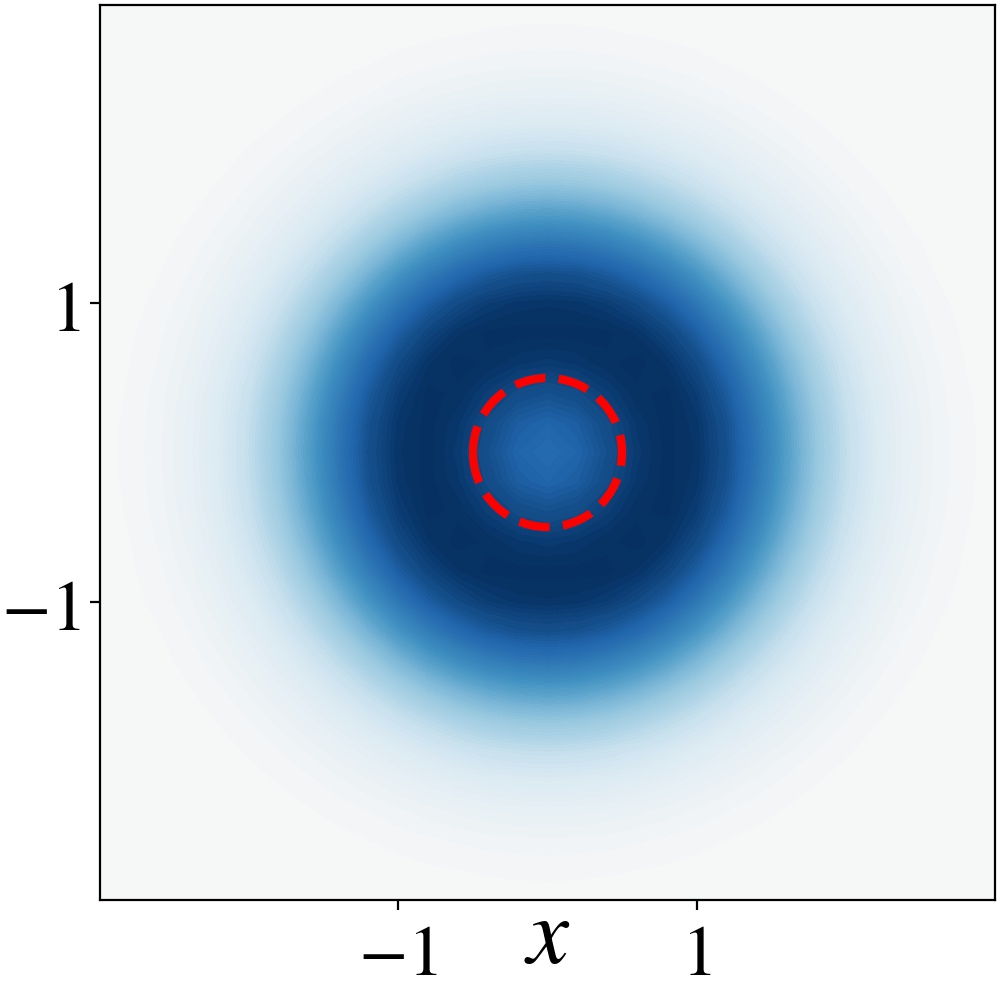}
    \caption{$A_c=0.5$}
    \label{}
    \end{subfigure}
    \hfill
    \begin{subfigure}[b]{0.32\linewidth}
    \includegraphics[width=1\linewidth]{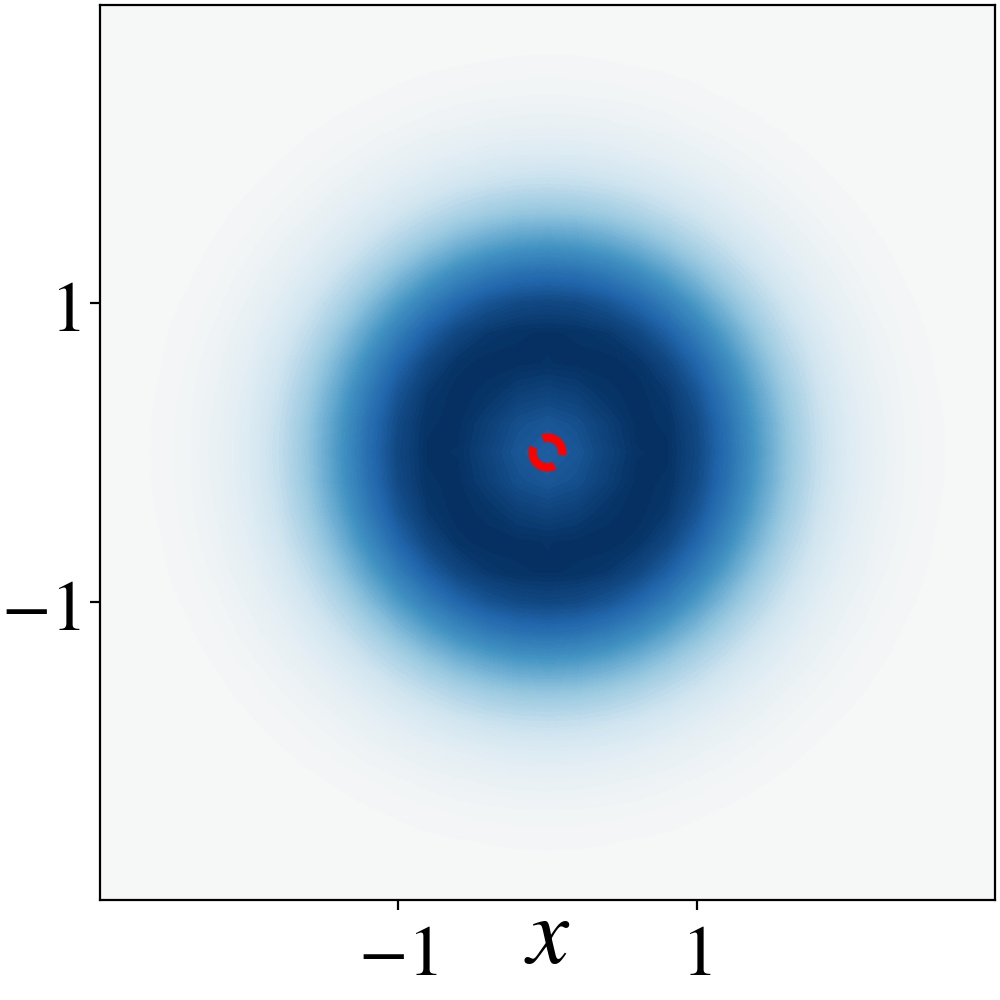}
    \caption{$A_c=0.1$}
    \label{}
    \end{subfigure}
    \hfill
    \begin{subfigure}[b]{0.32\linewidth}
    \includegraphics[width=1\linewidth]{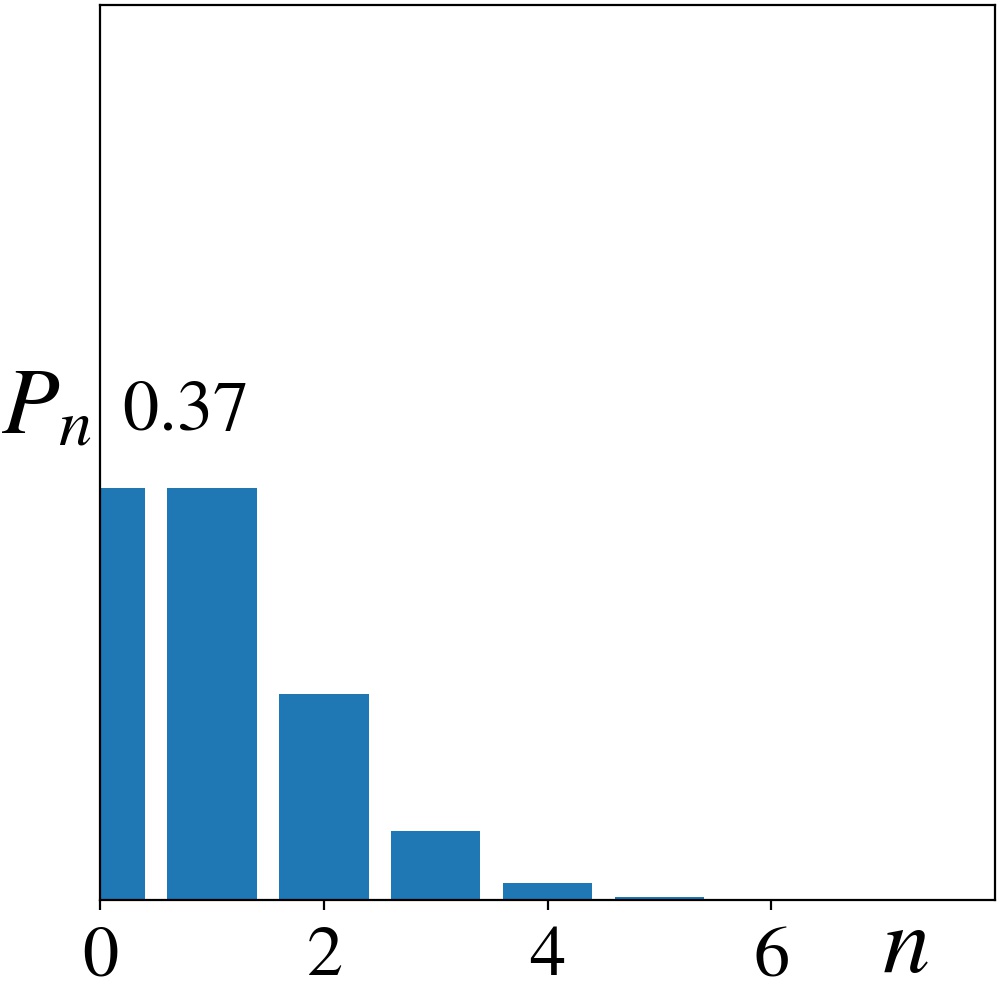}
    \caption{$A_c=1$}
    \label{}
    \end{subfigure}
    \hfill
    \begin{subfigure}[b]{0.32\linewidth}
    \includegraphics[width=1\linewidth]{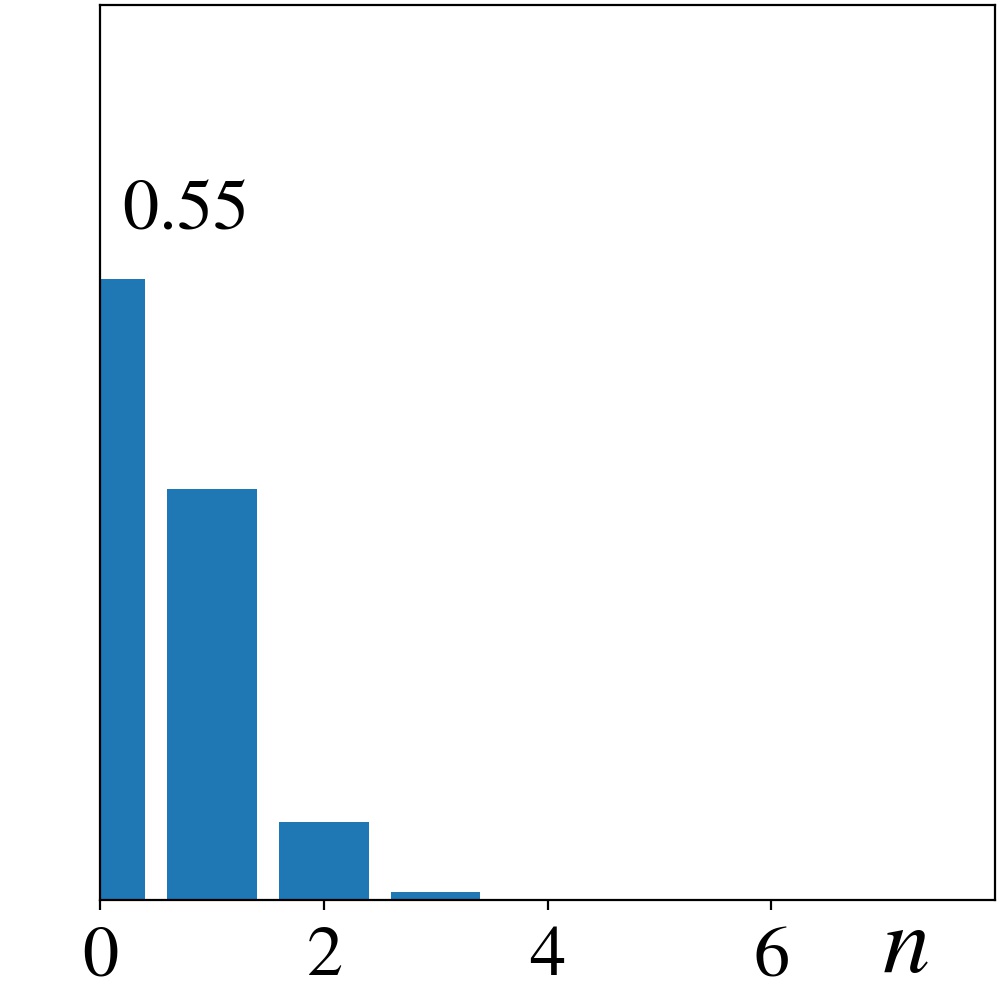}
    \caption{$A_c=0.5$}
    \label{}
    \end{subfigure}
    \hfill
    \begin{subfigure}[b]{0.32\linewidth}
    \includegraphics[width=1\linewidth]{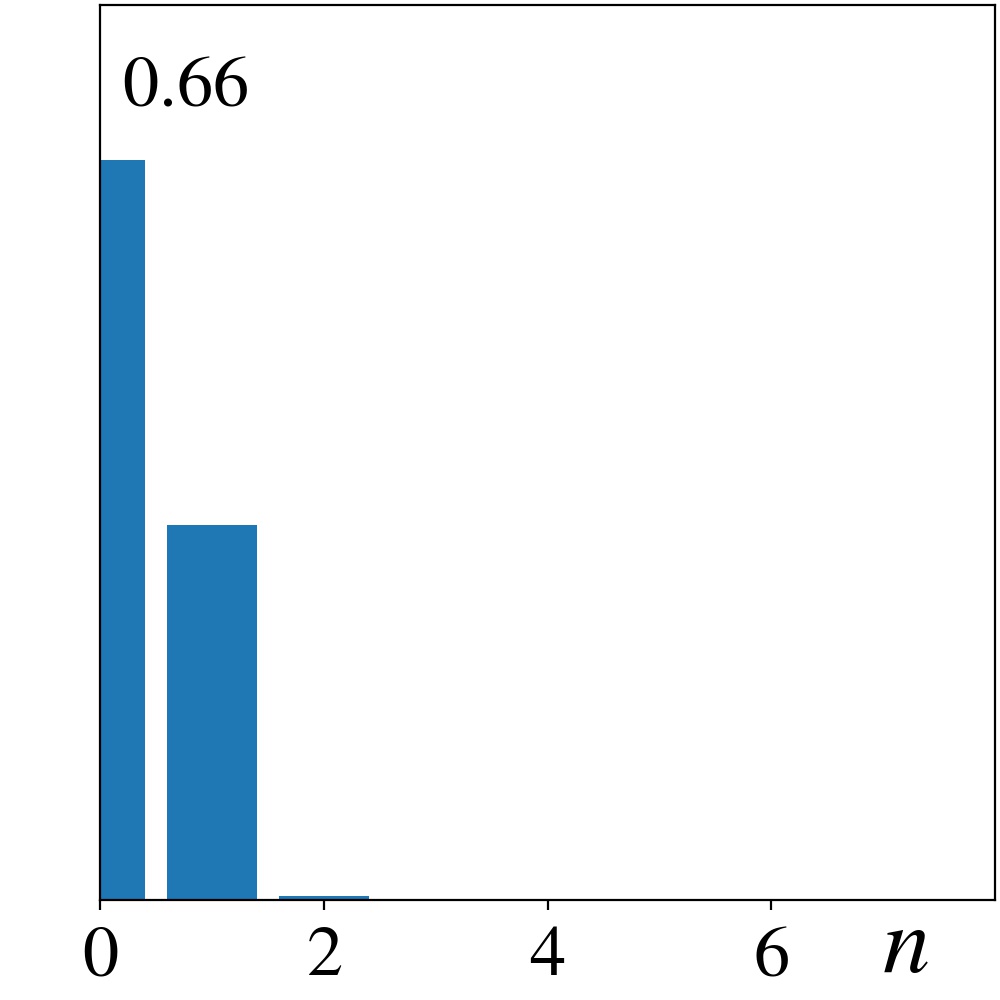}
    \caption{$A_c=0.1$}
    \label{}
    \end{subfigure}
    \caption{Same as in Fig.~\ref{fig:wig_fock1}, but for values of $\gamma_2$ approaching the quantum regime.  
    Dashed red circles in the top panels all have radius $A_c$, indicating that as one approaches the quantum regime, the amplitude of the limit cycle saturates at a value slightly below 1, measured in units of $x_0$, even as the classical amplitude $A_c$ tends to zero. As expected, very few Fock states are occupied in the quantum regime.}
    \label{fig:wig_fock2}
\end{figure}

Figures \ref{fig:wig_fock1} and \ref{fig:wig_fock2} show the steady-state Wigner functions and Fock-state distributions of the quantum RvdP limit-cycle of Eq.~\eqref{Eq:scaledmaster}, for different values of $A_c$.  Figure~\ref{fig:wig_fock1} shows that when coming from the classical regime, by reducing $A_c$ from $8$ down to $2$, the radius of the limit cycle is approximately $A_c$, and relatively many Fock states are populated. On the other hand, Figure~\ref{fig:wig_fock2} shows that when entering the quantum regime, as $A_c$ is lowered further from $1$ down to $0.1$, only a few Fock states are populated, and the radius of the limit cycle does not get much smaller than $x_{\rm zp} = x_0/\sqrt{2}$.

To see this more quantitatively, we follow Steiner~\cite{steiner16} and sum all the even, or alternatively all the odd, rate equations~\eqref{Eq:pn} for the Fock-state probabilities, thereby telescopically eliminating all the two-phonon transitions, and finding that in the steady state
\begin{equation}\label{Eq:sum-even}
    \Gamma_1\left(P_1-2P_2+3P_3\ldots\right)
    = K_1\left(P_0-2P_1+3P_2\ldots\right).
\end{equation}
In the quantum limit of large nonlinear damping $\gamma_2$, or small $A_c$, and low temperature $k_\textrm{B}T\ll\hbar\omega$, both $\Gamma_1$ and $K_1$, defined in Eq.~\eqref{Eq:scaled-rates}, tend to zero with corrections of order $\gamma_2^{-1}$, while $K_2$ tends to zero with corrections of order $\gamma_2^{-1}$ or $\exp{-2\hbar\omega/k_\textrm{B}T}$. An inspection of the first few rate equations~\eqref{Eq:pn} then shows that all $P_n$, with $n>1$, are smaller than $P_0$ and $P_1$ at least by an order of $\gamma_2^{-1}$ or $\exp{-2\hbar\omega/k_\textrm{B}T}$. Neglecting all these higher states, with $n>1$, in Eq.~\eqref{Eq:sum-even} then yields a relation between the occupation probabilities of the remaining two lowest states, given by
\begin{equation}
    P_0 = P_1\left(2 + \frac{\Gamma_1}{K_1}\right)
    \equiv P_1\left(2 + R\right).
\end{equation}
Thus, with $\Tr{\rho}=1$, in the low temperature quantum limit, with $\gamma_2\to\infty$, we find that the density matrix becomes
\begin{equation} \label{Eq:quantum-rho}
    \rho=\frac{2+R}{3+R}\ketbra{0} + \frac{1}{3+R}\ketbra{1}
    +\order{\gamma_2^{-1},e^{-2\hbar\omega/k_\textrm{B}T}},
\end{equation}
where the temperature-dependent ratio
\begin{equation}\label{Eq:Gamma-Kappa-Ratio}
    R=\frac{\Gamma_1}{K_1} = \frac{\bar{n}(\Delta_1)+\left[1+\bar{n}(\omega)\right]r}
    {\left[1+\bar{n}(\Delta_1)\right]+\bar{n}(\omega)r},
\end{equation}
tends to the bare ratio $r\equiv\gamma_1/\kappa_1$ of the linear damping rate to the pumping rate, when $T\to 0$. Also note that $R$ approaches 1 as $T$ increases, it is equal to 1 if and only if $r=1$ at arbitrary $T$, and it approaches $\exp{-\hbar\Delta_1/k_\textrm{B}T}$ for fixed $T$ as $r$ tends to zero.

\begin{figure}
\hfill
    \begin{subfigure}[b]{0.32\linewidth}
    \includegraphics[width=1\linewidth]{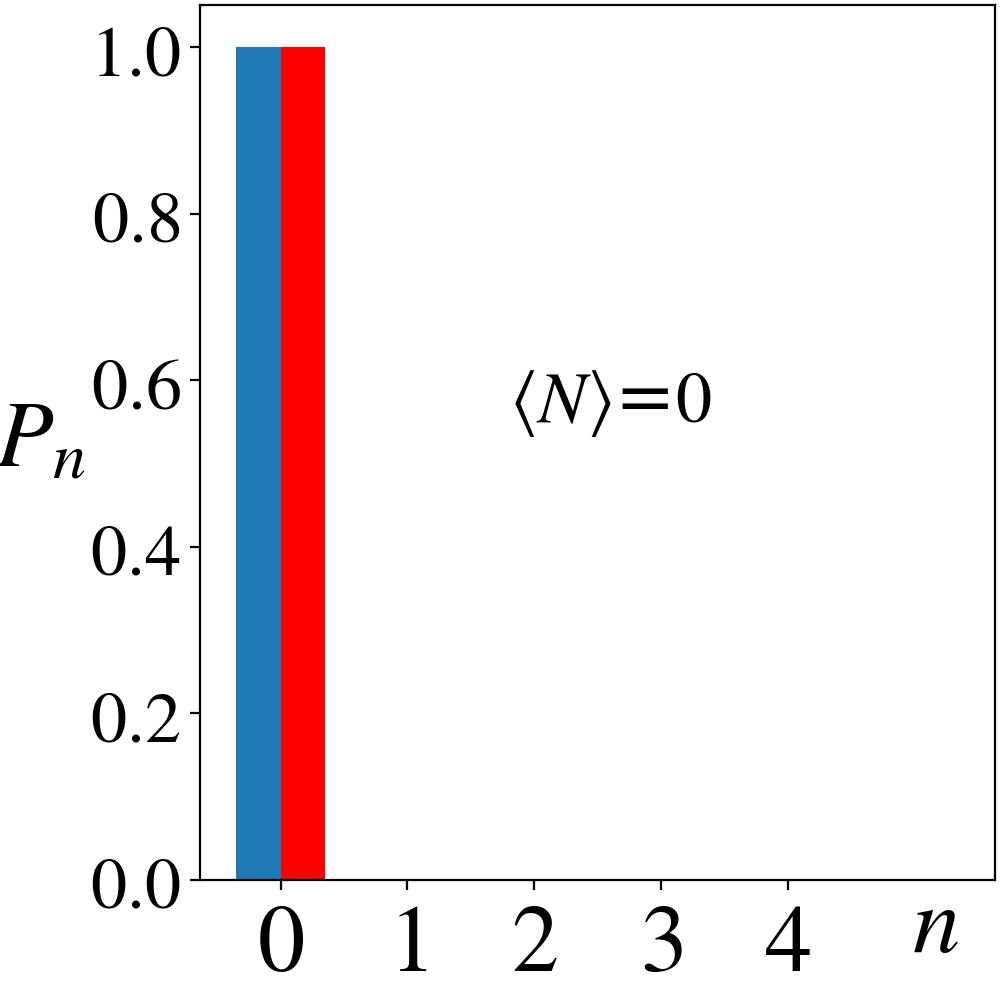}
    \caption{$\kappa_1=0$}
    \label{}
    \end{subfigure}
    \hfill
    \begin{subfigure}[b]{0.32\linewidth}
    \includegraphics[width=1\linewidth]{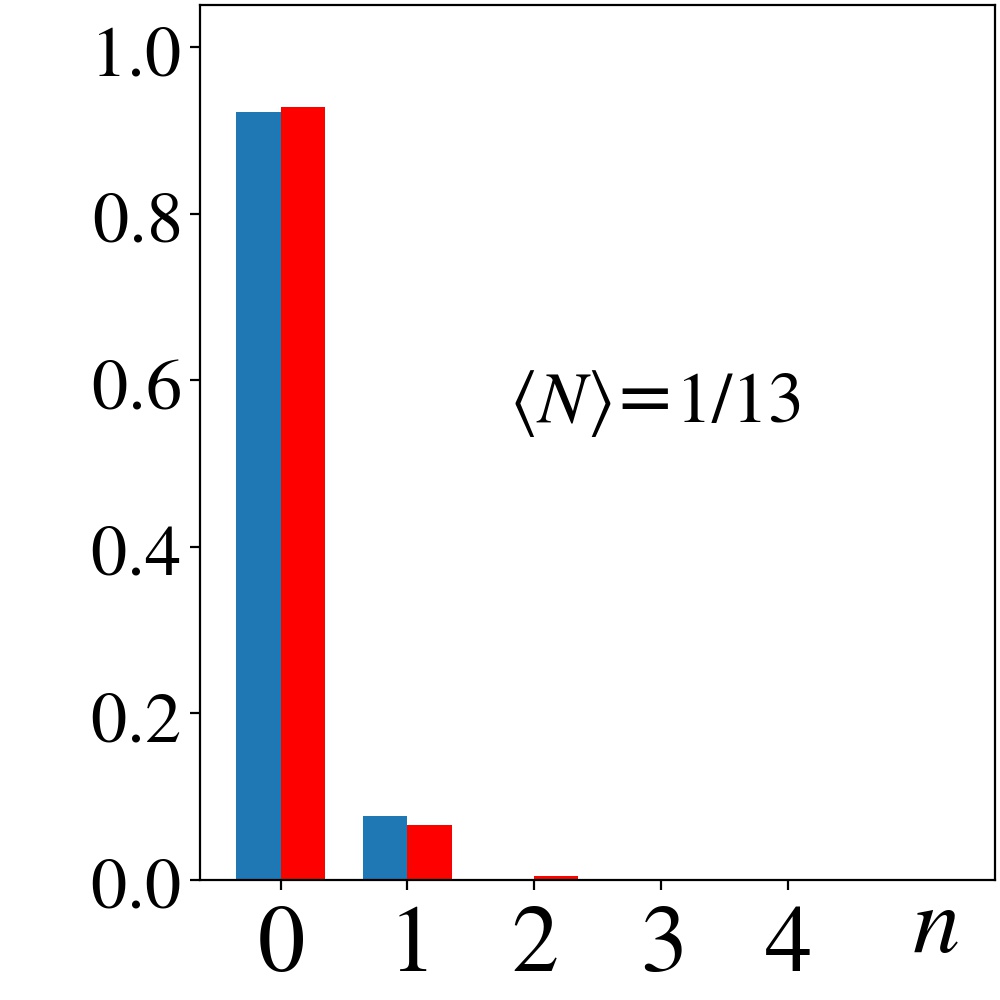}
    \caption{$\kappa_1=0.1$}
    \label{}
    \end{subfigure}
    \hfill
    \begin{subfigure}[b]{0.32\linewidth}
    \includegraphics[width=1\linewidth]{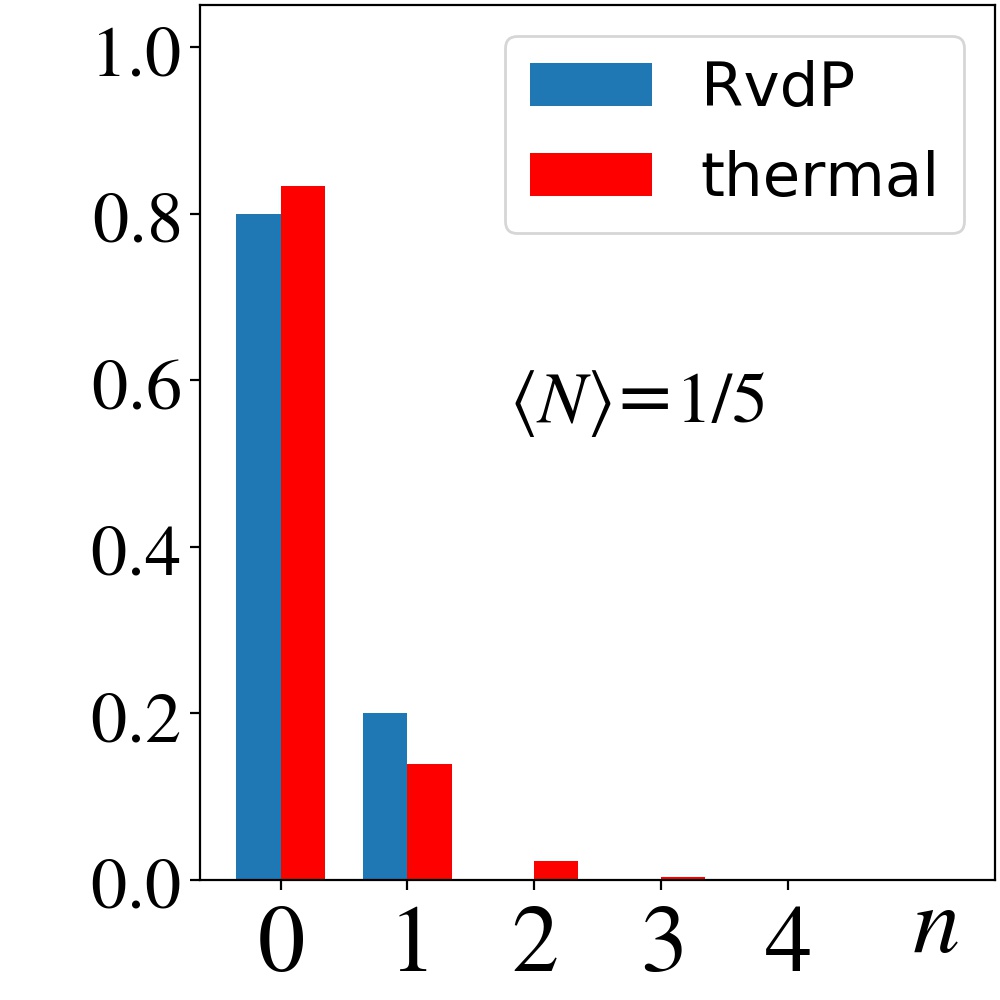}
    \caption{$\kappa_1=0.5$}
    \label{}
    \end{subfigure}
    \hfill
    \begin{subfigure}[b]{0.32\linewidth}
    \includegraphics[width=1\linewidth]{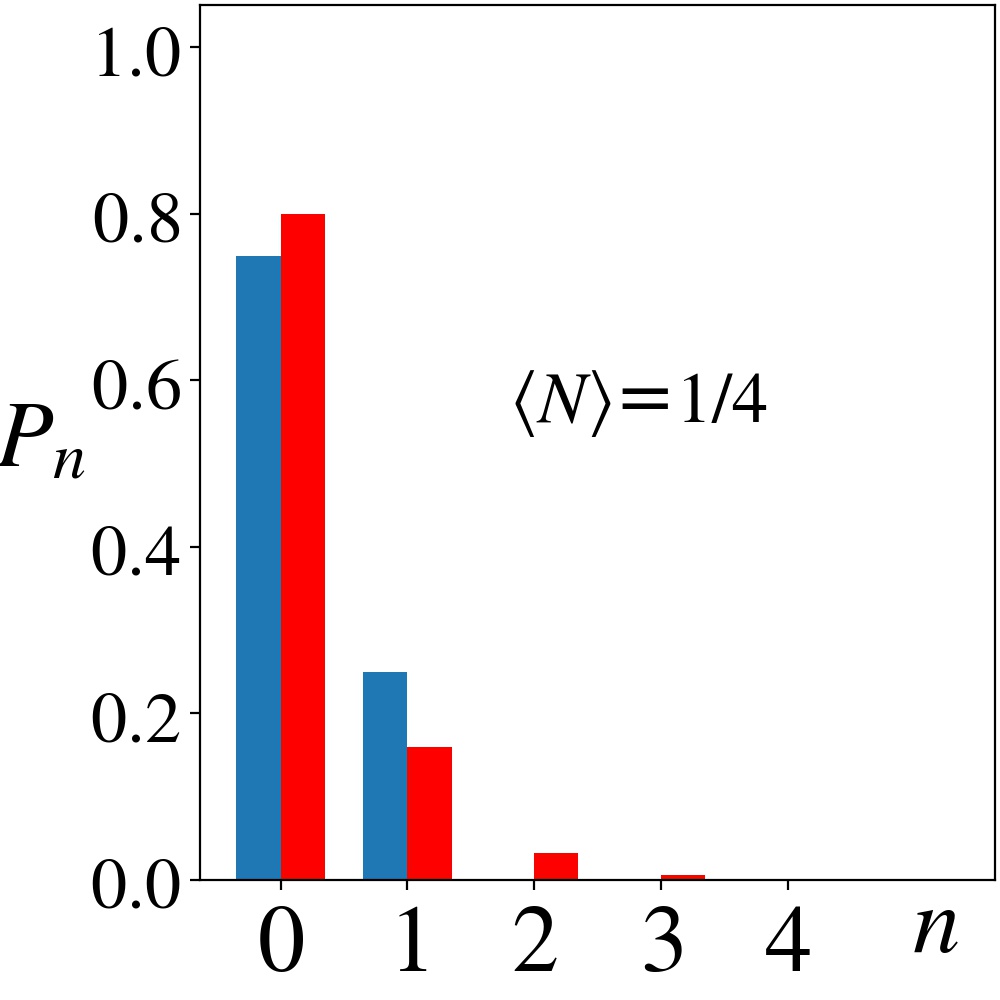}
    \caption{$\kappa_1=1$}
    \label{}
    \end{subfigure}
    \hfill
    \begin{subfigure}[b]{0.32\linewidth}
    \includegraphics[width=1\linewidth]{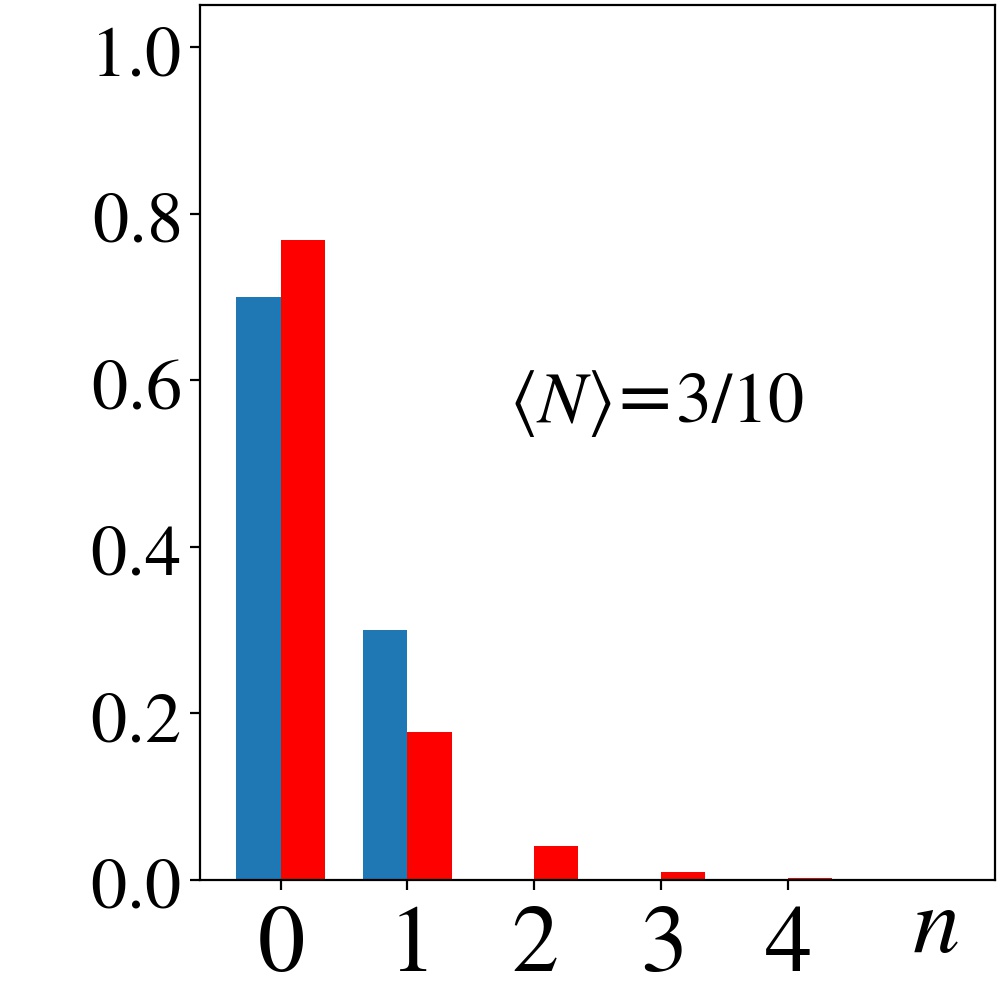}
    \caption{$\kappa_1=3$}
    \label{}
    \end{subfigure}
    \hfill
    \begin{subfigure}[b]{0.32\linewidth}
    \includegraphics[width=1\linewidth]{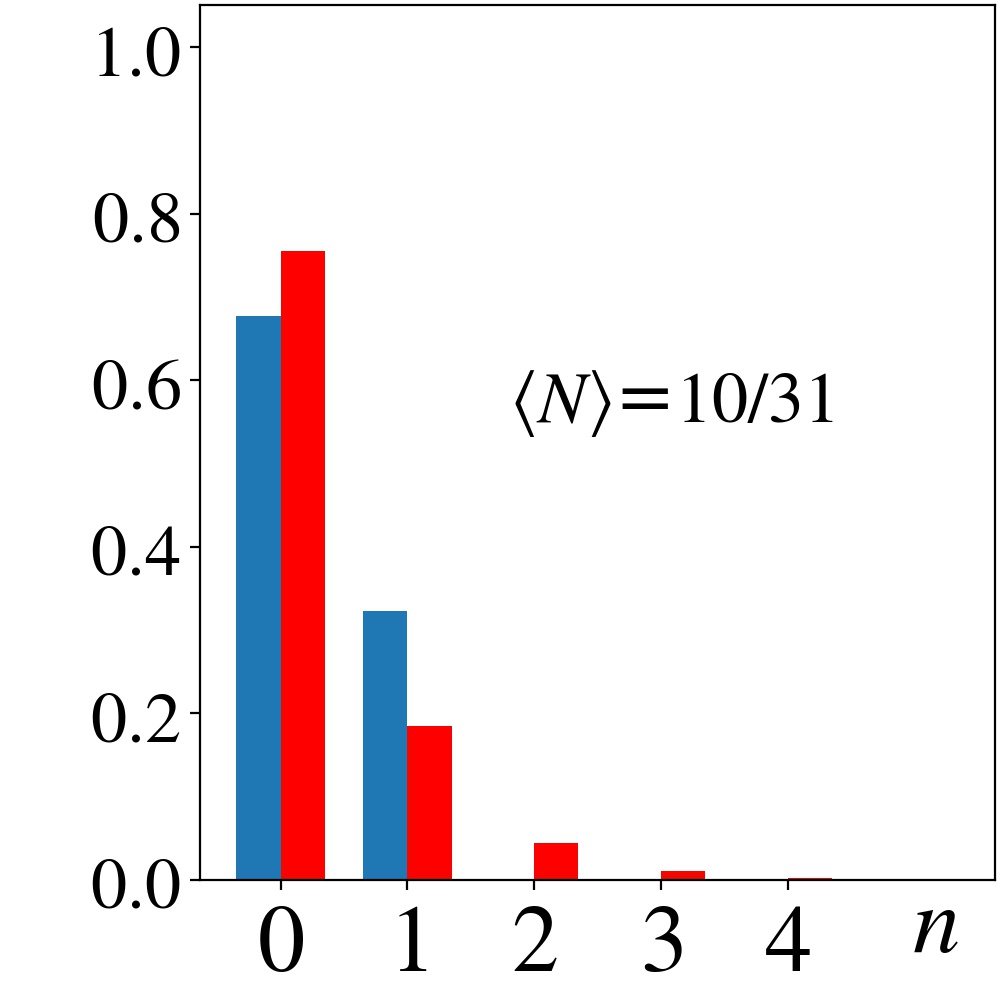}
    \caption{$\kappa_1=10$}
    \label{}
    \end{subfigure}
    \caption{Numerically obtained zero-temperature steady-state Fock distributions of the RvdP oscillator in the quantum limit, with $\gamma_1=1$ and $\gamma_2=10^5$, compared with thermal distributions with the same average phonon number $\expval{N}$, for different values of $\kappa_1$. The values of $\expval{N}$ for $\gamma_2\to\infty$ are specified inside each panel. The numerical values of $\expval{N}$, obtained with $\gamma_2=10^5$, deviate from the predicted values for infinite $\gamma_2$ in Eq.~\eqref{Eq:quantum-N} only to within $\order{10^{-5}}$ as expected. In this limit only the $\ket{0}$ and $\ket{1}$ states are occupied. As $\kappa_1$ increases, $\expval{N} \to 1/3$, and the RvdP distribution deviates further away from the thermal one.}
    \label{fig:fock}
\end{figure}

\begin{figure}
\hfill
    \begin{subfigure}[b]{0.34\linewidth}
    \includegraphics[width=1\linewidth]{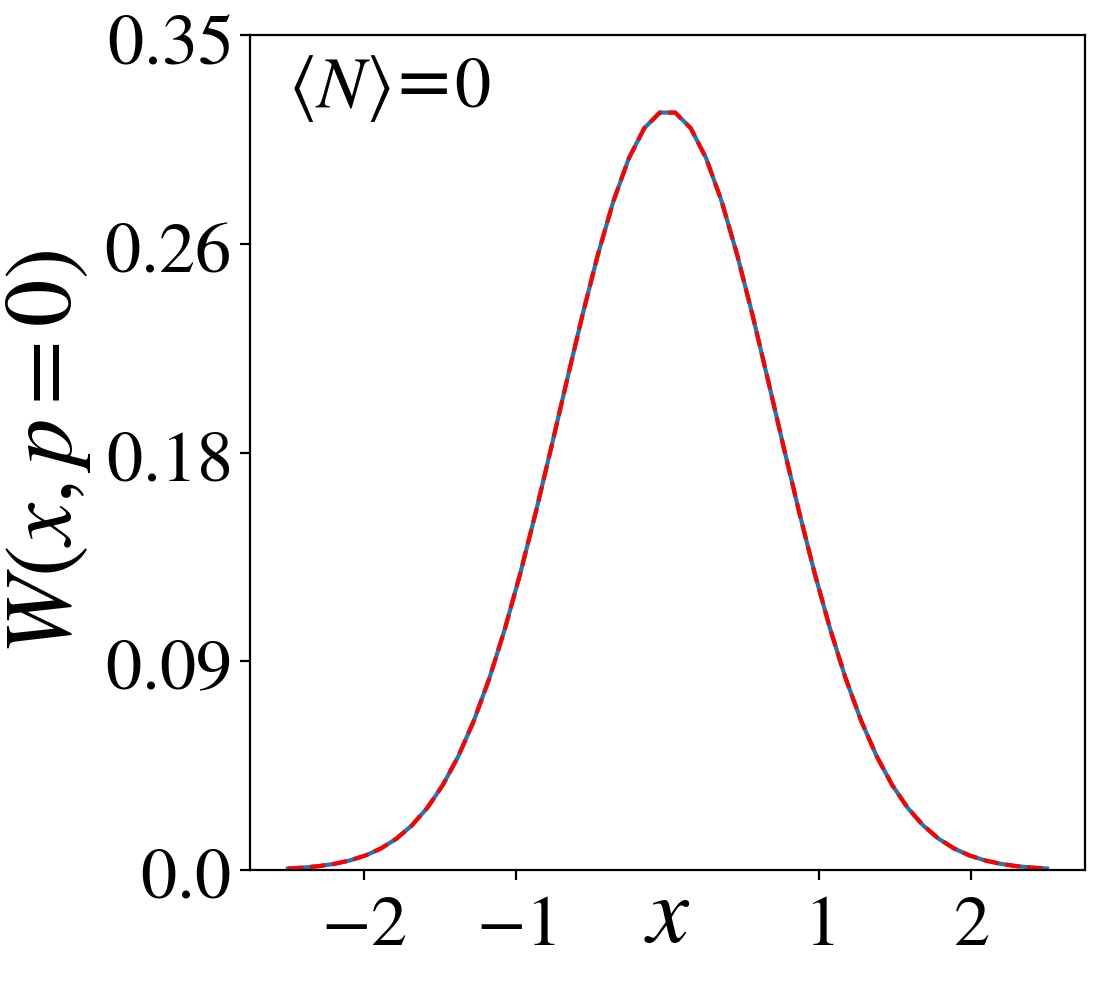}
    \caption{$\kappa_1=0$}
    \label{}
    \end{subfigure}
    \hfill
    \begin{subfigure}[b]{0.31\linewidth}
    \includegraphics[width=1\linewidth]{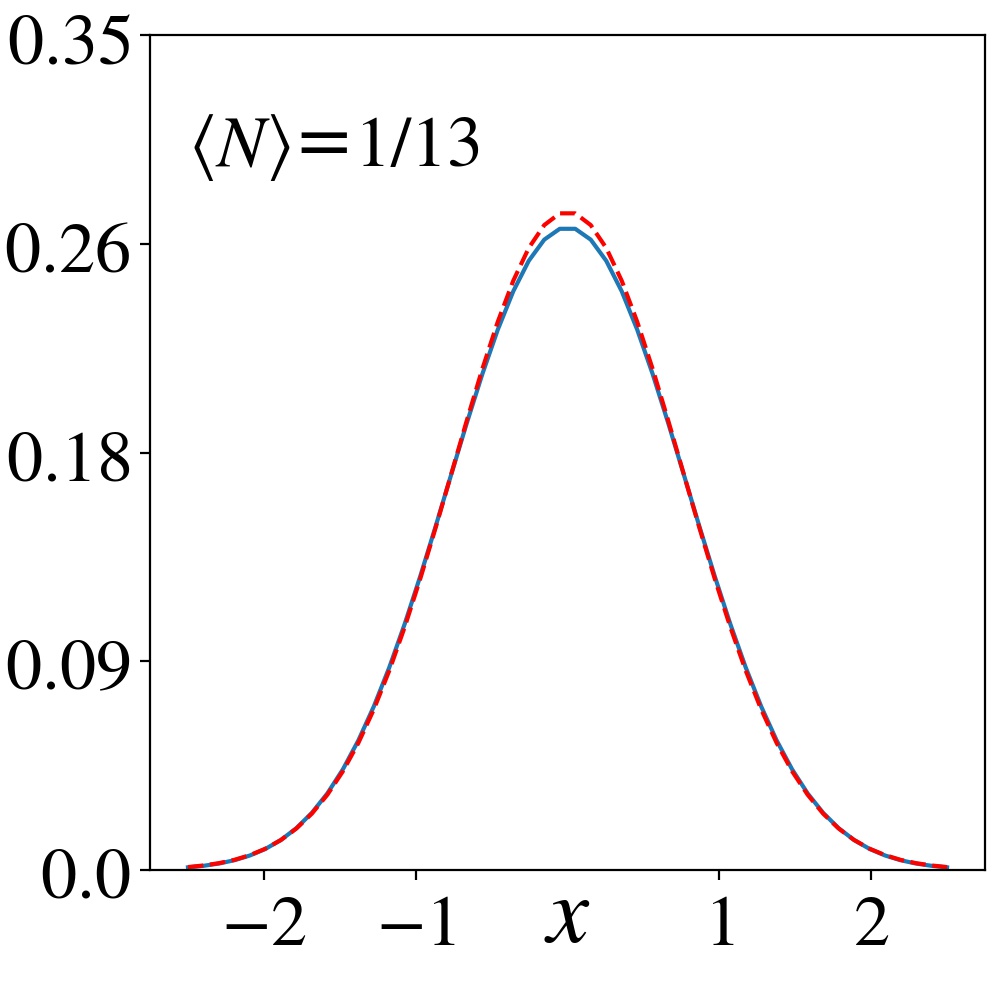}
    \caption{$\kappa_1=0.1$}
    \label{}
    \end{subfigure}
    \hfill
    \begin{subfigure}[b]{0.31\linewidth}
    \includegraphics[width=1\linewidth]{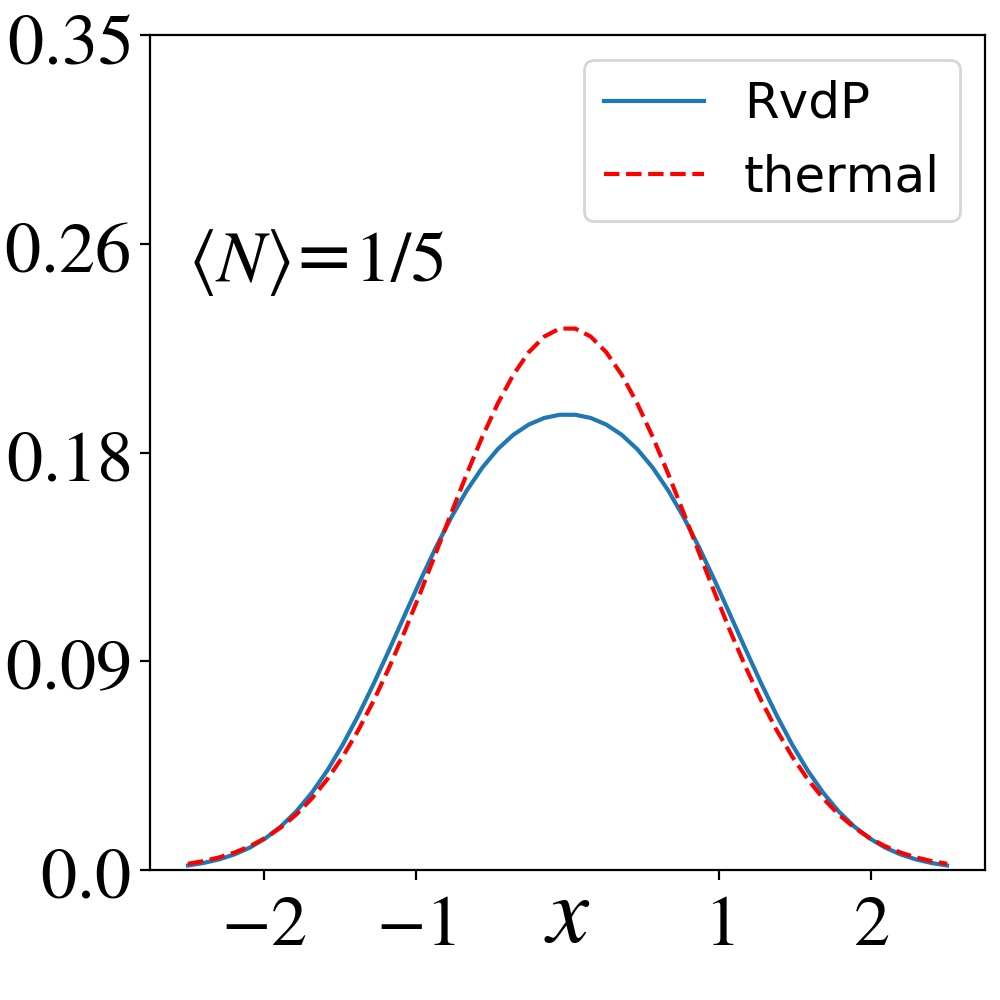}
    \caption{$\kappa_1=0.5$}
    \label{}
    \end{subfigure}
    \hfill
    \begin{subfigure}[b]{0.34\linewidth}
    \includegraphics[width=1\linewidth]{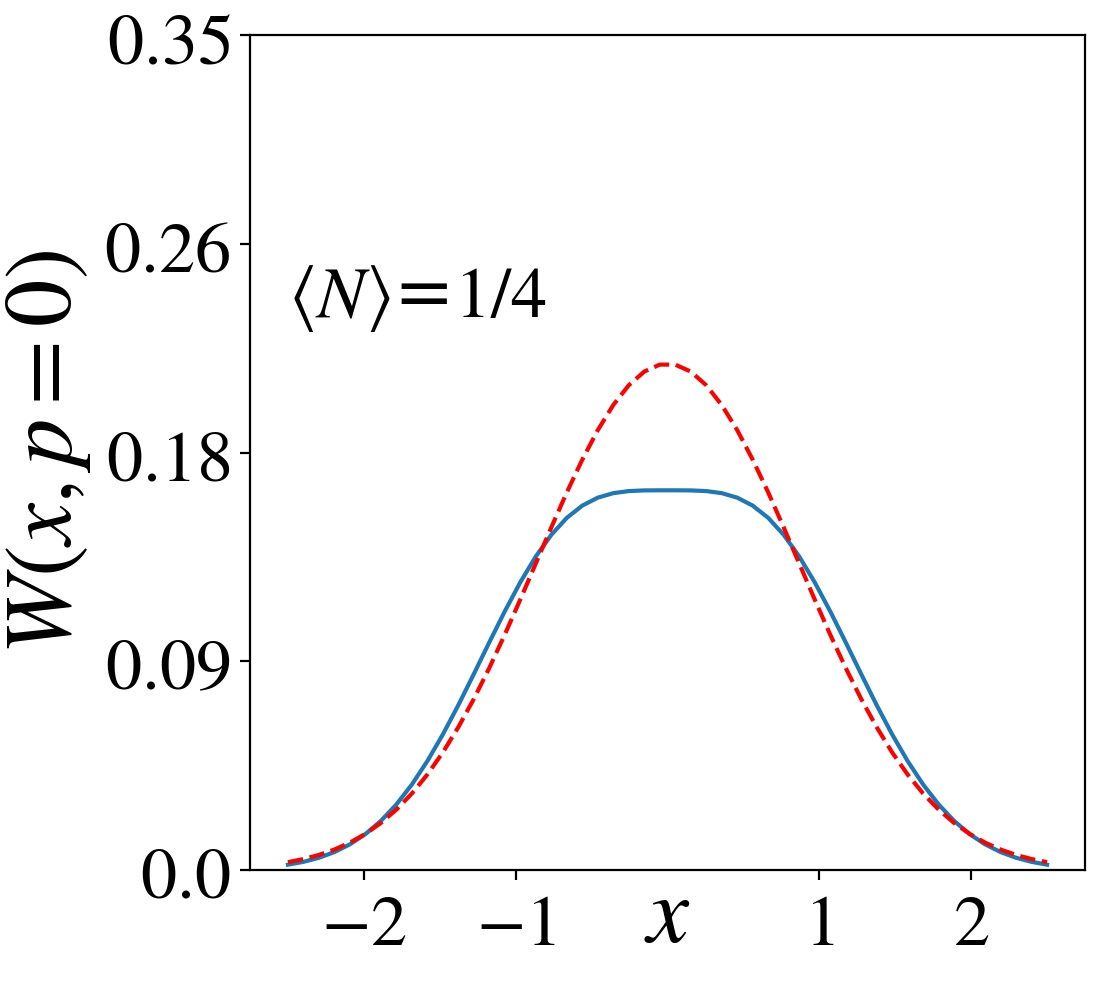}
    \caption{$\kappa_1=1$}
    \label{}
    \end{subfigure}
    \hfill
    \begin{subfigure}[b]{0.31\linewidth}
    \includegraphics[width=1\linewidth]{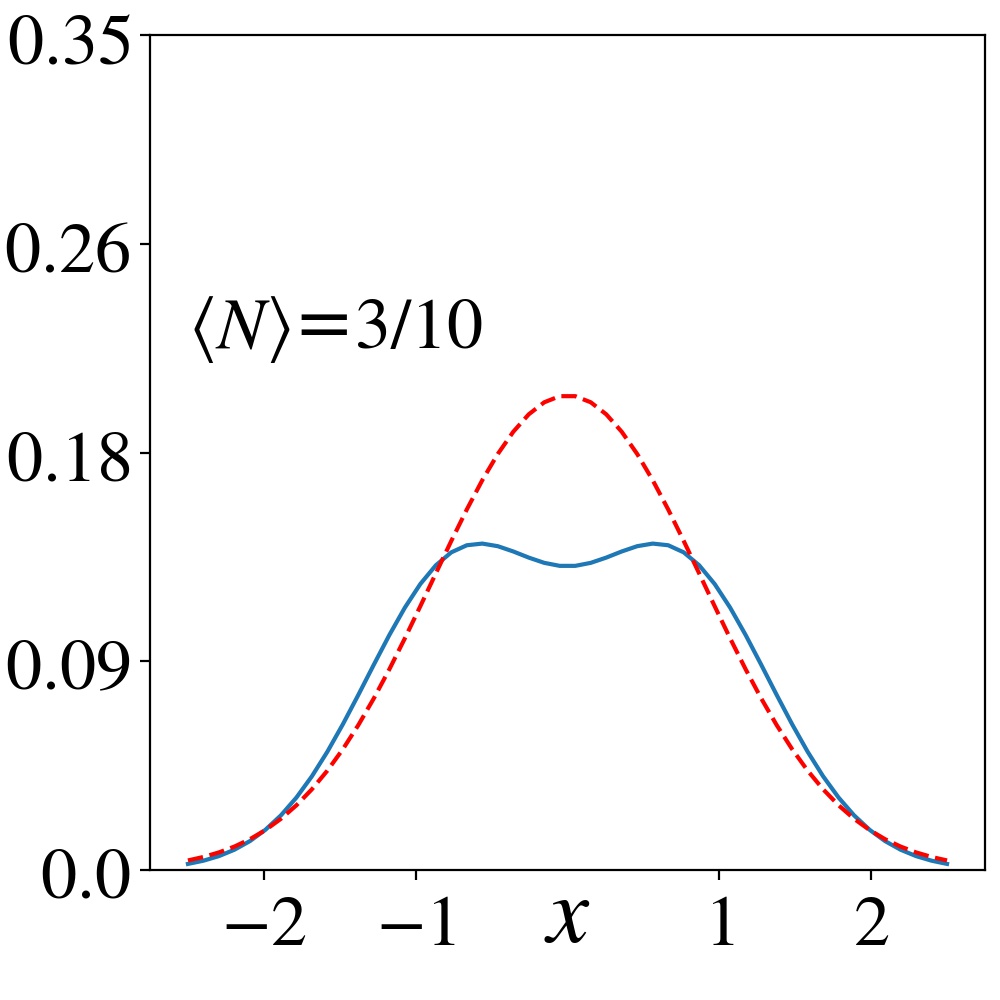}
    \caption{$\kappa_1=3$}
    \label{}
    \end{subfigure}
    \hfill
    \begin{subfigure}[b]{0.31\linewidth}
    \includegraphics[width=1\linewidth]{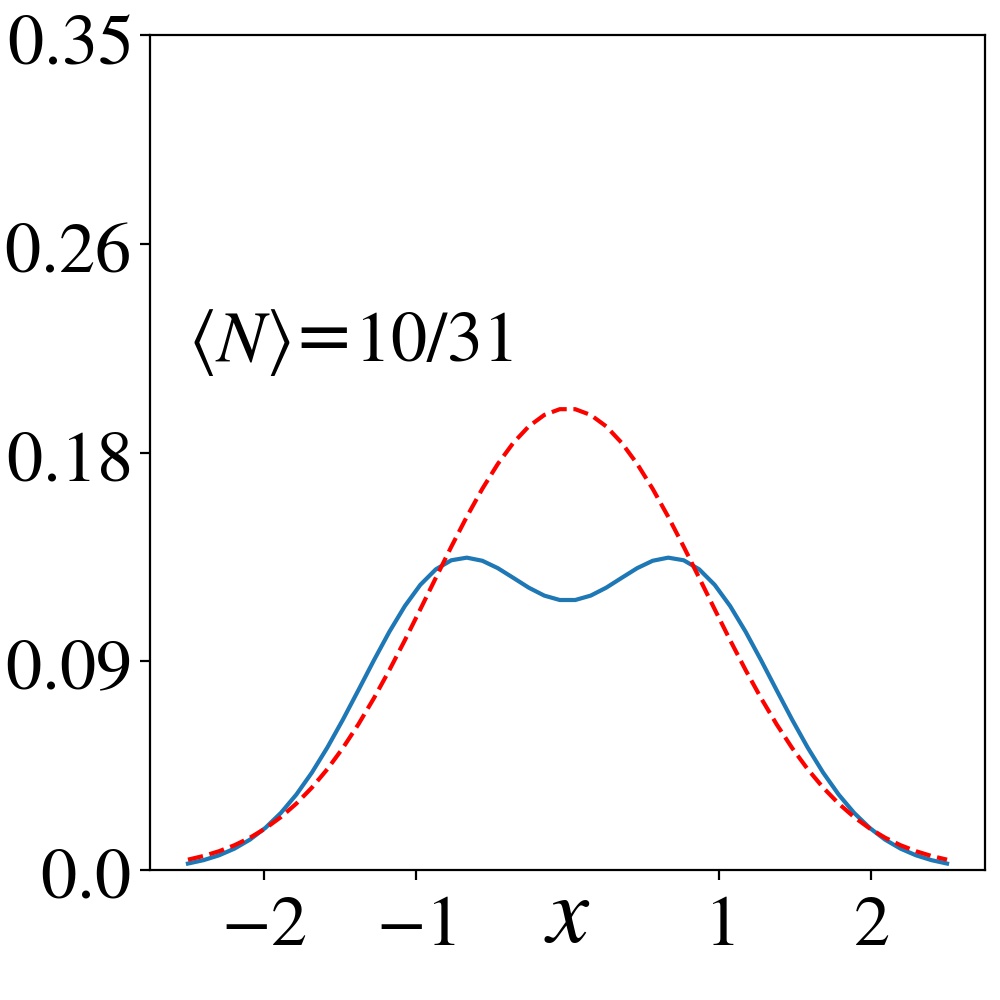}
    \caption{$\kappa_1=10$}
    \label{}
    \end{subfigure}
    \caption{Cross sections through the zero-temperature rotationally symmetric Wigner functions of the RvdP oscillator in the quantum limit (in solid blue), with  $\gamma_1=1$ and $\gamma_2=10^5$, as compared to those of thermal states (in dashed red) with the same average phonon occupation $\expval{N}$. The values of $\expval{N}$ for $\gamma_2\to\infty$ are specified inside each panel. The parameters used here are the same as in Fig.~\ref{fig:fock}. The Wigner functions exhibit a Hopf bifurcation at $\kappa_1=\gamma_1$, reminiscent of a continuous phase transition.}
    \label{fig:wigner}
\end{figure}

As was previously understood, in this limit only the $\ket{0}$ and $\ket{1}$ states are occupied, because all phonons in any other state are immediately annihilated by the infinitely strong nonlinear damping. But, contrary to the zero-temperature result of previous authors~\cite{lee13,lee14,walter14,walter15,davis18,ishibashi17,scarlatella19,amitai18,lorch16}, who take $\gamma_1=0$ above the bifurcation, we find that the actual occupation depends on the ratio $r=\gamma_1/\kappa_1$, and is not universal.
This is demonstrated numerically in Fig.~\ref{fig:fock}, where we compare the steady-state zero-temperature Fock-state distributions of the RvdP oscillator with thermal distributions for the same average phonon occupation, which according to Eq.~\eqref{Eq:quantum-rho} is given by
\begin{equation} \label{Eq:quantum-N}
\expval{N} 
\xrightarrow[\gamma_2 \to \infty]{} P_1 = \frac{1}{3+{\Gamma_1/K_1}} 
\xrightarrow[\ T \to 0\ ]{} \frac{1}{3+{\gamma_1/\kappa_1}}.
\end{equation}
Cross sections through the corresponding Wigner functions for the same parameter values are shown in Fig.~\ref{fig:wigner}, where one can observe the onset of the bifurcation at $\expval{N} = 1/4$ as the mean phonon number $\expval{N}$ gradually increases from $0$ to $1/3$. Note the quantitative differences between the Wigner functions that appear even below the bifurcation.

In the case of the quantum RvdP oscillator, we choose to associate the amplitude $A$ of the limit-cycle oscillations with the maxima $|\alpha|_{\max}$ of its circular Wigner function, which we evaluate either numerically or using Eq.~\eqref{Eq:Wigner-Fockstate}, while recalling the factor of $\sqrt{2}$ which arises from the definition of Eq.~\eqref{Eq:a-adag-def}. For the extreme quantum-limit steady-state density matrix of Eq.~\eqref{Eq:quantum-rho} this yields
\begin{equation}\label{Eq:Wigner-q}
    W_q(\alpha,\alpha^*) = \frac{1}{ \pi}\frac{1}{3+\Gamma_1/K_1} \left(4|\alpha|^2+1+ \frac{\Gamma_1}{K_1}\right)e^{-2 |\alpha|^2},
\end{equation}
whose maximum determines the limit-cycle amplitude
\begin{equation}\label{Eq:W_peak}
    A^2=2|\alpha|_{\max}^2 
    = \frac{1}{2}\left(1-\frac{\Gamma_1}{K_1}\right)
    \xrightarrow[\ T \to 0\ ]{} \frac{1}{2}\left(1-\frac{\gamma_1}{\kappa_1}\right),
\end{equation}
where in the zero-temperature limit, the ratio $R=\Gamma_1/K_1$ appearing in Eqs.~\eqref{Eq:Wigner-q} and \eqref{Eq:W_peak} is replaced by $r=\gamma_1/\kappa_1$. Note that the bifurcation occurs at $R=1$, which according to Eq.~\eqref{Eq:Gamma-Kappa-Ratio} happens if and only if $r=1$ regardless of the temperature. In the case of finite $\gamma_2$, we expect these expressions to have corrections of $\order{\gamma_2^{-1}}$, as higher Fock states become populated. 

\begin{figure}
    \begin{subfigure}[b]{0.48\linewidth}
    \includegraphics[width=1\linewidth]{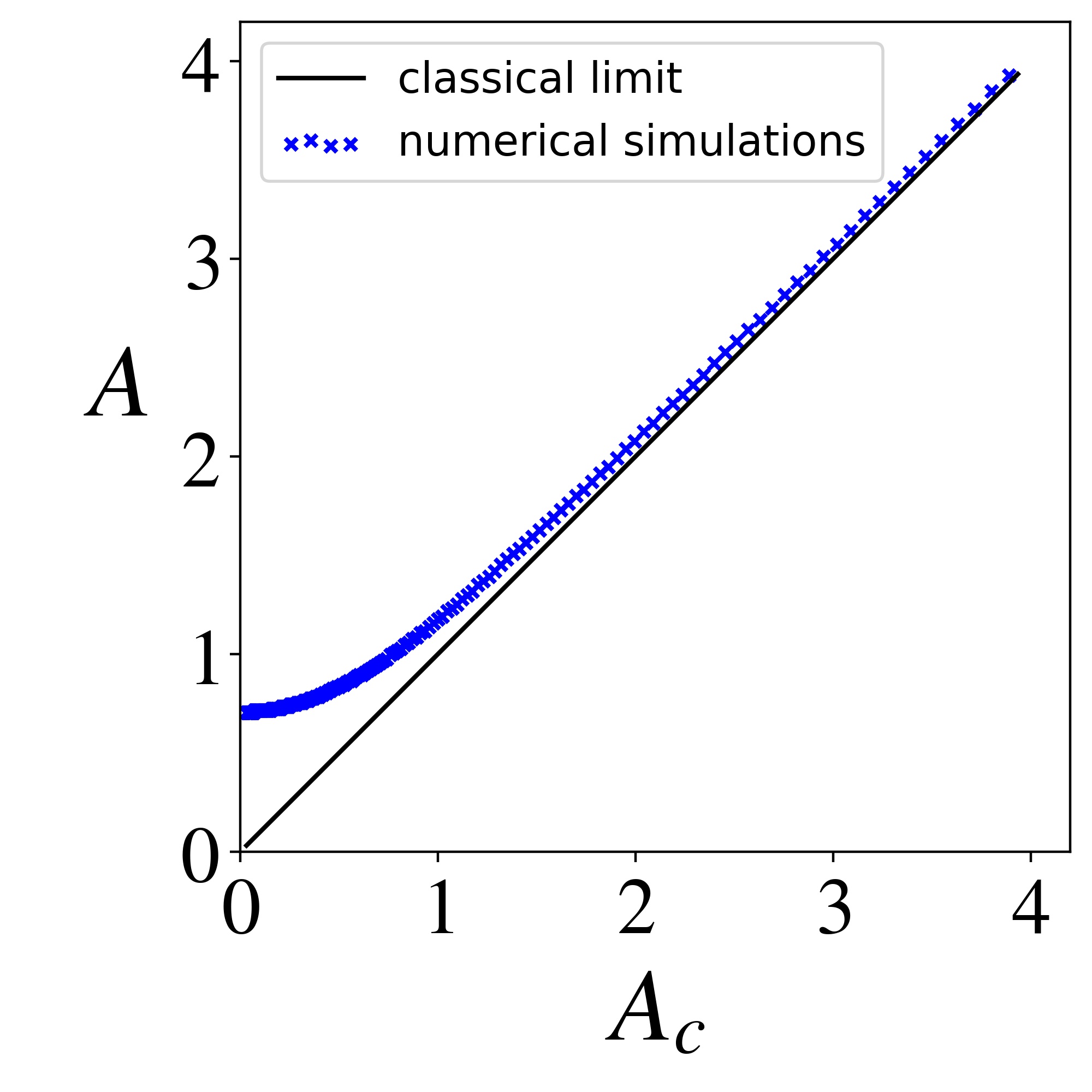}
    \caption{$r=0$}
    \label{}
    \end{subfigure}
    \hfill
    \begin{subfigure}[b]{0.48\linewidth}
    \includegraphics[width=1\linewidth]{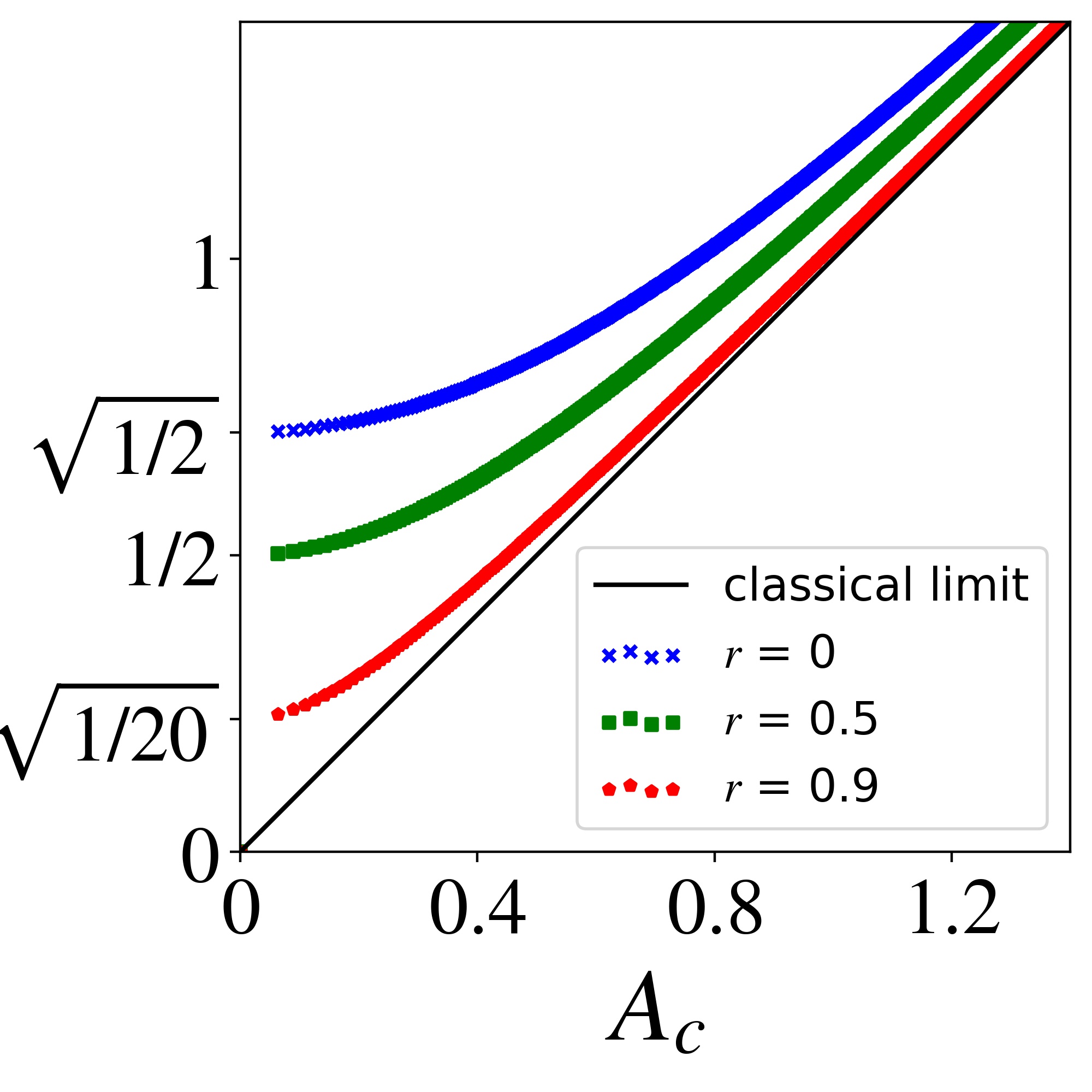}
    \caption{$r=0,0.5$, and $0.9$}
    \label{}
    \end{subfigure}
    \caption{Zero-temperature amplitude $A$ of the quantum RvdP limit cycle, calculated numerically with $\epsilon=1$, as $\gamma_2\to\infty$, plotted as a function of $A_c=\sqrt{\epsilon/\gamma_2}$ which tends to 0, and showing the $r$-dependent saturation predicted by Eq.~\eqref{Eq:W_peak}. (a) without linear damping $\gamma_1=0$, $r=0$, and (b) for different values of the ratio $r=\gamma_1/\kappa_1$. }
    \label{fig:q_bif}
\end{figure}

In the zero-temperature quantum limit, the Wigner functions still exhibit a clear bifurcation to self-oscillations with an amplitude that grows continuously from zero, as the $\kappa_1=\gamma_1$ threshold is crossed. Nevertheless, the nature of this bifurcation is quite different from the classical Hopf bifurcation. In the classical regime, one expects the amplitude of steady-state oscillations to scale as the square root of the reduced pumping, $A_c = \sqrt{\epsilon/\gamma_2}$, where $\epsilon=\kappa_1-\gamma_1$, and therefore for the oscillations to die out for infinite nonlinear damping (unless the pumping rate $\kappa_1$ is infinite as well). This is shown by a straight black line in Fig.~\ref{fig:q_bif}. However, in the quantum regime, the $\ket{1}$ state is protected from nonlinear damping, which enables the oscillator to undergo a bifurcation into self-oscillations, at an amplitude given by Eq.~\eqref{Eq:W_peak}, even when the nonlinear damping is infinitely strong. The linear pumping rate $\kappa_1$ need only be large compared to the linear damping rate $\gamma_1$. This is purely a quantum effect. Accordingly, as we noticed earlier in Fig.~\ref{fig:wig_fock2}, as $\gamma_2$ tends to infinity rather than decaying to zero as $\sqrt{\epsilon/\gamma_2}$,  the zero-temperature steady-state amplitude saturates at $\sqrt{(1-r)/2}=\sqrt{\epsilon/2\kappa_1}$. This is demonstrated numerically by the colored curves in Fig.~\ref{fig:q_bif} for a few values of the ratio $r$.

\begin{figure}
    \begin{subfigure}[b]{0.48\linewidth}
    \includegraphics[width=1\linewidth]{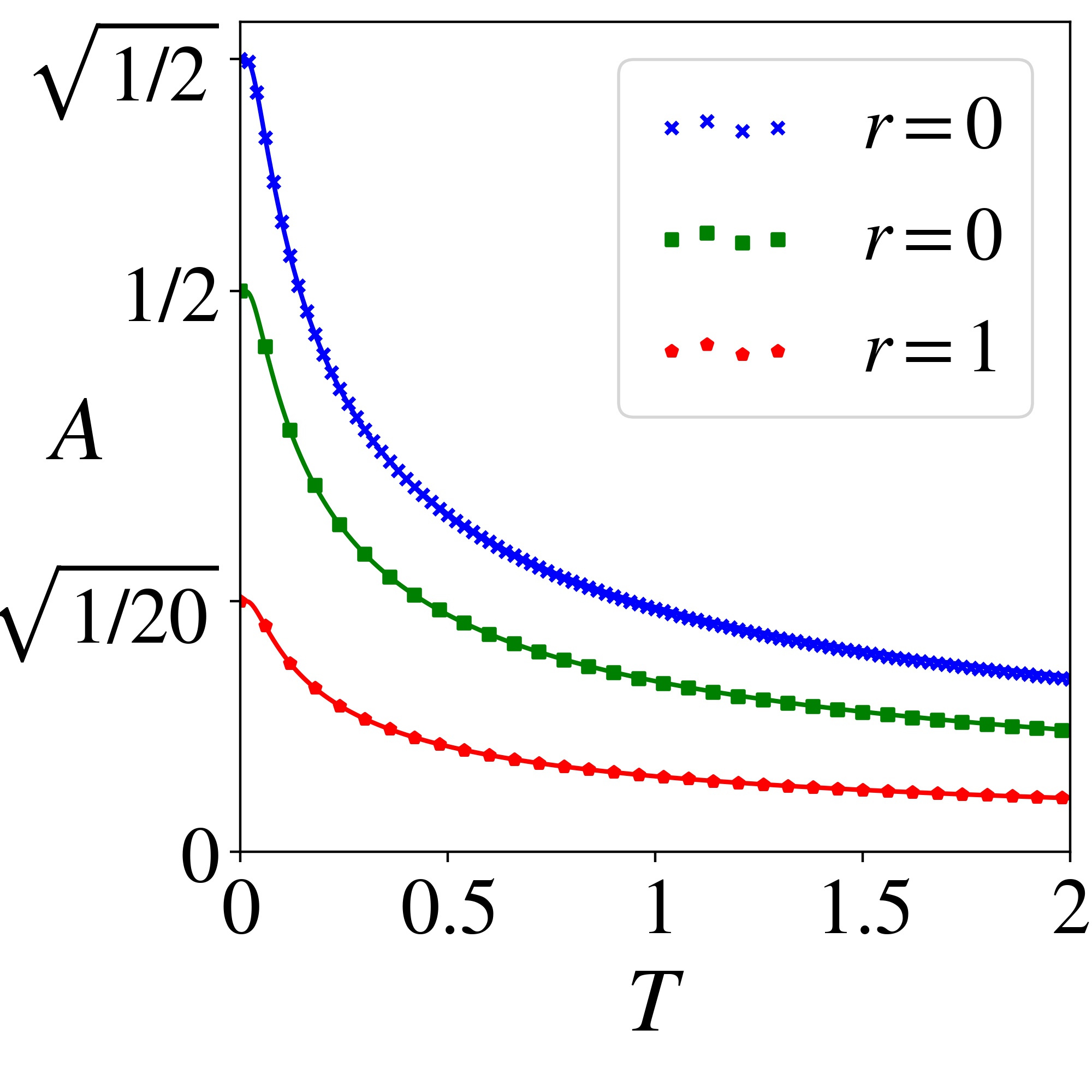}
    \caption{$\Delta_1 = 0.1$}
    \label{}
    \end{subfigure}
    \hfill
    \begin{subfigure}[b]{0.48\linewidth}
    \includegraphics[width=1\linewidth]{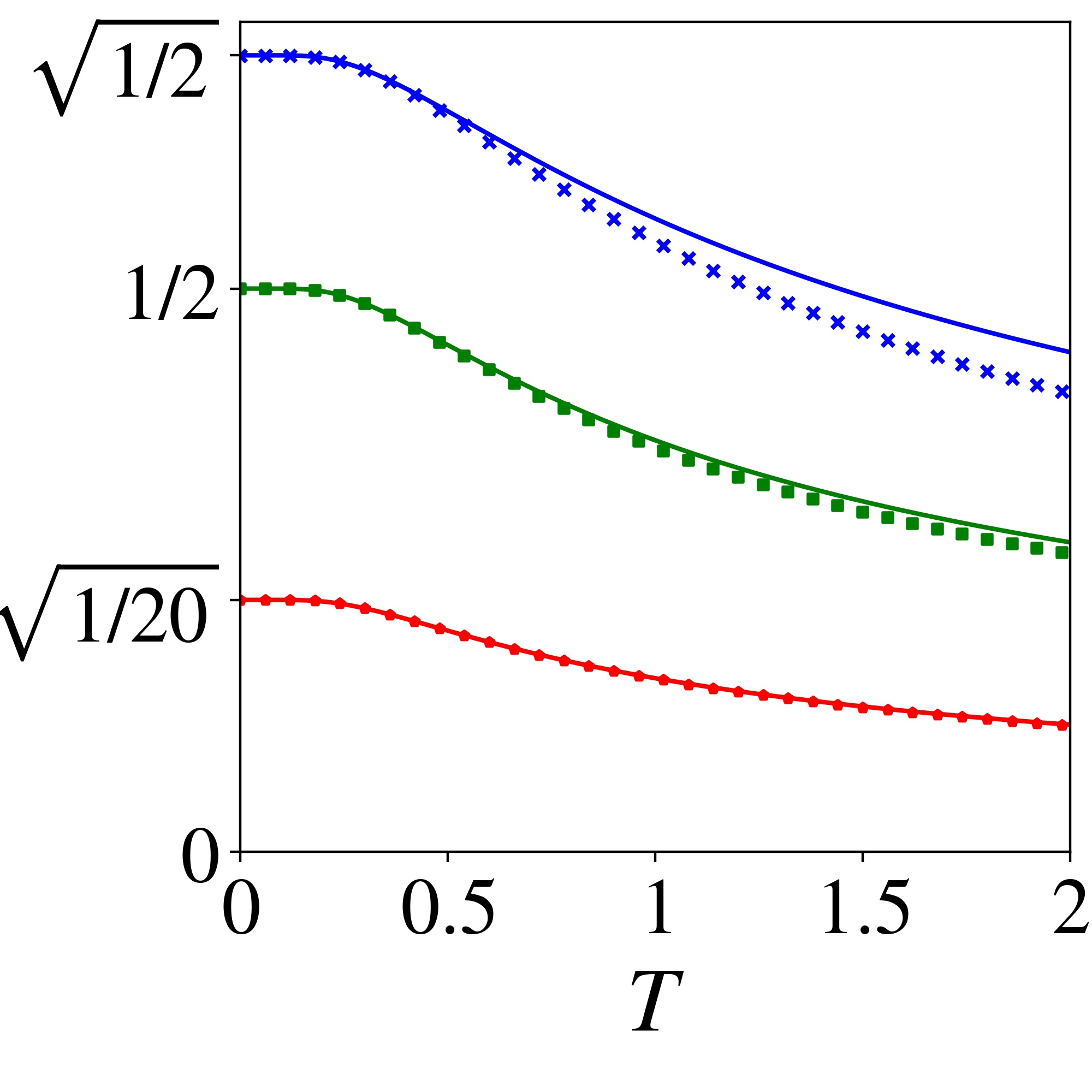}
    \caption{$\Delta_1 = 1$}
    \label{}
    \end{subfigure}
    \caption{Amplitude $A$ of the RvdP limit cycle as a function of temperature, for different values of $r$ in the quantum limit, with $\kappa_1=1$ and $\gamma_2=10^5$, and for pump detunings of (a) $\Delta_1=0.1$, and (b) $\Delta_1=1$. Numerical values (scattered points), obtained by solving the steady-state master equation~\eqref{Eq:scaledmaster}, are compared with the approximate expression of Eq.~\eqref{Eq:W_peak} (solid lines), showing good agreement at low temperatures, particularly for small detuning. As the temperature increases, and $R$ approaches 1, the amplitude decreases to zero.}
    \label{fig:approx_and_num_amps}
\end{figure}

This quantum effect is somewhat smeared out when temperature is turned on and the amplitude saturates at $\sqrt{(1-R)/2}$, rather than $\sqrt{(1-r)/2}$, decreasing with temperature towards zero, as $R$ increases from $r$ towards 1. This is confirmed numerically in Fig.~\ref{fig:approx_and_num_amps}, showing the oscillation amplitude in the quantum limit decaying to zero as the temperature increases. As expected, the approximate expression of Eq.~\eqref{Eq:W_peak} holds better at low temperatures and for small pump detuning $\Delta_1$.

\begin{figure}
    \begin{subfigure}[b]{0.48\linewidth}
    \includegraphics[width=1\linewidth]{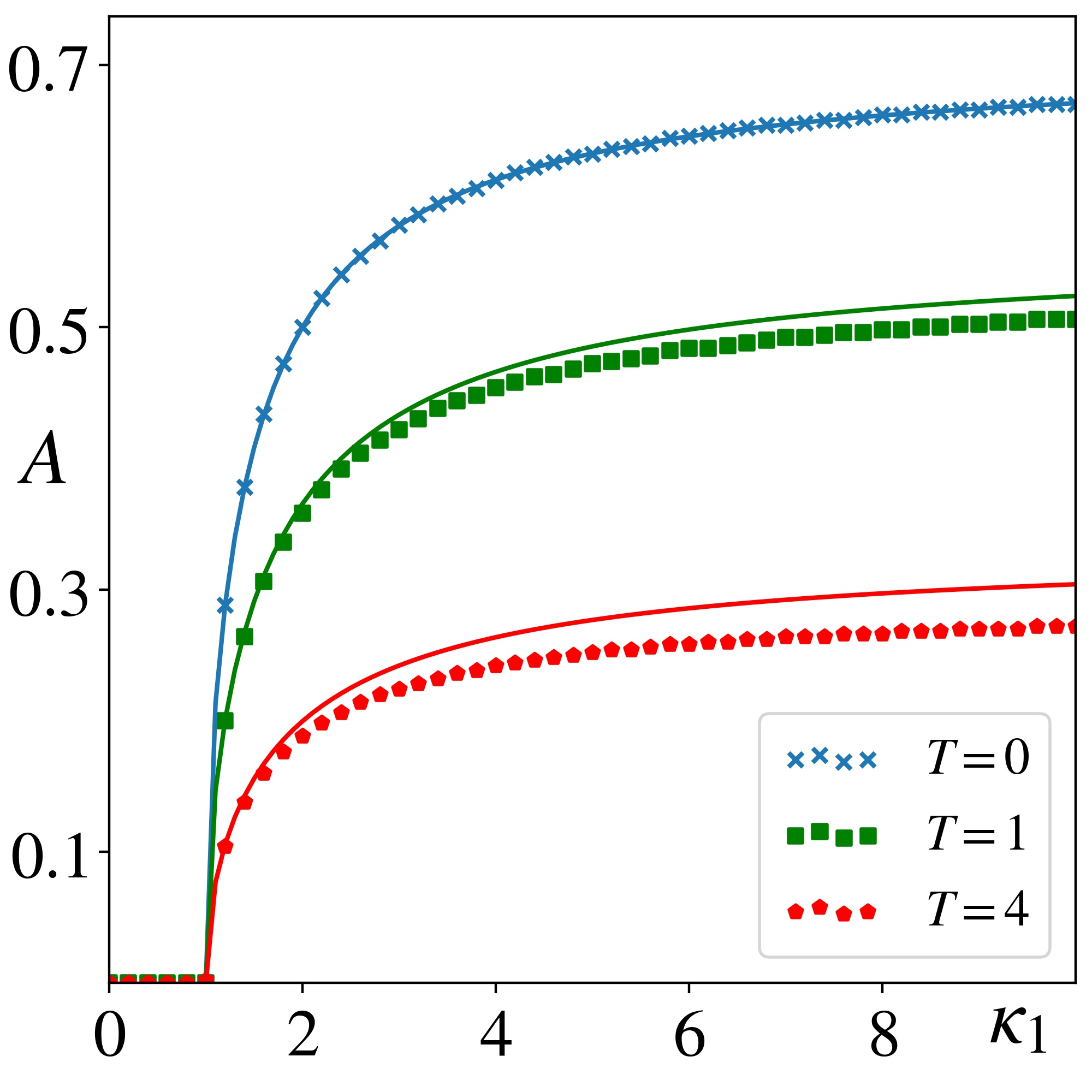}
    \caption{$\gamma_2=10^5$}
    \label{}
    \end{subfigure}
    \hfill
    \begin{subfigure}[b]{0.48\linewidth}
    \includegraphics[width=1\linewidth]{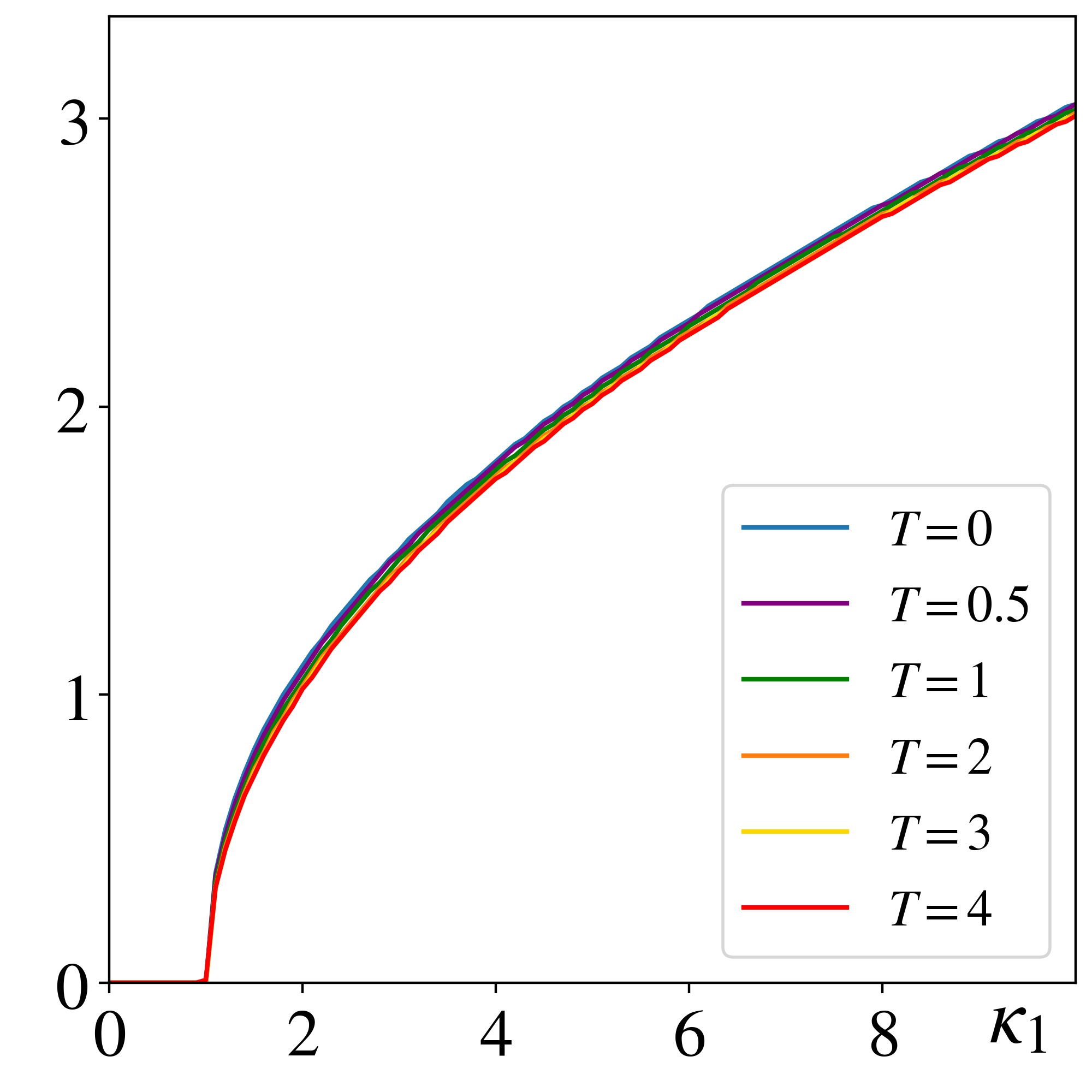}
    \caption{$\gamma_2=1$}
    \label{}
    \end{subfigure}
    \caption{Amplitude $A$ of the RvdP limit cycle with $\gamma_1=1$ as a function of $\kappa_1=1/r$ for different temperatures, in (a) the quantum limit with $\gamma_2=10^5$, and (b) the classical limit with $\gamma_2=1$. The temperature seems to have no effect on the overall shape of the curves in the classical limit, whereas in the quantum limit it causes the amplitude to saturate at lower values as $\kappa_1$ increases or $r$ decreases. Numerical values, obtained by solving the steady-state master equation~\eqref{Eq:scaledmaster}, are compared in panel (a) to solid lines showing the infinite $\gamma_2$, low temperature, approximate solution of Eq.~\eqref{Eq:W_peak}.}
    \label{fig:saturation}
\end{figure}

\begin{figure}
\hfill
    \begin{subfigure}[b]{0.34\linewidth}
    \includegraphics[width=1\linewidth]{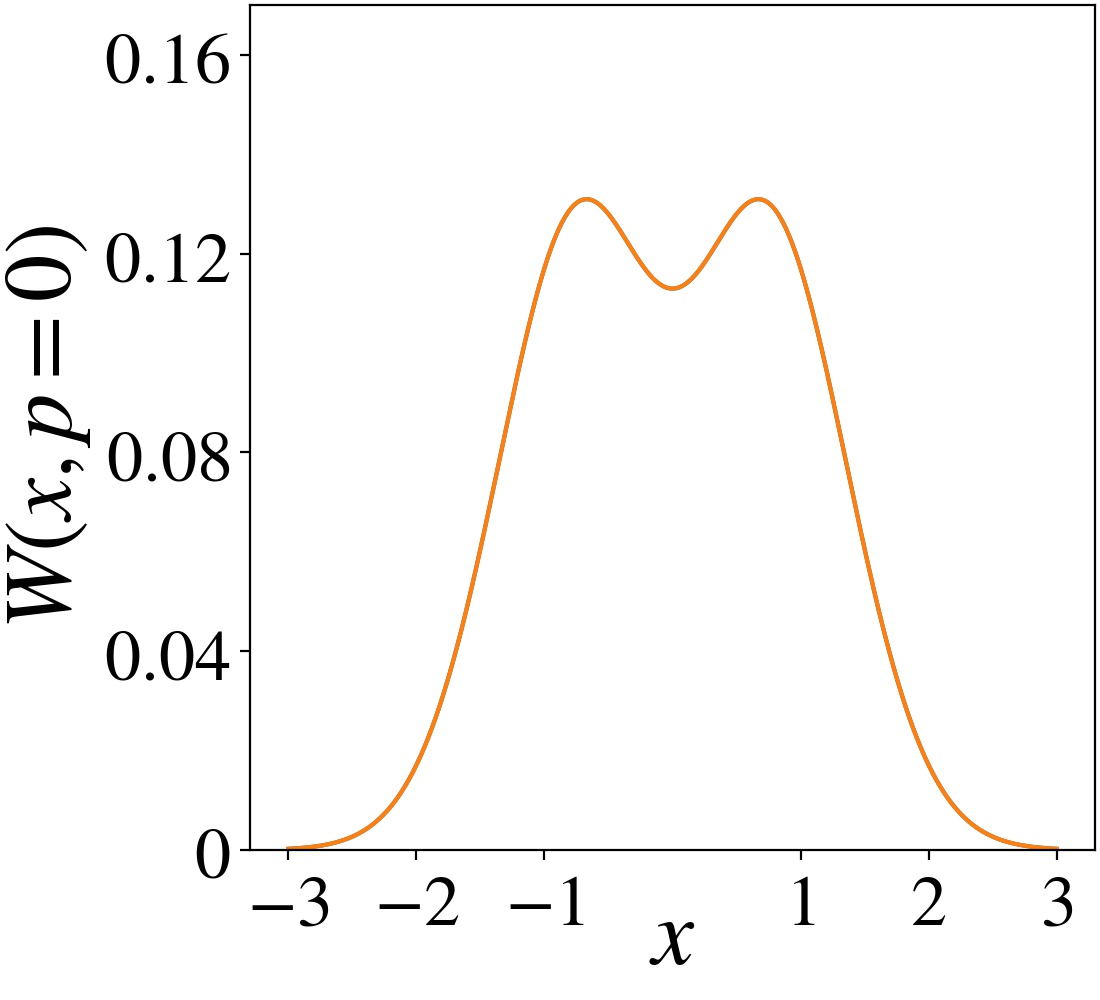}
    \caption{$T=0$}
    \label{}
    \end{subfigure}
    \hfill
    \begin{subfigure}[b]{0.31\linewidth}
    \includegraphics[width=1\linewidth]{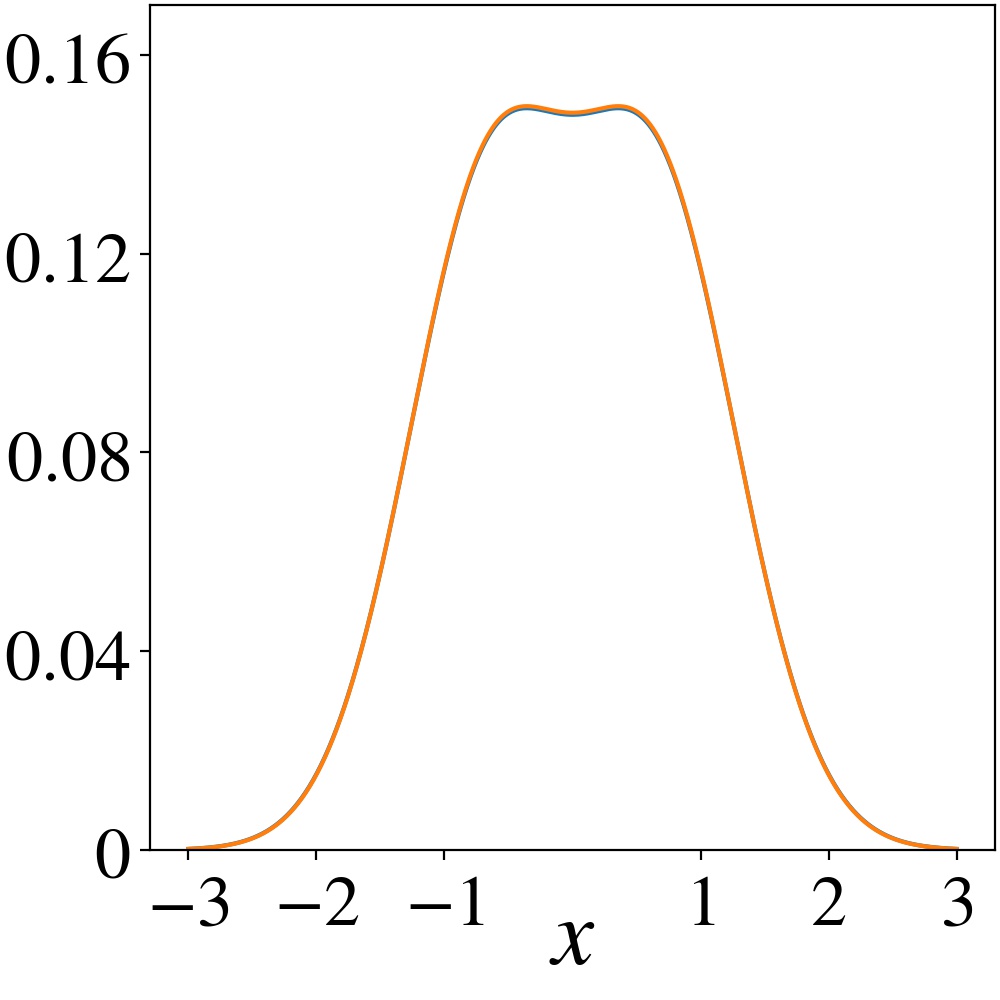}
    \caption{$T=0.3$}
    \label{}
    \end{subfigure}
    \hfill
    \begin{subfigure}[b]{0.31\linewidth}
    \includegraphics[width=1\linewidth]{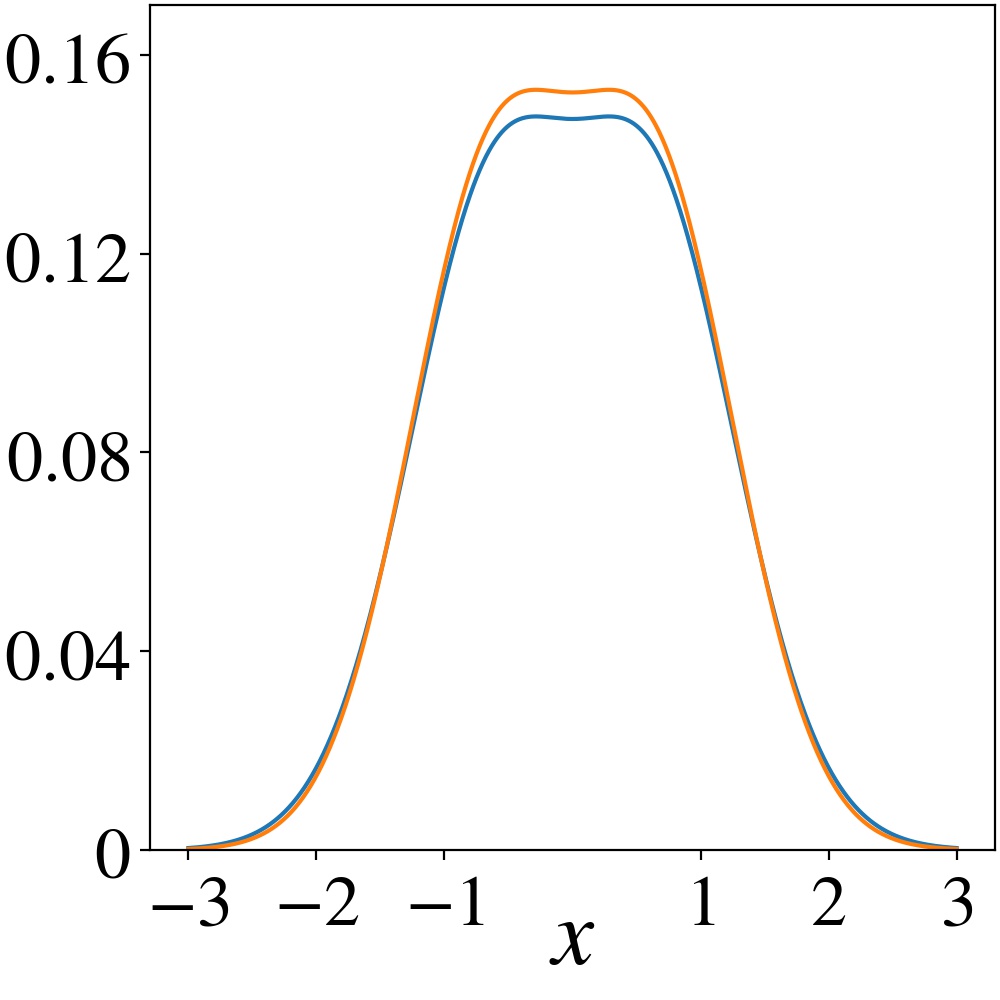}
    \caption{$T=0.5$}
    \label{}
    \end{subfigure}
    \hfill
    \begin{subfigure}[b]{0.31\linewidth}
    \includegraphics[width=1\linewidth]{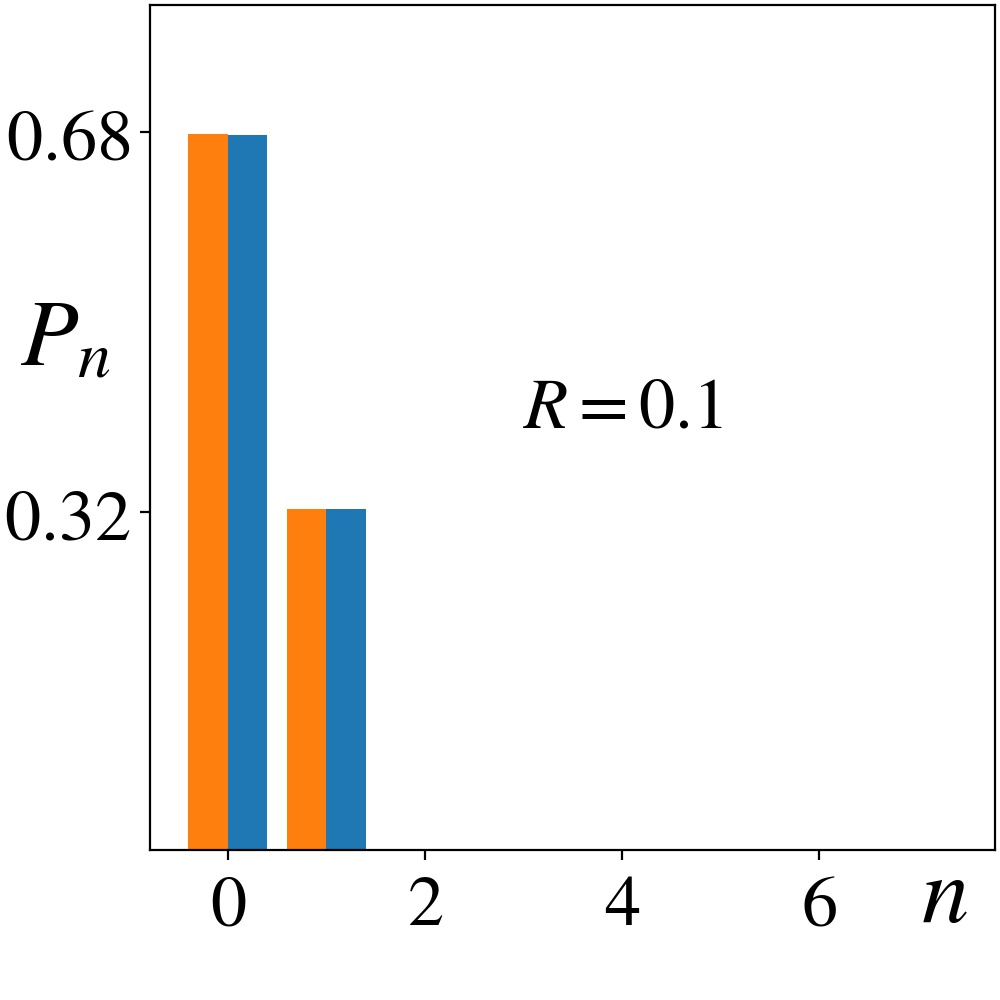}
    \caption{$T=0$}
    \label{}
    \end{subfigure}
    \hfill
    \begin{subfigure}[b]{0.31\linewidth}
    \includegraphics[width=1\linewidth]{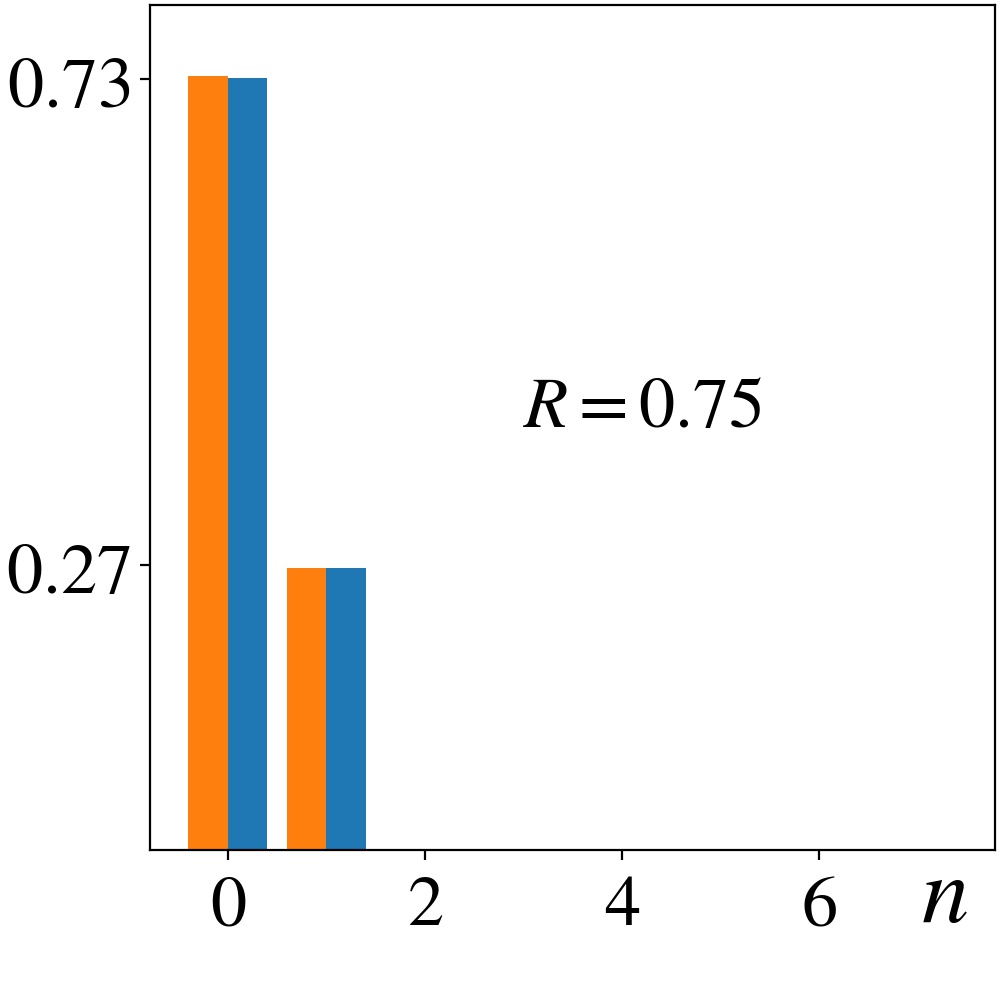}
    \caption{$T=0.3$}
    \label{}
    \end{subfigure}
    \hfill
    \begin{subfigure}[b]{0.31\linewidth}
    \includegraphics[width=1\linewidth]{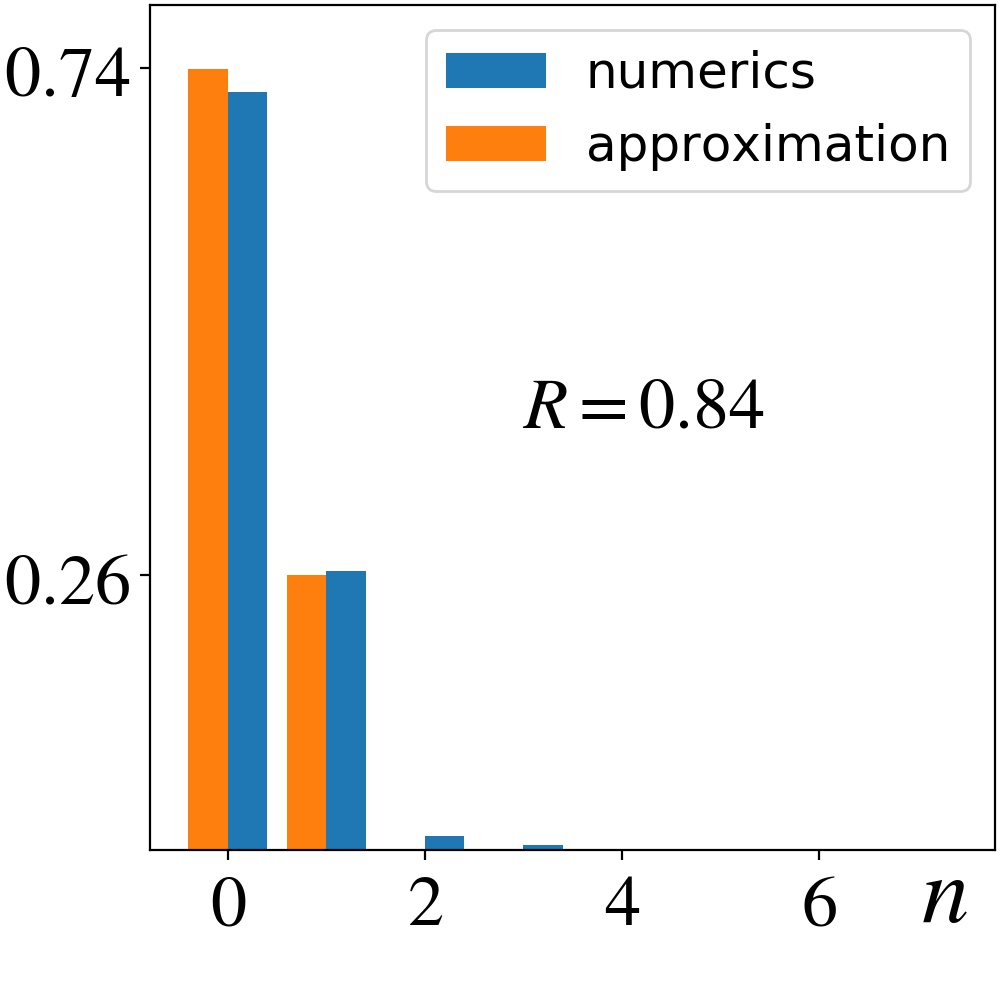}
    \caption{$T=0.5$}
    \label{}
    \end{subfigure}
    \caption{Wigner function cross sections and Fock-state distributions in the quantum limit, with $\gamma_2=10^5$,  $\kappa_1=10$, $\gamma_1=1$, and $\Delta_1=0.1$, for different temperatures. The approximate solutions~\eqref{Eq:W_peak} are shown in orange alongside exact numerical solutions in blue, showing good agreement. At low temperatures only the $\ket{0}$ and $\ket{1}$ states are populated, with nonzero occupation probabilities in the $\ket{2}$ and $\ket{3}$ states beginning to appear at $T=0.5$. Tick values on the vertical axes correspond to the approximate values in orange.}
    \label{fig:approx_and_num_cross}
\end{figure}

In the limit of $r\to 0$, as $\kappa_1$ increases or $\gamma_1$ decreases, the infinite-$\gamma_2$ oscillation amplitude tends to
\begin{equation}\label{Eq:W_peak_r=0}
    A^2
    = \frac{1}{2}\left(1-\frac{\Gamma_1}{K_1}\right)
    \xrightarrow[\ r \to 0\ ]{} \frac{1}{2}\left(1-e^{-\frac{\hbar\Delta_1}{k_\textrm{B}T}}\right),
\end{equation}
with an exponential dependence on temperature. This is demonstrated in Fig.~\ref{fig:saturation}(a) for $\gamma_2=10^5$, while Fig.~\ref{fig:saturation}(b)  shows essentially no temperature dependence of the amplitude in the classical limit with $\gamma_2=1$. A closer inspection of this exponential temperature dependence for $r=0.1$ is shown in Fig.~\ref{fig:approx_and_num_cross}, where we plot the Fock-state distributions and Wigner-function cross sections, for $T\leq0.5$. One can see how the increase in temperature gradually smears out the limit cycle. On one hand, as can be infered form Eq.~\eqref{Eq:quantum-rho}, the increase in $R$ causes an increase of the occupation probability $P_0$ of the $\ket{0}$ state, while at the same time increasing the neglected corrections of $\order{\exp{-2\hbar\omega/k_\textrm{B}T}}$ in the form of nonzero occupation probabilities of the $\ket{2}$ and $\ket{3}$ states.

\section{Conclusions}
\label{Sec:Conclusion}

We have studied a collection of master equations that yield quantum limit cycles in their steady-state dynamics. They all describe a simple harmonic oscillator, interacting with the environment through a combination of Lindblad operators, responsible for linear and nonlinear damping and energy injection, or pumping, in the form of single-phonon or double-phonon emission and absorption processes. We have established the correct correspondence between these quantum master equations and their classical counterparts, noting that the commonly used quantum model---which is symmetric under phase-space rotations and therefore always yields circular limit-cycles---is often mistaken to be the ``van der Pol (vdP) oscillator'', even though it actually corresponds to the classical ``Rayleigh-van der Pol (RvdP) oscillator''. We have also noted that, in all cases, the correspondence holds only for oscillations just above the bifurcation, namely, only to first order in the bifurcation parameter $\epsilon$.

We have analyzed a generalized version of the quantum RvdP limit cycle, applicable to a broad range of physical systems, such as nanomechanical oscillators, optical oscillators or lasers, electronic or superconducting oscillating circuits, and cold ions. We have obtained an exact analytical solution to the master equation in its steady state for arbitrary temperature, and considered its small-amplitude quantum limit---obtained by increasing the nonlinear damping rate---in some detail. A number of features emerge in this quantum regime, some of which were previously overlooked. Most important is the fact that, at $T=0$, the $\ket{1}$ state of the quantum oscillator is protected from nonlinear damping. One therefore still obtains limit-cycle oscillations, even with an infinite nonlinear damping rate, yet these quantum limit cycles are strongly affected by both the linear damping and the pumping rates, and are not universal as previously believed. We show that whereas in the classical regime it is only the difference between the linear pumping and the linear damping rates that affects the zero-temperature dynamics, in the quantum regime the ratio of the two rates plays a significant role as well, as they each contribute an independent source of spontaneous quantum processes. We have also described the effect of temperature in smearing out these nonclassical bifurcations.  

We have performed a numerical comparison between classical and quantum dynamics of the different models, showing perfect correspondence---where expected---between the quantum Wigner functions and the corresponding classical phase-space distributions. The agreement holds not only for the steady-state limit cycle dynamics, but for the transients as well, whereby an initial oscillating coherent state first quickly relaxes, or drifts, to the expected amplitude, and only then slowly diffuses around the limit cycle losing its initial phase. Deviations between the two occur in the quantum regime, as just mentioned above, where rather than decaying to zero as nonlinear damping increases, the quantum limit-cycle is protected, with its amplitude saturating at around zero-point motion, at a value that depends on the ratio of the linear pumping and damping rates. Deviations also occur far above the bifurcation, where the quantum and classical models no longer agree with each other. It should be emphasized that the Wigner functions that describe all the limit cycles are ``essentially classical'', developing no negative regions for any choice of parameters. This is a well-known property of the simple harmonic oscillator, which persists in these open systems, as long as the oscillator is linear~\cite{katz07, *katz08} and is uncoupled to additional oscillators or other degrees of freedom.

Our results should provide a firmer theoretical basis for ongoing studies of physical phenomena such as quantum entrainment and synchronization, and more generally, nonequilibrium nonlinear quantum dynamics involving self-sustained oscillators. We hope that our analytical results could be tested experimentally in the near future, where they should provide better tools with which to analyze the measured data.

\begin{acknowledgments}
The authors thank Moshe Goldstein and Haim Suchowski for their insightful comments on an earlier version of this manuscript. RL thanks Nadav Steiner for his contribution~\cite{steiner16} during early stages of this work, and Mark Dykman for fruitful and inspiring discussions. This research was supported by the U.S.-Israel Binational Science Foundation (BSF) under Grant No.~2012121. 
\end{acknowledgments}

\bibliography{RvdP}

\end{document}